\documentclass[twocolumn,aps,prd,preprintnumbers,showpacs,superscriptaddress,nofootinbib,amsmath,amssymb,floats,floatfix,showkeys,notitlepage]{revtex4}

\usepackage{lipsum}
\usepackage{graphicx}
\usepackage{subfigure}
\usepackage{palatino}
\usepackage{changes}
\usepackage{hyperref}
\hypersetup{colorlinks=true,linkcolor=blue,urlcolor=blue,citecolor=blue}
\usepackage[toc,page]{appendix}
\usepackage[normalem]{ulem}
\usepackage{adjustbox}
\usepackage{latexsym}
\usepackage{amsmath}
\usepackage{amssymb}
\usepackage{amsfonts}
\usepackage{times}
\usepackage{dcolumn}
\usepackage{bm}
\usepackage{tikz}
\usepackage{bigints}
\usepackage{array,tabularx,multirow}
\usepackage[tracking=true]{microtype}
\SetTracking{}{500}
\SetTracking{encoding={*}, shape=sc}{40}
\UseRawInputEncoding 
\allowdisplaybreaks

\newcommand{\be}{\begin{eqnarray}}
\newcommand{\ee}{\end{eqnarray}}

\renewcommand{\d}{\mbox{${\rm d}$}}

\newcommand{\bea}{\begin{eqnarray}}\newcommand{\eea}{\end{eqnarray}}
\newcommand{\brr}{\begin{array}}\newcommand{\err}{\end{array}}
\newcommand{\bit}{\begin{itemize}}\newcommand{\eit}{\end{itemize}}
\newcommand{\ben}{\begin{enumerate}}\newcommand{\een}{\end{enumerate}}

\newcommand{\ba}{\begin{array}}
\newcommand{\ea}{\end{array}}

\graphicspath{{./images/} }

\newcolumntype{M}[1]{>{\centering\arraybackslash}m{#1}}
\newcolumntype{N}{@{}m{0pt}@{}}

\def\d{{\mathrm d}}

\begin{document} \sloppy

\title{Testing dynamical torsion effects on the charged black hole's shadow, deflection angle and greybody with M87* and Sgr. A* from EHT}

\author{Reggie C. Pantig}
\email{rcpantig@mapua.edu.ph}
\affiliation{Physics Department, Map\'ua University, 658 Muralla St., Intramuros, Manila 1002, Philippines}
\author{Ali \"Ovg\"un}
\email{ali.ovgun@emu.edu.tr}
\affiliation{Physics Department, Eastern Mediterranean University, Famagusta, 99628 North
Cyprus via Mersin 10, Turkey.}

\begin{abstract}
Poincar\'e Gauge's theory of gravity is the most noteworthy alternative extension of general relativity that has a correspondence between spin and spacetime geometry. In this paper, we use Reissner-Nordstrom-de Sitter and anti-de Sitter solutions, where torsion $\tau$ is added as an independent field, to analyze the weak deflection angles $\hat{\alpha}$ of massive and null particles in finite distance regime. We then apply $\hat{\alpha}$ to determine the Einstein ring formation in M87* and Sgr. A* and determine that relative to Earth's location from these black holes, massive torsion effects can provide considerable deviation, while the cosmological constant's effect remains negligible. Furthermore, we also explore how the torsion parameter affects the shadow radius perceived by both static and co-moving (with cosmic expansion) observers in a Universe dominated by dark energy, matter, and radiation. Our findings indicate that torsion and cosmological constant parameters affect the shadow radius differently between observers in static and co-moving states. We also show how the torsion parameter affects the luminosity of the photonsphere by studying the shadow with infalling accretion. The calculation of the quasinormal modes, greybody bounds, and high-energy absorption cross-section are also affected by the torsion parameter considerably.
\end{abstract}

\date{\today}


\maketitle


\section{Introduction} \label{sec1}
Einstein's theory of General Relativity (GR) is the most beautiful of all existing physical theories which have passed all tests of time \cite{Einstein:1916vd}. After Einstein had shown that light should be bent by gravity in GR, Sir Arthur Eddington tested this assumption at the 1919 Total Eclipse of the Sun \cite{Dyson:1920cwa}. GR also predicts the most important compact object there is - black holes (BHs) whose features not only fascinated theoretical physicists, but also those in the field of experimental physics. Recently, the "chirp" due to the gravitational field produced by the coalescence of two BHs is observed in LIGO/VIRGO collaboration  \cite{LIGOScientific:2016aoc} and shadow images of the BHs at the center of M87 and Milky Way galaxies are captured by the Event Horizon Telescope (EHT) \cite{EventHorizonTelescope:2019dse,EventHorizonTelescope:2022xnr}.

The correspondence between the geometry of spacetime and the energy-momentum of matter is one of the most fascinating foundations of GR.  Energy-momentum tensor is the source of the gravitational field and can be written in terms of the curvature tensor. To extend the GR, one can consider the intrinsic angular momentum of matter as an alternative source of the interactions. The gauge approach in field theory has been studied by early works of Weyl \cite{weyl}, Cartan \cite{cartan1986manifolds}, Fock \cite{fock}, and later by Utiyama \cite{Utiyama}, Sciama \cite{sciama1962analogy}, and Kibble \cite{Kibble:1961ba}. For more details, one can check the review paper on the development of gauge gravity based on the Poincar\'e symmetry group \cite{Blagojevic:2012bc} which is one of the extensions of the GR known as Poincar\'e Gauge theory of gravity (PGTG). PGTG has an asymmetric affine connection that stands within a Riemann-Cartan manifold (combine the curvature with torsion) which provides a well-defined correspondence between spin and the geometry of spacetime \cite{Hehltorsion}. In theory, the source of the torsion is a non-trivial spin density tensor. Recently, Cembranos and Valcarcel have found a new exact solution charged black hole solution with massive torsion in a PGTG containing higher-order corrections quadratic in the curvature tensor \cite{Cembranos:2017pcs}.

In this paper, we aim to explore the torsion effects in combination with the cosmological constant. To our knowledge, very few studies were carried out concerning BHs with a torsion parameter. The torsion effects on black holes can have different physical interpretations depending on how it was derived from the Einstein field equation \cite{Blagojevic:2021pqp,Obukhov:2020hlp,Bahamonde:2020fnq}. For example, in Ref. \cite{he2022}, the torsion charge's effect on the black hole shadow was studied and explored. Torsion effects are also important to information \cite{gkigkitzis2019}. Long-ranged limit due to the black hole in torsion bigravity was also considered \cite{nikiforova2020}. When it comes to the cosmological constant parameter $\Lambda$, a well-known black hole spacetime that incorporates the de Sitter spacetime ($+\Lambda$) is called the Kottler metric \cite{kottler1918}. Driven by the cosmological constant, the black hole shadow with such an influence was studied in \cite{Perlick:2018,bisnovatyi2018}. Different treatments can also be found in Ref. \cite{firouzjaee2019}. Recently, the formalism that involves $\Lambda$ to the shadow perceived by a co-moving observer was applied in McVittie spacetime \cite{Tsupko2020}. Interestingly, it allowed the examination of the approximate angular radius of the shadow under the effect of which type of Universe we live in (dark energy, matter, or radiation-dominated). The effect of the cosmological constant under different field theories was also considered by other authors (See Ref. \cite{maluf2021}). We observe that they lack the anti-de Sitter analysis, and we aim in this paper to provide an exploration of it.

The black hole shadow is so important in the study of BH physics since it can tell us the mass and size of the BH, how the spacetime is distorted near it, and how matter and radiation fall near the event horizon - a region of no escape. It was Synge in 1966 who first theoretically studied the shadow of the simplest solution to the Einstein field equation - the Schwarzschild metric \cite{Synge1966}, and in 1979, Luminet proved the expression for the angular radius of the shadow \cite{Luminet1979}. Since then, the formalism was applied to a number of static and spherically symmetric black hole spacetime models to explore how certain parameters induce deviations from the known black hole shadow size \cite{Herdeiro:2021lwl,Cunha:2019hzj,Cunha:2019ikd,Cunha:2018acu,Cunha:2016wzk,Vincent:2016sjq,Afrin:2021imp,  Jha:2021bue,  Khodadi:2021gbc, Khodadi:2020jij, Kumar:2018ple,  Kumar:2020owy,  Zeng:2020dco,  He:2022yse, Dokuchaev:2020wqk,Vagnozzi:2022moj, Roy:2021uye, Vagnozzi:2019apd, Allahyari:2019jqz,Bambi:2019tjh,Meng:2022kjs,Chen:2022lct,Chen:2022nbb,Wang:2022kvg,Bronzwaer:2021lzo,Falcke:1999pj,Atamurotov:2013sca,Abdujabbarov:2015xqa,Wei:2019pjf,Wei:2018xks,Abdolrahimi:2015rua,Adair:2020vso,Abdolrahimi:2015kma,Konoplya:2020bxa,Konoplya:2019sns,Konoplya:2019xmn,Shaikh:2018lcc,Shaikh:2019fpu,Rahaman:2021web,Belhaj:2020rdb,Belhaj:2020okh,Belhaj:2021rae,Chakhchi:2022fls,Perlick:2018iye,Perlick:2021aok,Pantig2020b,Sokoliuk:2022owk}. Pantig et al. \cite{PANTIG2022168722,Pantig:2022toh} have studied the possible effects of dark matter on a Schwarzschild black hole with the correction of extended uncertainty principle on black hole shadow and also the effect of dark matter on the weak deflection angle by black holes at the galactic center. \"Ovg\"un has shown \cite{Ovgun:2021ttv} that a confining charge gives a significant contribution to the shadow of the black hole with confining electric potential in scalar-tensor description of regularized 4-dimensional Einstein-Gauss-Bonnet gravity. Okyay and \"Ovg\"un have studied the nonlinear electrodynamics effects on the black hole shadow \cite{Okyay:2021nnh}. The black hole shadow in symmergent gravity has been investigated in \cite{Cimdiker:2021cpz}. Moreover, Pantig and \"Ovg\"un have studied the effect of Dehnen dark matter halo profile on a black hole in an ultra-faint dwarf galaxy \cite{Pantig:2022whj}. Slowly rotating Kerr-like black hole in bumblebee gravity has been studied by Kuang and \"Ovg\"un in \cite{Kuang:2022xjp}. Authors provide some constraints on parameters of NLE using the observations of M87* and Sgr. A* from EHT \cite{Uniyal:2022vdu}. \"Ovg\"un et al. have shown the shadow cast and deflection angle of Kerr-Newman-Kasuya spacetime \cite{Ovgun:2018tua}, and also \cite{Ovgun:2020gjz} studied the shadow of a black hole in Einstein-Cartan-Kibble-Sciama  theory of gravity.

Gravitational lensing is a powerful tool to test gravity \cite{Virbhadra:1999nm,Bozza:2002zj,Hasse:2001by,Perlick:2003vg,He:2020eah}. There are several commonly used calculation methods of gravitational lensing and the most used method for the investigation of the deflection angle in weak field limits is the one proposed by Gibbons and Werner as an alternative method for the asymptotically flat black holes \cite{Gibbons:2008rj}. First, they obtained the optical metric of the black hole spacetime and then applied the Gauss-Bonnet theorem (GBT). Then, Werner extended it for stationary spacetimes and calculated the deflection angle of the Kerr black hole in weak fields \cite{Werner_2012}. Nowadays, there are many applications of this method on black holes, wormholes, and other spacetimes \cite{Ovgun:2018fnk,Ovgun:2019wej,Ovgun:2018oxk,Javed:2019kon,Javed:2019rrg,Javed:2019ynm,Javed:2020lsg,Javed:2019qyg,Ovgun:2018fte,Javed:2019jag,Ishihara:2016vdc,Takizawa:2020egm,Ono:2019hkw,Ishihara:2016sfv,Ono:2017pie,Li:2020dln,Li:2020wvn,Pantig:2020odu}. In the framework of dynamical torsion, our one of the aim is to investigate the effect of dynamical torsion on the weak deflection angle of both massive and null particles by applying the Gauss-bonnet theorem.

Here is how the paper is organized: In Sect. \ref{sec2} we introduce the metric of the black hole with massive torsion. In Sect. \ref{sec3}, we study how the torsion parameter affects the deflection angle both for massive and null particles. Sect. \ref{sec4}-\ref{sec5} is dedicated to the analysis of the null geodesic and shadow cast for static and co-moving observers. The shadow with infalling spherical accretion is explored in Sect. \ref{sec6}. The eikonal (geometric optics) limit of quasinormal modes will be analyzed in Sect. \ref{sec7}. We also explored the greybody factors and high energy absorption cross-section using the Sinc approximation in Sect. \ref{sec8}. We formulate our conclusion in Sect. \ref{sec9} and state future research directions. Throughout the paper, we will use the natural units $G=c=1$ and the signature ($-,+,+,+$).

\section{Black holes in Quadratic Poincar\'e gauge gravity model with massive torsion} \label{sec2}
Classical solutions of the Poincar\'e gauge help us to better understand the influence of torsion on gravitational dynamics. In this section, we focus our attention on the static and spherically
symmetric solution found by Cembranos and Valcarcel \cite{Cembranos:2017pcs}, in which the electric charge of
the standard RN metric is “imitated” by a torsion-induced parameter $\kappa$.

The general gravitational action related to the original Poincar\'e gauge model with three independent quadratic scalar invariants of torsion (representing the mass terms) is given as \cite{Cembranos:2017pcs}:
\begin{align}
    &S=\frac{1}{16 \pi} \int d^{4} x \sqrt{-g}\Bigg[\mathcal{L}_{m}-\tilde{R}-\frac{1}{4}\left(d_{1}+d_{2}\right) \tilde{R}^{2} \nonumber\\
    &-\frac{1}{4}\left(d_{1}+d_{2}+4 c_{1}+2 c_{2}\right) \tilde{R}_{\lambda \rho \mu \nu} \tilde{R}^{\mu \nu \lambda \rho}+c_{1} \tilde{R}_{\lambda \rho \mu \nu} \tilde{R}^{\lambda \rho \mu \nu} \nonumber\\
    &+c_{2} \tilde{R}_{\lambda \rho \mu \nu} \tilde{R}^{\lambda \mu \rho v}+d_{1} \tilde{R}_{\mu \nu} \tilde{R}^{\mu \nu}+d_{2} \tilde{R}_{\mu \nu} \tilde{R}^{v \mu} \nonumber\\
    &+\alpha T_{\lambda \mu \nu} T^{\lambda \mu \nu}+\beta T_{\lambda \mu \nu} T^{\mu \lambda \nu}+\gamma T^{\lambda}{ }_{\lambda v} T_{\mu}^{\mu}\Bigg],
\end{align}
where $\alpha, \beta, \gamma, c_{1}, c_{2}, d_{1},$ and $d_{2}$ are constant parameters and in the absence of matter, i.e. $\mathcal{L}_{m} = 0$.

It is noted that tilde stands for magnitudes with torsion and without tilde for torsion-free quantities.
 
%

Note that the curvature tensor is

\begin{equation}\tilde{R}^{\lambda}\,_{\rho \mu \nu}=\partial_{\mu}\tilde{\Gamma}^{\lambda}\,_{\rho \nu}-\partial_{\nu}\tilde{\Gamma}^{\lambda}\,_{\rho \mu}+\tilde{\Gamma}^{\lambda}\,_{\sigma \mu}\tilde{\Gamma}^{\sigma}\,_{\rho \nu}-\tilde{\Gamma}^{\lambda}\,_{\sigma \nu}\tilde{\Gamma}^{\sigma}\,_{\rho \mu}\,,\end{equation}

where the affine connection (with torsion-free Levi-Civita connection and a contortion component) is: 
\begin{equation}
\tilde{\Gamma}^{\lambda}\,_{\mu \nu} = {\Gamma}^{\lambda}\,_{\mu \nu} + K^{\lambda}\,_{\mu \nu}\,.
\end{equation}

Moreover, the field strength tensors are derived from the gauge connection of the Poincar\'e group $ISO (1,3)$:

\begin{equation}A_{\mu}=e^{a}\,_{\mu}P_{a}+\omega^{a b}\,_{\mu}J_{a b}\,,\end{equation}
where $e^{a}\,_{\mu}$ stands for the vierbein field:

\begin{equation}
g_{\mu \nu}=e^{a}\,_{\mu}\,e^{b}\,_{\nu}\,\eta_{a b}\,,
\end{equation}
and $\omega^{a b}\,_{\mu}$ the spin connection of a Riemann-Cartan manifold
\begin{equation}
\omega^{a b}\,_{\mu}=e^{a}\,_{\lambda}\,e^{b \rho}\,\tilde{\Gamma}^{\lambda}\,_{\rho \mu}+e^{a}\,_{\lambda}\,\partial_{\mu}\,e^{b \lambda}\,\
\end{equation}
and corresponding  ISO(1, 3) gauge field strength tensors are written as:

\begin{equation}F^{a}\,_{\mu \nu}=e^{a}\,_{\lambda}\,T^{\lambda}\,_{\nu \mu}\,,\end{equation}

\begin{equation}F^{a b}\,_{\mu \nu}=e^{a}\,_{\lambda}e^{b}\,_{\rho}\,\tilde{R}^{\lambda \rho}\,_{\mu \nu}\,,\end{equation}

Afterward one can write torsion into three components: a trace vector $T_{\mu}$, an axial vector $S_{\mu}$ and a traceless and also pseudotraceless tensor $q^{\lambda}\,_{\mu \nu}$ as follows:

\begin{equation}
T^{\lambda}\,_{\mu \nu}=\frac{1}{3}\left(\delta^{\lambda}\,_{\nu}T_{\mu}-\delta^{\lambda}\,_{\mu}T_{\nu}\right)+\frac{1}{6}\,g^{\lambda \rho}\varepsilon\,_{\rho \sigma \mu \nu}S^{\sigma}+q^{\lambda}\,_{\mu \nu}\,,
\end{equation}
with $\varepsilon\,_{\rho \sigma \mu \nu}$.

In this paper, we focus our attention on the static and spherically symmetric solution found by Cembranos and Valcarcel \cite{Cembranos:2017pcs}, in which the electric charge of the standard RN metric is “imitated” by a massive torsion-induced parameter in quadratic Poincar\'e gauge gravity model as follows:
\begin{equation} \label{metric}
    d s^{2}=-f(r) d t^{2}+\frac{d r^{2}}{f(r)}+r^{2}\left(d \theta^{2}+\sin ^{2} \theta d \varphi^{2}\right),
\end{equation}
where the metric lapse function reads
\begin{equation} \label{lapsefunc}
    f(r)=1-\frac{2 m}{r}+\frac{\tau^{2}+Q^2}{r^{2}} - \frac{\Lambda r^2}{3},
\end{equation}
where, the new contribution is proportional to the square of the new parameter $\tau$ (for simplicity, we let the constant parameter $d_1$ to be absorbed by the torsion parameter $\kappa$ and write $\tau = \sqrt{d_1} \kappa$). Note that the parameter of $\kappa$ determines the intensity of the strength tensor corresponding to the torsion \cite{Cembranos:2017pcs,Cembranos:2016gdt}.

In addition, we wrote the combined effect of the electric and magnetic charges as $Q^2 = q_\text{e}^2 + q_\text{m}^2$. As one will also notice, the metric is generalized to include the cosmological constant.
\begin{figure*}
    \centering
    \includegraphics[width=0.48\linewidth]{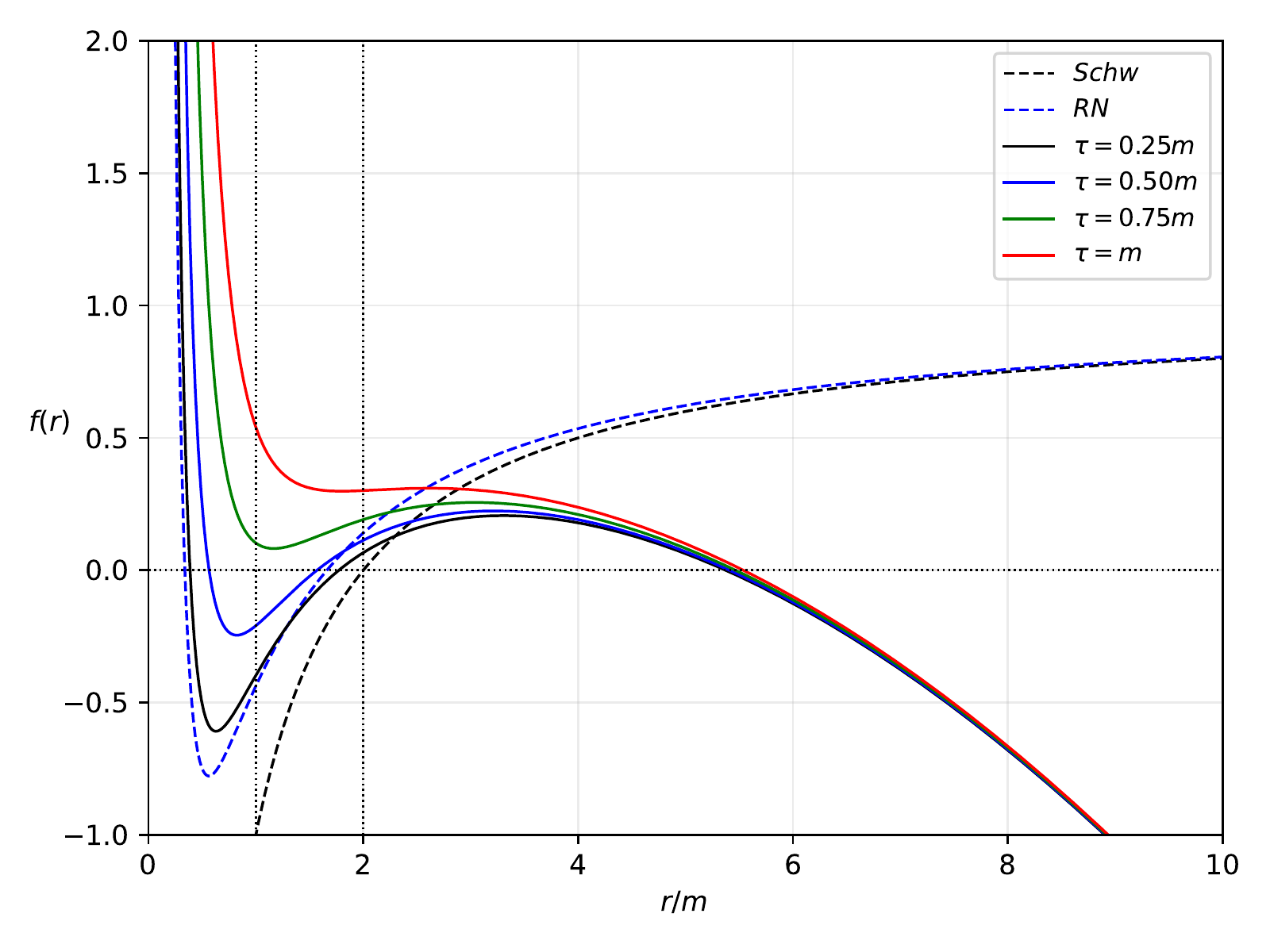}
    \includegraphics[width=0.48\linewidth]{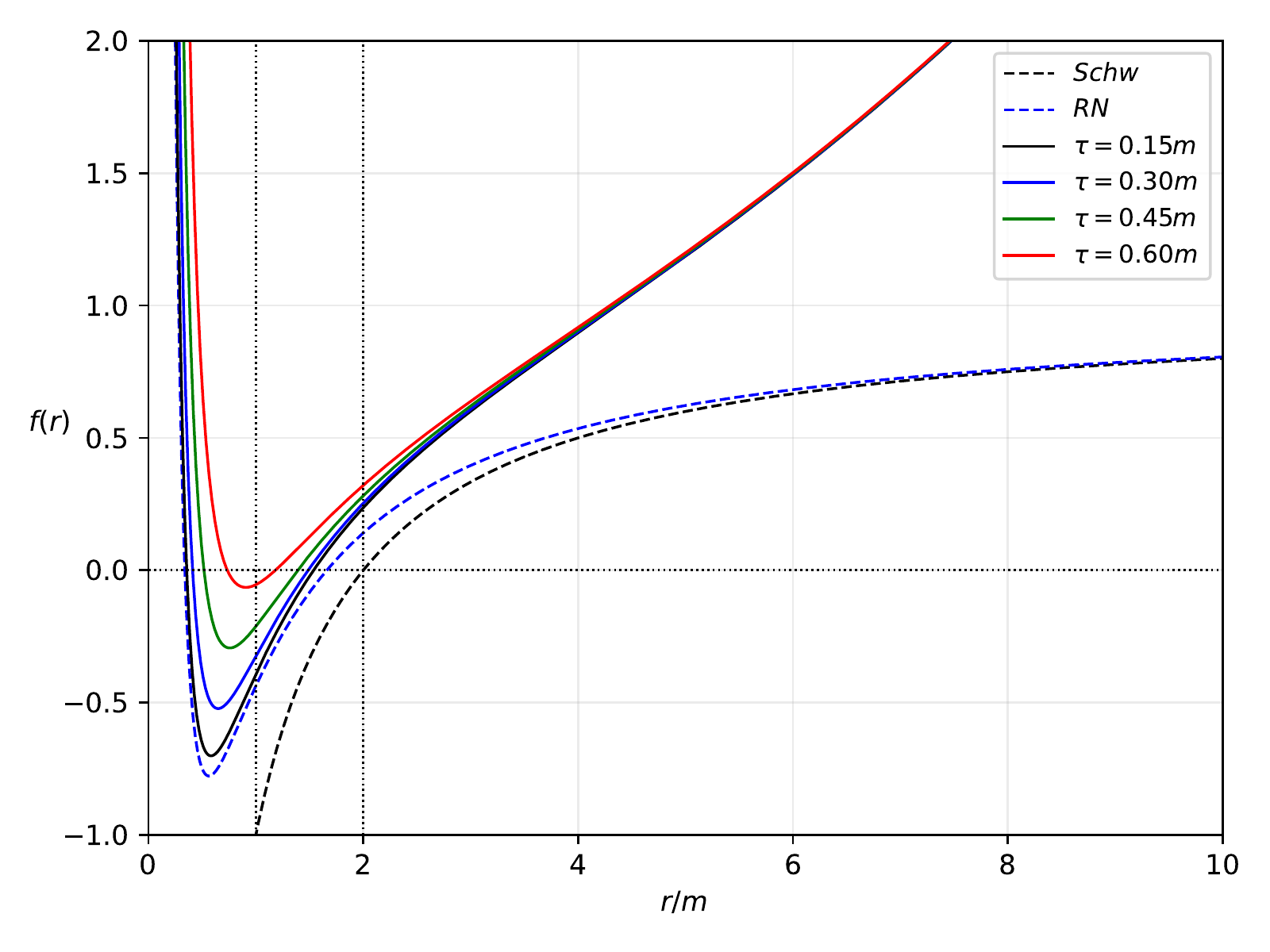}
    \caption{The lapse function $f(r)$ behavior in response to different values of the torsion parameter $\tau$. Here, we take $m=1$, $Q = 0.75m$, and scaled the cosmological constant to $\Lambda = 0.0675 \text{ m}^{-2}$. The left and right figures correspond to de Sitter (dS) case $+\Lambda$, and anti-de Sitter (AdS) case $-\Lambda$ respectively.}
    \label{lapse}
\end{figure*}

The outer horizon is located at the larger root of $f(r)=0$ and this can be visualized through Fig. \ref{lapse}. We can also see in the figure how the dS and AdS cases are different in terms of null boundary formation. Continuing, one can calculate the following 4-velocity via
\begin{equation}
    u=u^{t} \partial_{t},
\end{equation}
and then the normalization condition is satisfied by
\begin{equation}
    1=u^{\mu} u_{\mu}
\end{equation}
where $u^{t}=\frac{1}{\sqrt{g_{t t}}}$. Since the metric components are functions of $r$ and $\theta$, particle acceleration $a_{p}^{\mu}$ is obtained from 
$$
a^{\mu}=-g^{\mu v} \partial_{v} \ln u^{t} .
$$
Surface gravity $(\kappa)$ is defined as follows
$$
dl_\text{prop}=\lim _{r \rightarrow r_\text{h}} \frac{\sqrt{a^{\mu} a_{\mu}}}{u^{t}}
$$
Hence, the surface gravity is obtained as follows to find the black hole temperature:
\begin{equation}
    \kappa:=\left.\frac{1}{2} \partial_{r} f^{2}\right|_{r_{+}}, \quad T:=\frac{\kappa}{2 \pi}.
\end{equation}
Then, a simple calculation of the horizon area yields
\begin{equation}
    A_\text{BH}=\int_{0}^{2 \pi} d \varphi \int_{0}^{\pi} \sqrt{-g} d \theta=4 \pi r_\text{h}.
\end{equation}
Hence, the entropy of a black hole becomes
\begin{equation}
    S_\text{BH}=\frac{A_\text{BH}}{4}=\pi r_\text{h}.
\end{equation}

\section{Weak Deflection Angle using Gauss-bonnet theorem} \label{sec3}
In this section, we present the calculation of the deflection angle of a massive particle around a charged black hole with dynamical torsion and a cosmological constant. For a static, spherically symmetric (SSS) spacetime,
\begin{align}
    ds^2&=g_{\mu \nu}dx^{\mu}dx^{\nu} \nonumber \\
    &=-A(r)dt^2+B(r)dr^2+C(r)d\Omega^2
\end{align}
the Jacobi metric reads
\begin{align}
    dl^2&=g_{ij}dx^{i}dx^{j} \nonumber \\
    &=[E^2-m^2A(r)]\left(\frac{B(r)}{A(r)}dr^2+\frac{C(r)}{A(r)}d\Omega^2\right),
\end{align}
where $d\Omega^2=d\theta^2+r^2\sin^2\theta$ is the line element of the unit two-spheres, and $E$ is the energy of the massive particle defined by
\begin{equation} \label{en}
    E = \frac{\mu}{\sqrt{1-v^2}},
\end{equation}
where $v$ is the particle's velocity. In the equatorial plane then where $\theta = \pi/2$, the Jacobi metric is reduced to
\begin{equation} \label{eJac}
    dl^2=(E^2-\mu^2A(r))\left(\frac{B(r)}{A(r)}dr^2+\frac{C(r)}{A(r)}d\phi^2\right)
\end{equation}
without loss of generality. The determinant of the Jacobi metric above can also be easily calculated as
\begin{equation}
    g=\frac{B(r)C(r)}{A(r)^2}(E^2-\mu^2 A(r))^2.
\end{equation}
Next, we will use these equations to find the weak deflection angle using the Gauss-Bonnet theorem (GBT), originally stated as
\begin{equation} \label{eGBT}
    \iint_DKdS+\sum\limits_{a=1}^N \int_{\partial D_{a}} \kappa_{\text{g}} d\ell+ \sum\limits_{a=1}^N \theta_{a} = 2\pi\chi(D).
\end{equation}
Here, the Gaussian curvature $K$ describing the domain $D$ is a freely orientable $2D$ curved surface $S$ with infinitesimal area element $dS$. The boundary of $D$ are given by $\partial D_{\text{a}}$ (a=$1,2,..,N$), and the geodesic curvature $\kappa_{\text{g}}$ is integrated over the path $d\ell$ along a positive convention. Also, $\theta_\text{a}$ is the jump angle, which $\chi(D)$ is the Euler characteristic where in our case is equal to $1$ since $D$ is in a non-singular region.

It was shown by \cite{Li:2020wvn} that in a SSS spacetime with no asymptotic flatness, Eq. \eqref{eGBT} can be written as
\begin{equation} \label{eIshi}
    \hat{\alpha} = \iint_{D}KdS + \phi_{\text{RS}},
\end{equation}
where $r_\text{co}$ is the radius of the particle's circular orbit, and S and R are the radial positions of the source and receiver respectively. These are the integration domains, and we note that the infinitesimal curve surface $dS$ is given by
\begin{equation}
    dS = \sqrt{g}drd\phi.
\end{equation}
Furthermore, $\phi_\text{RS}$ is the coordinate position angle between the source and the receiver defined as $\phi_\text{RS} = \phi_\text{R}-\phi_\text{S}$, which can be found through the iterative solution of
\begin{align}
    F(u) &= \left(\frac{du}{d\phi}\right)^2  \nonumber\\
    &= \frac{C(u)^2u^4}{A(u)B(u)}\Bigg[\left(\frac{E}{J}\right)^2-A(u)\left(\frac{1}{J^2}+\frac{1}{C(u)}\right)\Bigg],
\end{align}
where we have used the substitution $r = 1/u$ and the angular momentum of the massive particle
\begin{equation}
    J = \frac{\mu v b}{\sqrt{1-v^2}},
\end{equation}
where $b$ is the impact parameter. With Eq. \eqref{metric}, we find
\begin{align}
    &F(u) = \frac{E^2-1}{J^2}-u^2-u^2\left(\frac{1}{J^2}+u^2\right)(\tau^2+Q^2) \nonumber\\
    &+\left(\frac{1}{J^2 u^2}+1\right)\frac{\Lambda}{3} + \left(\frac{1}{J^2}+u^2\right)2mu.
\end{align}
Doing the iteration method, we find
\begin{equation} \label{orb}
    u(\phi) = \frac{\sin(\phi)}{b}+\frac{1+v^2\cos^2(\phi)}{b^2v^2}m - \frac{(\tau^2+Q^2)}{2v^2 b^3} + \frac{\Lambda b}{6v^2}.
\end{equation}

In terms of affine connections, the Gaussian curvature $K$ is defined as
\begin{align}
    K&=\frac{1}{\sqrt{g}}\left[\frac{\partial}{\partial\phi}\left(\frac{\sqrt{g}}{g_{rr}}\Gamma_{rr}^{\phi}\right)-\frac{\partial}{\partial r}\left(\frac{\sqrt{g}}{g_{rr}}\Gamma_{r\phi}^{\phi}\right)\right] \nonumber \\
    &=-\frac{1}{\sqrt{g}}\left[\frac{\partial}{\partial r}\left(\frac{\sqrt{g}}{g_{rr}}\Gamma_{r\phi}^{\phi}\right)\right]
\end{align}
since $\Gamma_{rr}^{\phi} = 0$ for Eq. \eqref{eJac}. If in a certain spacetime there is an analytical solution for the $r_\text{co}$, then we have the relation
\begin{equation} \label{gct}
    \int_{r_\text{co}}^{r(\phi)} K\sqrt{g}dr = -\frac{A(r)\left(E^{2}-A(r)\right)C'-E^{2}C(r)A(r)'}{2A(r)\left(E^{2}-A(r)\right)\sqrt{B(r)C(r)}}\bigg|_{r = r(\phi)}
\end{equation}
since
\begin{equation}
    \left[\int K\sqrt{g}dr\right]\bigg|_{r=r_\text{co}} = 0.
\end{equation}
The prime denotes differentiation with respect to $r$. The weak deflection angle for non-asymptotic spacetime is then \cite{Li:2020wvn},
\begin{align} \label{ewda}
    \hat{\alpha} &= \int^{\phi_\text{R}}_{\phi_\text{S}} \left[-\frac{A(r)\left(E^{2}-A(r)\right)C'-E^{2}C(r)A(r)'}{2A(r)\left(E^{2}-A(r)\right)\sqrt{B(r)C(r)}}\bigg|_{r = r(\phi)}\right] \times \nonumber\\
    & d\phi + \phi_\text{RS}.
\end{align}
Using Eq. \eqref{orb} in Eq. \eqref{gct}, we find
\begin{align} \label{gct2}
    &\left[\int K\sqrt{g}dr\right]\bigg|_{r=r_\phi} = -\frac{\left(2E^{2}-1\right)m(\cos(\phi_\text{R})-\cos(\phi_\text{S}))}{\left(E^{2}-1\right)b} \nonumber\\
    &-\frac{\left(3E^{2}-1\right)((\tau^2+Q^2))\left[\phi_{RS}-\frac{(\sin(2\phi_\text{R})-\sin(2\phi_\text{S})}{2}\right]}{4\left(E^{2}-1\right)b^{2}} \nonumber\\
    &+\frac{\left(1+E^{2}\right)b^{2}\Lambda(\cot(\phi_\text{R})-\cot(\phi_\text{S}))}{6\left(E^{2}-1\right)} -\phi_{RS} \nonumber\\
    &+ \mathcal{O}[m(\tau^2+Q^2),m\Lambda,\Lambda (\tau^2+Q^2),m(\tau^2+Q^2)\Lambda],
\end{align}
where we retained $E$ to avoid the clutter caused by the velocity $v$, and chose only the dominant terms. To find the expression for $\phi$, we use Eq. \eqref{orb} and solve it. For the source and receiver respectively, we find
\begin{align} \label{s}
    &\phi_\text{S} =\arcsin(bu)+\frac{m\left[v^{2}\left(b^{2}u^{2}-1\right]-1\right)}{bv^{2}\sqrt{1-b^{2}u^{2}}} \nonumber\\
    &+\frac{(\tau^2+Q^2)}{2b^{2}v^{2}\sqrt{1-b^{2}u^{2}}}-\frac{b^{2}\Lambda}{3\sqrt{2}v^{2}\sqrt{2-2b^{2}u^{2}}} \nonumber \\
    &+ \mathcal{O}[m(\tau^2+Q^2),m\Lambda,\Lambda (\tau^2+Q^2),m(\tau^2+Q^2)\Lambda],
\end{align}
\begin{align} \label{r}
    &\phi_\text{R} =\pi -\arcsin(bu)-\frac{m\left[v^{2}\left(b^{2}u^{2}-1\right]-1\right)}{bv^{2}\sqrt{1-b^{2}u^{2}}} \nonumber\\
    &-\frac{(\tau^2+Q^2)}{2b^{2}v^{2}\sqrt{1-b^{2}u^{2}}}+\frac{b^{2}\Lambda}{3\sqrt{2}v^{2}\sqrt{2-2b^{2}u^{2}}} \nonumber \\
    &+ \mathcal{O}[m(\tau^2+Q^2),m\Lambda,\Lambda (\tau^2+Q^2),m(\tau^2+Q^2)\Lambda].
\end{align}
Based on this, we write $\phi_\text{RS} = \pi - 2\phi_\text{S}$. Now, we take note of the following relations:
\begin{align}
    \cos(\pi-\phi_\text{S})&=-\cos(\phi_\text{S}), \nonumber\\
    \cot(\pi-\phi_\text{S})&=-\cot(\phi_\text{S}), \nonumber\\
    \sin(\pi-\phi_\text{S})&=\sin(\phi_\text{S}).
\end{align}
The last property makes the sine terms in Eq. \eqref{gct2} cancel. We find $\cos(\phi_\text{S})$ as
\begin{align} \label{cs}
    &\cos(\phi_\text{S}) = \sqrt{1-b^{2}u^{2}}-\frac{mu\left[v^{2}\left(b^{2}u^{2}-1\right)-1\right]}{\sqrt{v^{2}\left(1-b^{2}u^{2}\right)}} \nonumber\\
    &-\frac{(\tau^2+Q^2)u}{\sqrt{2}\sqrt{bv^{2}\left(1-b^{2}u^{2}\right)}}+\frac{b^{3}u\Lambda}{6\sqrt{2}v^{2}\sqrt{1-b^{2}u^{2}}} \nonumber \\
    &+ \mathcal{O}[m(\tau^2+Q^2),m\Lambda,\Lambda (\tau^2+Q^2),m(\tau^2+Q^2)\Lambda],
\end{align}
and $\cot(\phi_\text{S})$ as
\begin{align} \label{cr}
    &\cot(\phi_\text{S}) = \frac{\sqrt{1-b^{2}u^{2}}}{bu}+\frac{m\left[v^{2}(-b^{2}u^{2}+1)+1\right]}{b^{3}u^{2}v^{2}\sqrt{1-b^{2}u^{2}}} \nonumber\\
    &-\frac{(\tau^2+Q^2)}{2b^{4}u^{2}v^{2}\sqrt{1-b^{2}u^{2}}}+\frac{\Lambda}{6\sqrt{2}u^{2}v^{2}\sqrt{1-b^{2}u^{2}}} \nonumber \\
    &+ \mathcal{O}[m(\tau^2+Q^2),m\Lambda,\Lambda (\tau^2+Q^2),m(\tau^2+Q^2)\Lambda].
\end{align}
By plugging Eqs. \eqref{s}-\eqref{cr} in Eq. \eqref{ewda}, we finally obtain (after also using Eq. \eqref{en})
\begin{align} \label{ewda_exact}
    &\hat{\alpha} \sim \frac{m\left(v^{2}+1\right)}{bv^{2}}\left(\sqrt{1-b^{2}u_\text{R}^{2}}+\sqrt{1-b^{2}u_\text{S}^{2}}\right) \nonumber\\
    &-\frac{(\tau^2+Q^2)\left(v^{2}+2\right)}{4b^{2}v^{2}}\left[\pi-(\arcsin(bu_\text{R})+\arcsin(bu_\text{S}))\right] \nonumber\\
    &+\frac{b\Lambda\left(v^{2}-2\right)}{6v^{2}}\left(\frac{\sqrt{1-b^{2}u_\text{R}^{2}}}{u_\text{R}}+\frac{\sqrt{1-b^{2}u_\text{S}^{2}}}{u_\text{S}}\right) \nonumber\\
    &+\mathcal{O}[m(\tau^2+Q^2),m\Lambda,\Lambda (\tau^2+Q^2),m(\tau^2+Q^2)\Lambda],
\end{align}
which also involves the finite distance $u_\text{S}$ and $u_\text{R}$. We can approximate this form by making $u$ to be nearly zero and in such a case, $b^2u^2 \sim 0$ and we have
\begin{align} \label{ewda_approx}
    &\hat{\alpha} \sim \frac{2m\left(v^{2}+1\right)}{bv^{2}}-\frac{(\tau^2+Q^2) \pi\left(v^{2}+2\right)}{4b^{2}v^{2}} \nonumber\\
    &+\frac{b\Lambda\left(v^{2}-2\right)}{6v^{2}}\left(\frac{1}{u_\text{R}}+\frac{1}{u_\text{S}}\right) \nonumber\\
    &+\mathcal{O}[m(\tau^2+Q^2),m\Lambda,\Lambda (\tau^2+Q^2),m(\tau^2+Q^2)\Lambda].
\end{align}
Note also that for the case of null particles, where $v = 1$, the above reduces to a known expression:
\begin{align} \label{wdafin}
    &\hat{\alpha} \sim \frac{4m}{b}-\frac{3\pi(\tau^2+Q^2) }{4b^{2}}-\frac{b\Lambda}{6}\left(\frac{1}{u_\text{R}}+\frac{1}{u_\text{S}}\right) \nonumber\\
    &+\mathcal{O}[m(\tau^2+Q^2),m\Lambda,\Lambda (\tau^2+Q^2),m(\tau^2+Q^2)\Lambda].
\end{align}
The above is in agreement with the result in Ref. \cite{Ishihara2016}.
\begin{figure*}
    \centering
    \includegraphics[width=0.48\textwidth]{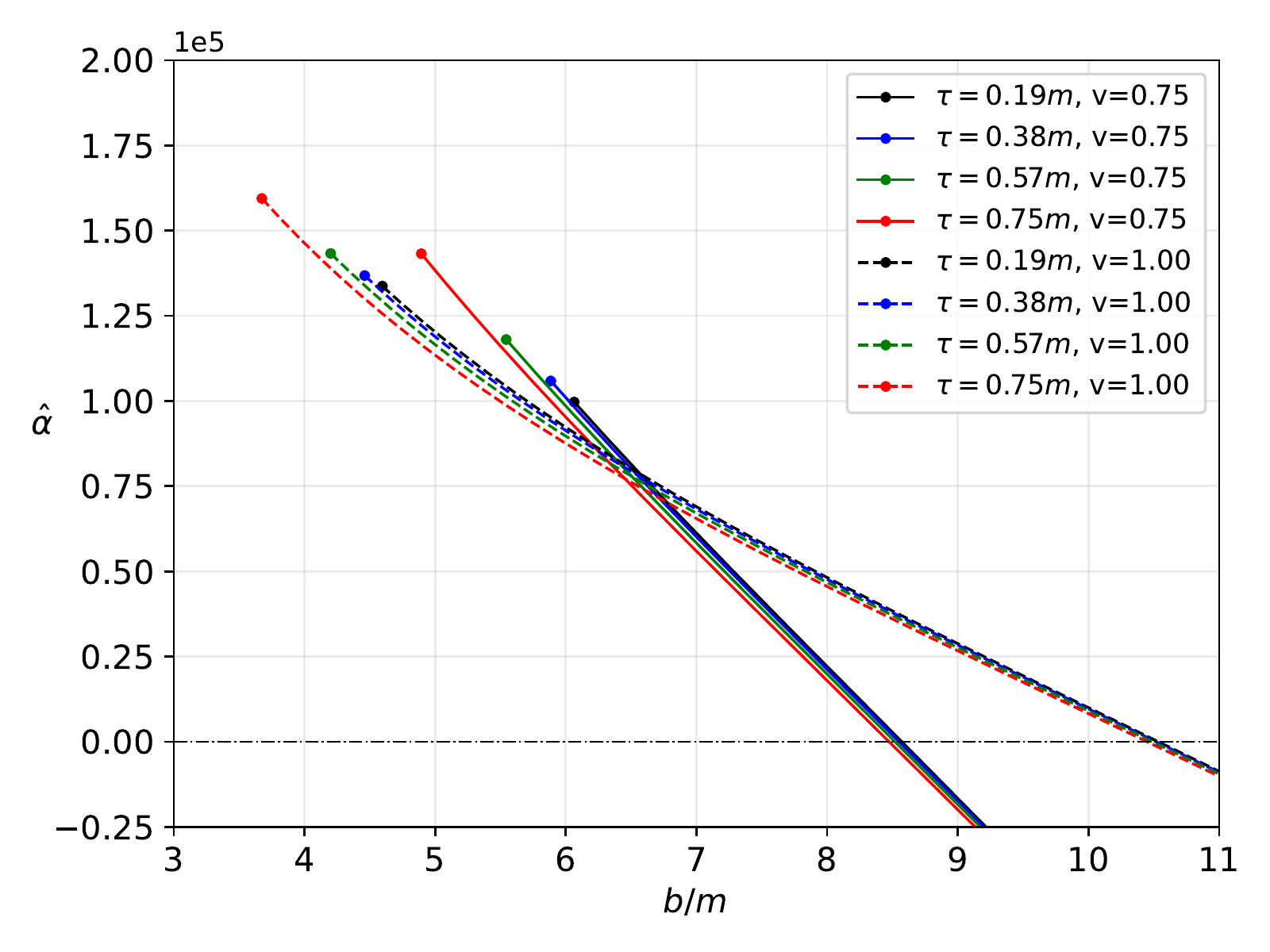}
    \includegraphics[width=0.48\textwidth]{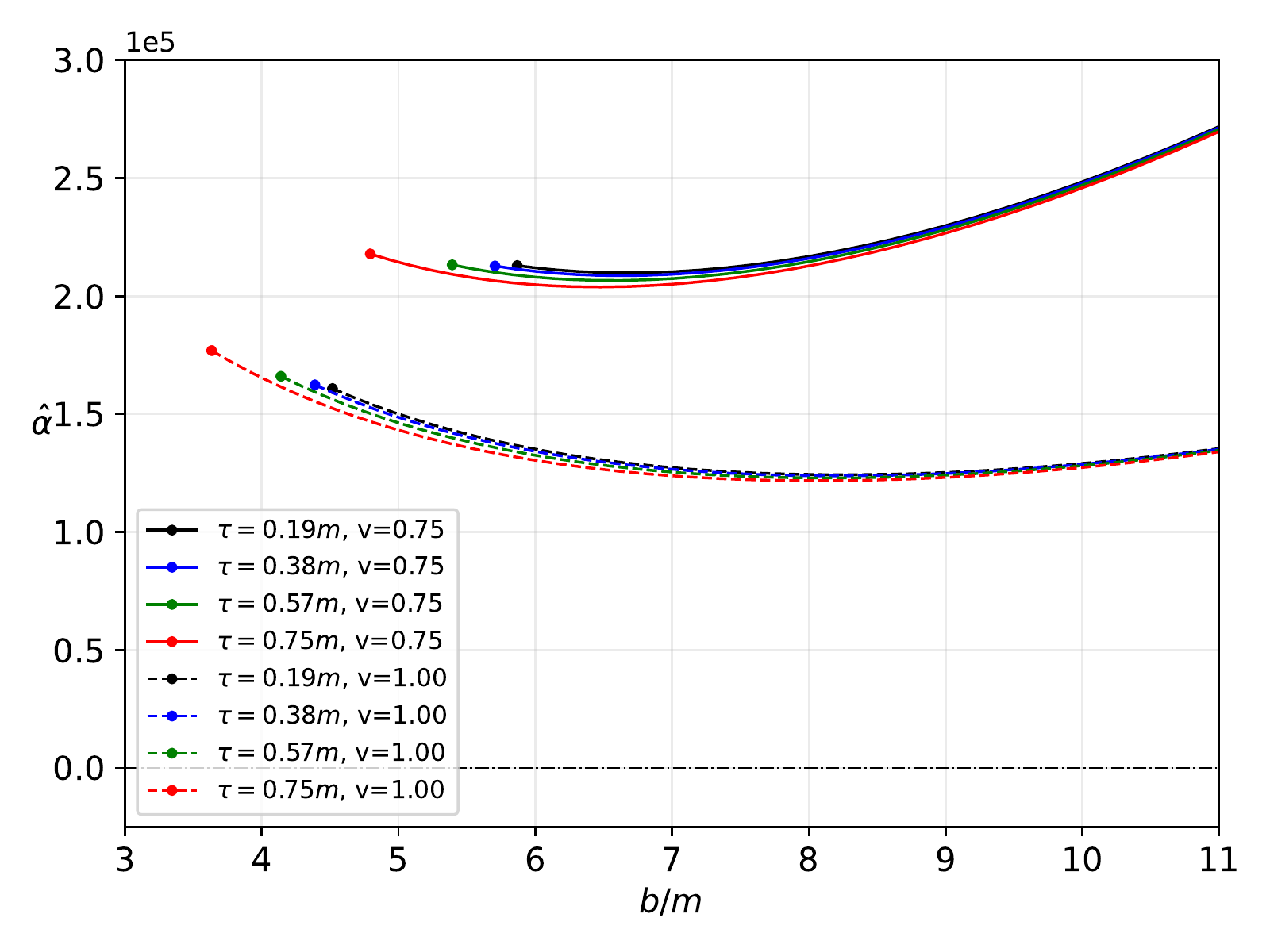}
    \caption{Comparison of weak deflection angle (Eq. \eqref{ewda_exact}) between massive particles (solid lines) and null particles (dashed line) as the torsion parameter changes. The dots correspond to the value of the critical impact parameter for each case. The finite distance of the source and receiver are also considered where it is assumed that $u_\text{S} = u_\text{R} = u$ and $u = 0.5b^{-1}$. Also in this plot (left), $Q = 0.75m$, and $\Lambda = 0.005 \text{ m}^{-2}$. The right figure is for $-\Lambda$. }
    \label{wda_new}
\end{figure*}
\begin{figure*}
    \centering
    \includegraphics[width=0.48\textwidth]{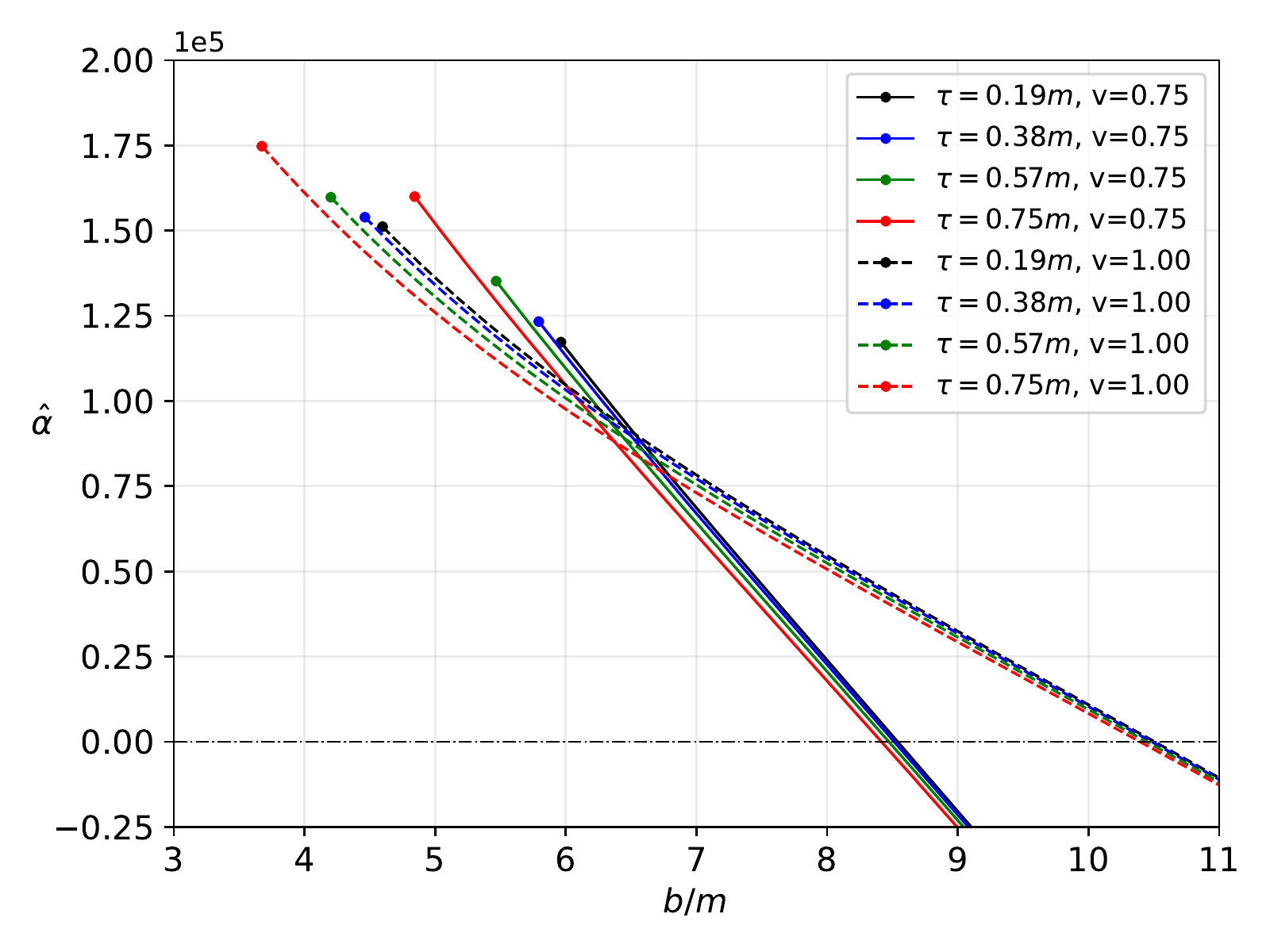}
    \includegraphics[width=0.48\textwidth]{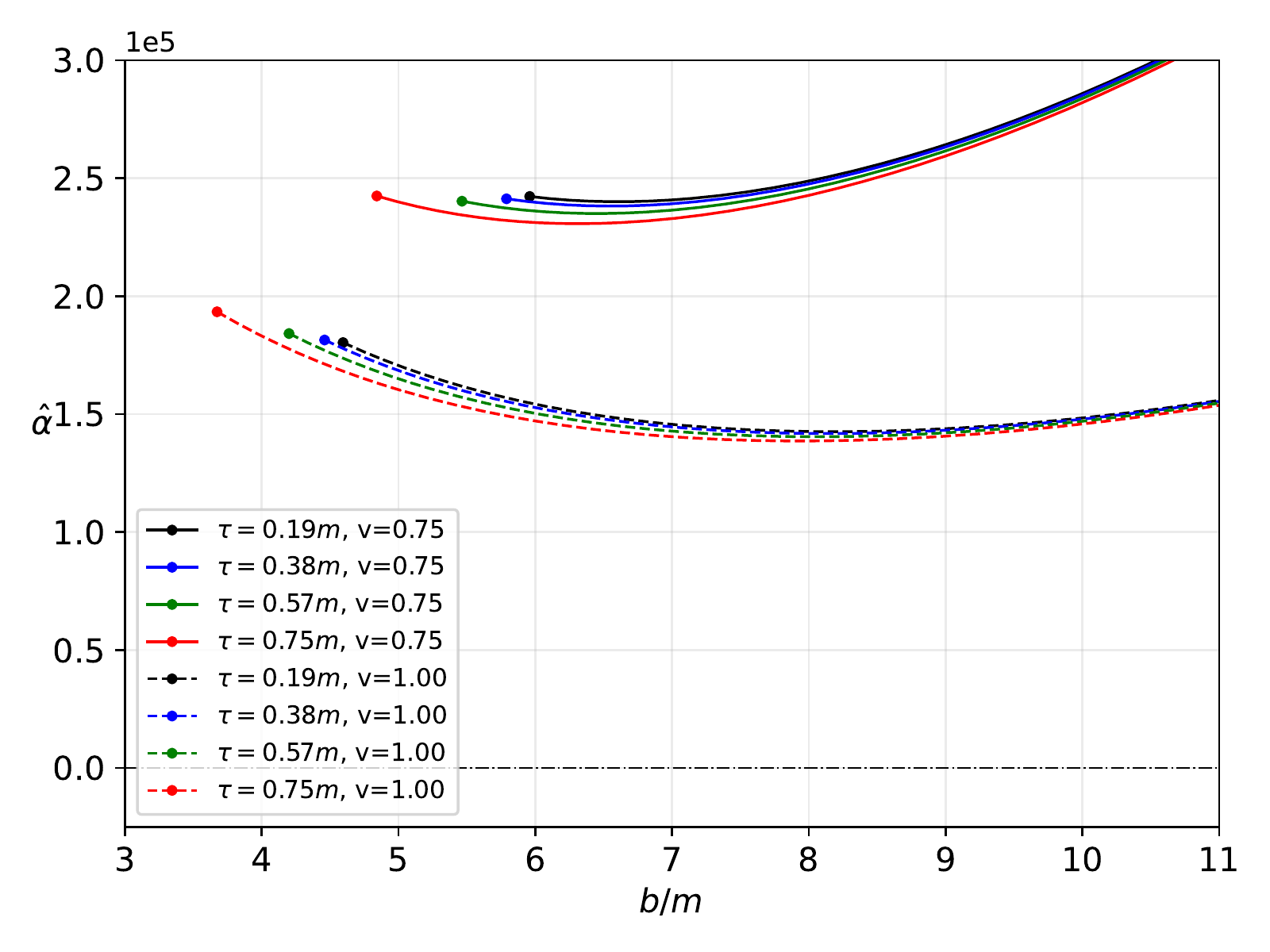}
    \caption{Comparison of weak deflection angle (Eq. \eqref{ewda_approx}) between massive particles (solid lines) and null particles (dashed line) as the torsion parameter changes. The dots correspond to the value of the critical impact parameter for each case. The finite distance of the source and receiver are also considered where it is assumed that $u_\text{S} = u_\text{R} = u$ and $u = 0.01b^{-1}$. Also in this plot (left), $Q = 0.75m$, and $\Lambda = 0.0001\text{ m}^{-2}$. The right figure is for $-\Lambda$.}
    \label{wda_new2}
\end{figure*}
In Figs. \ref{wda_new} and \ref{wda_new2}, we plotted the behavior of the weak deflection angle calculated by an observer near the black hole ($u = 0.5b^{-1}$) and far from the black hole $u = 0.01b^{-1}$. In each figure, the weak deflection angle of massive and null particles are also compared. We included the location's critical impact parameters to easily estimate the allowed values $\hat{\alpha}$. Let us start first with similar observations present in these plots. One can see the effect of the increasing value of the torsion parameter is to decrease the value of $\hat{\alpha}$. Also, the change due to the torsion parameter's influence diminishes as the impact parameter $b/m$ increases, as the cosmological constant dominates at large distances. Let us now compare the effect of $+\Lambda$ between near and remote receivers. First, we can notice that massive particles are more sensitive to the effect of $\Lambda$ since their deflection angle tends to zero first at large $b/m$ than the null particles. For remote observer, if $\Lambda = 0.005\text{ m}^{-2}$, it does not produce any $+\hat{\alpha}$, hence we lowered the scaled effect to $\Lambda = 0.0001\text{ m}^{-2}$. It indicates that $\hat{\alpha}$ itself becomes sensitive $\Lambda$ as we observe black holes at remote distances. Next is the AdS type which in general, we have the same observation as the dS type in terms of the black hole's distance from the receiver. The only key difference is that the $\hat{\alpha}$ does not approach zero as $b/m \to \infty$.

\begin{figure}
    \centering
    \includegraphics[width=0.48\textwidth]{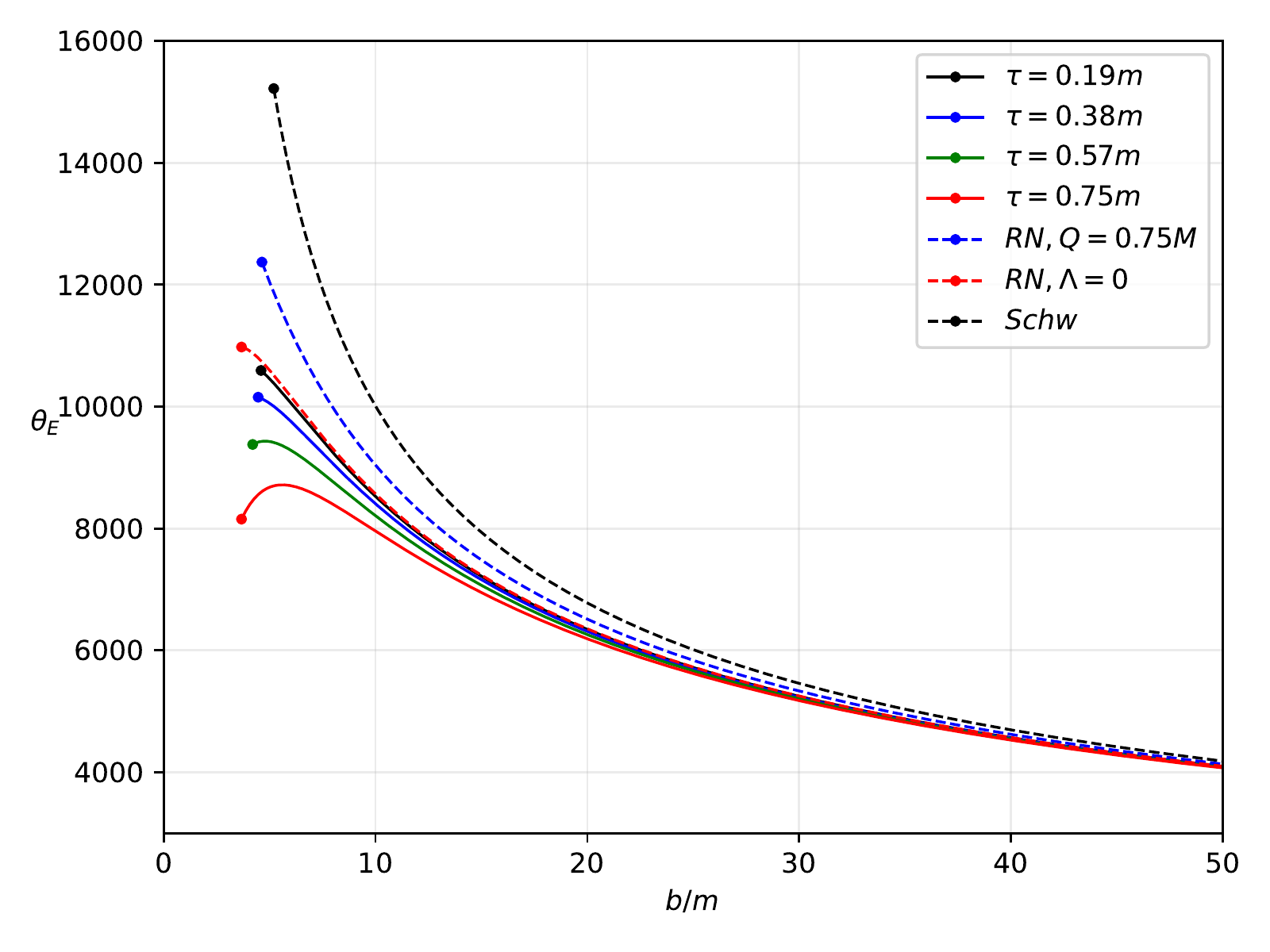}
    \caption{Plot (theoretical) of the Einstein ring formation in $\mu$as. Here, we set $Q = 0.75m$, $\Lambda = 0.0001\text{ m}^{-2}$, and $D_\text{S} = D_\text{R} = 100b$. The dots correspond to the value of the critical impact parameter for each case.}
    \label{eringplot}
\end{figure}

One observable that involves the weak deflection angle is the Einstein ring, which is very useful for astronomical observations. Let $D_\text{S}$ and $D_\text{R}$ be the position of the source and receiver respectively. The thin lens approximation implies that $D_\text{RS}=D_\text{S}+D_\text{R}$, and the position of the weak field images is given by
\begin{equation}
    D_\text{RS}\tan\beta=\frac{D_\text{R}\sin\theta-D_\text{S}\sin(\hat{\alpha}-\theta)}{\cos(\hat{\alpha}-\theta)}.
\end{equation}
When $\beta=0$, an Einstein ring is formed, and the above equation can be simplified into
\begin{equation}
    \theta_\text{E}\sim\frac{D_\text{S}}{D_\text{RS}}\hat{\alpha}.
\end{equation}
Finally, we can use the relation $b=d_\text{R}\sin\theta \sim d_\text{R}\theta$ and obtain
\begin{align} \label{ering}
    &\theta_\text{E} = \frac{-9\pi\epsilon D_\text{R}D_\text{S}}{4D_\text{R}\left(\Lambda+6\right)\left(D_\text{RS}\right)} \nonumber \\
    &+\frac{\sqrt{3}D_\text{R}D_\text{S}\left\{ 128\left(\Lambda+6\right)\left(D_\text{RS}\right)m+27\pi^{2}D_\text{R}D_\text{S}\epsilon^{2 }\right\} ^{1/2}}{4D_\text{R}\left(\Lambda+6\right)\left(D_\text{RS}\right)}
\end{align}
where the parameter $\epsilon=(\tau^2+Q^2)/b^2$. We plot Eq. \eqref{ering} in Fig. \ref{eringplot}, where it represents a source and a receiver that is close to the black hole.
The immediate effect of the torsion parameter is to further decrease the value of the angular radius of the Einstein ring, in comparison to the RN case. The effect is magnified for lower values of $b/m$ but diminishes as $b/m$ increases.

\begin{figure*}
    \centering
    \includegraphics[width=0.48\textwidth]{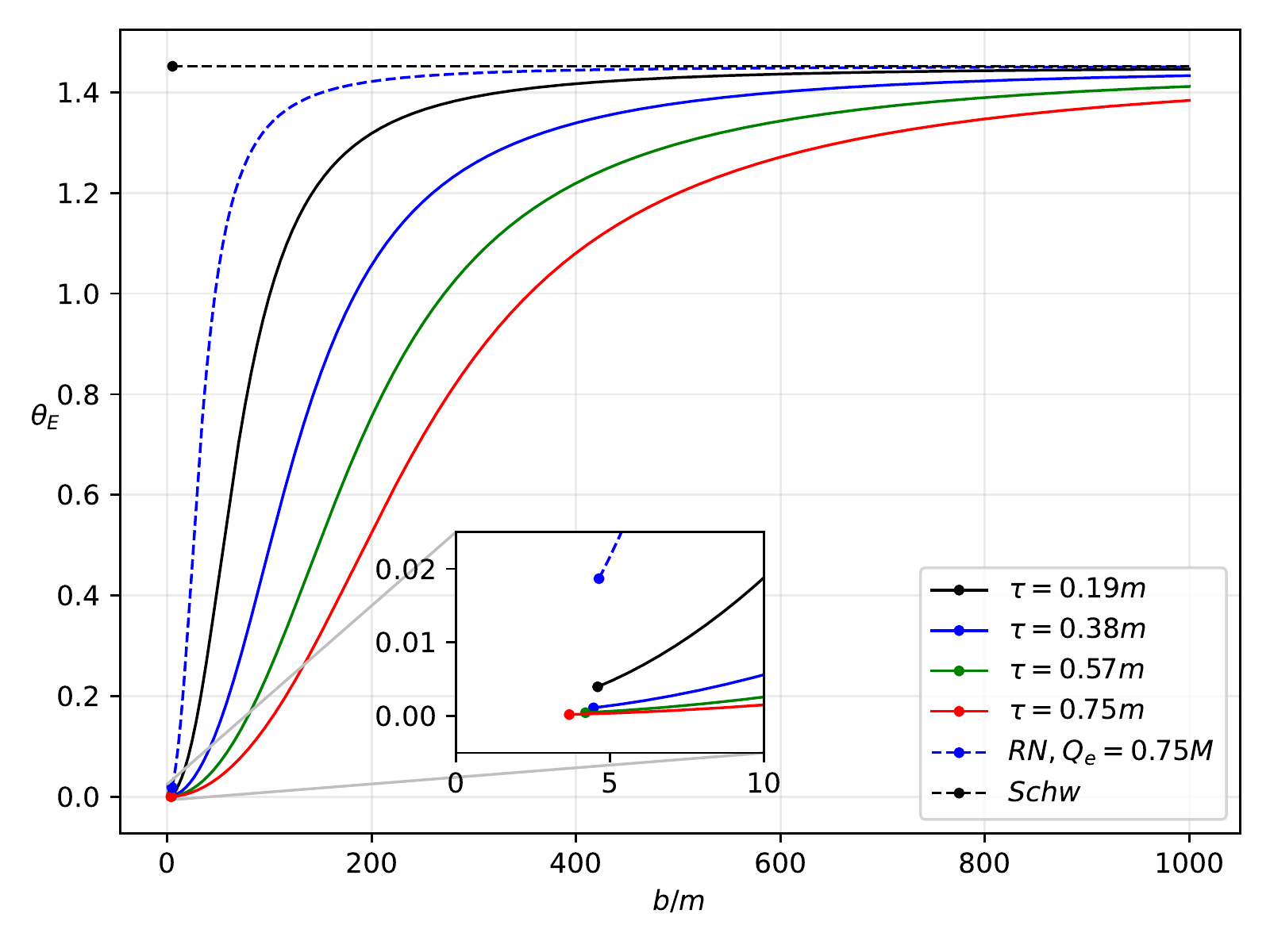}
    \includegraphics[width=0.48\textwidth]{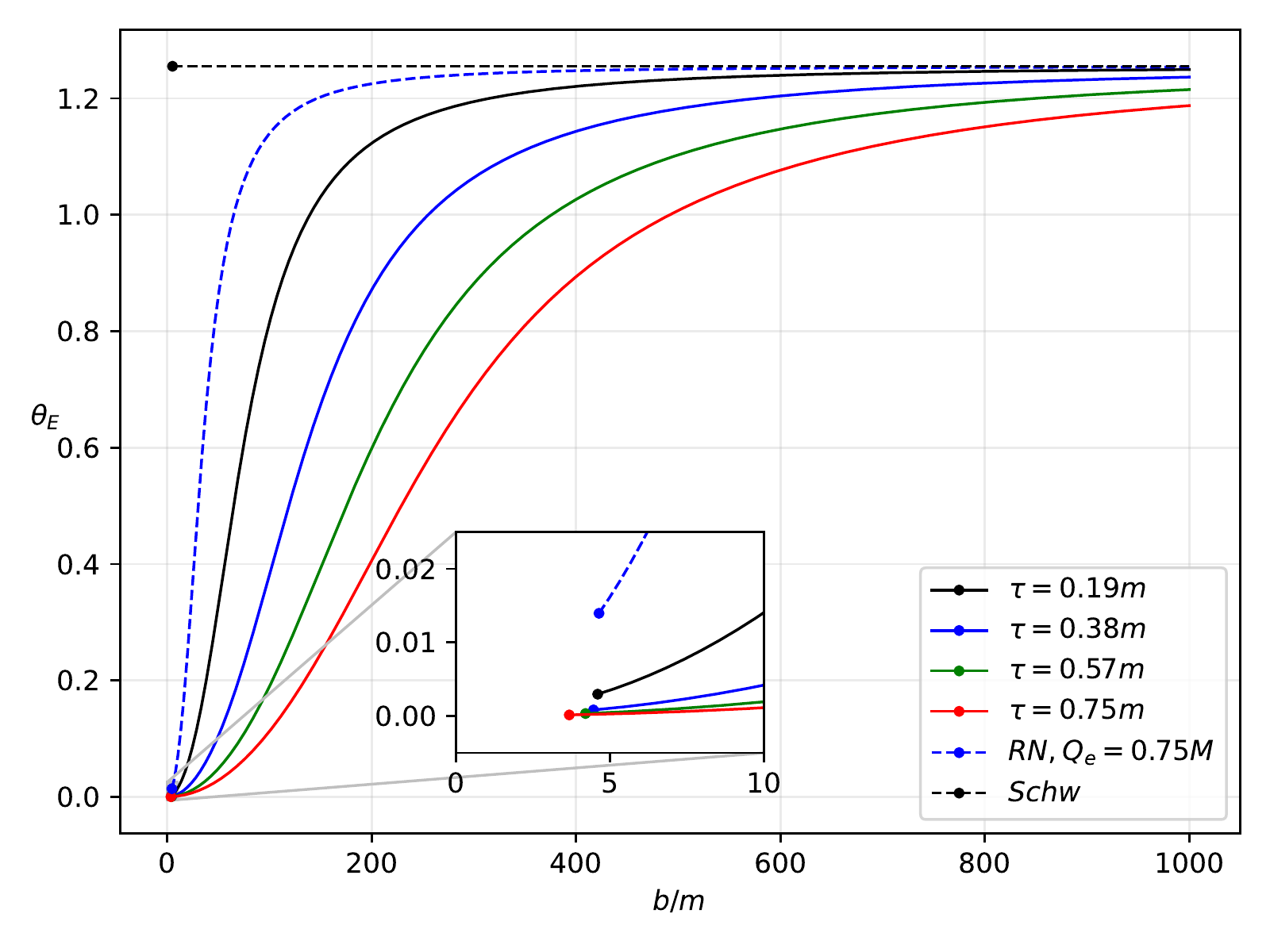}
    \caption{Plot of the Einstein ring formation for Sgr. A* (left), and M87* (right). Here, we assumed that $Q = 0.10m$, and $D_\text{S} = D_\text{R}$. The cosmological constant is taken as $\Lambda \sim 1.10^{-52} \text{ m}^{-2}$}
    \label{eringplot2}
\end{figure*}

Next, let us plot the Einstein ring due to the black hole Sgr. A* and M87*. See Fig. \ref{eringplot2}. For Sgr. A*, we used $m_\text{Sgr. A*} = 4.3$x$10^6 \: M_\odot$ and $D_\text{R} = 8.33$ kpc, and for M87*, $m_\text{M87*} = 6.5$x$10^9 \: M_\odot$ and $D_\text{R} = 16.8$ Mpc. Note that when $\tau = 0, Q = 0,$ and $\Lambda = 0$, $\theta_\text{E}^\text{Sgr. A*} = 1.453 \mu$as, and $\theta_\text{E}^\text{M87*} = 1.257 \mu$as.
First, we see a differing behavior of the Einstein ring from Fig. \ref{eringplot} because of the vast distance of Earth from these black holes. Moreover, using the actual value of the cosmological constant, its effect is vanishingly small in these cases. Nonetheless, we can see the effect of the torsion parameter on the behavior of the Einstein ring's angular radius. That is, as $\tau$ increases, $\theta_\text{E}$ decreases. The effect seems to be large, again, at smaller values of $b/m$ compared to the large values. It then gives the possibility of detecting the effect of torsion, even if $Q$ is small.

\section{Null Geodesics and Shadows Cast} \label{sec4}
\begin{figure}
    \centering
    \includegraphics[width=0.48\textwidth]{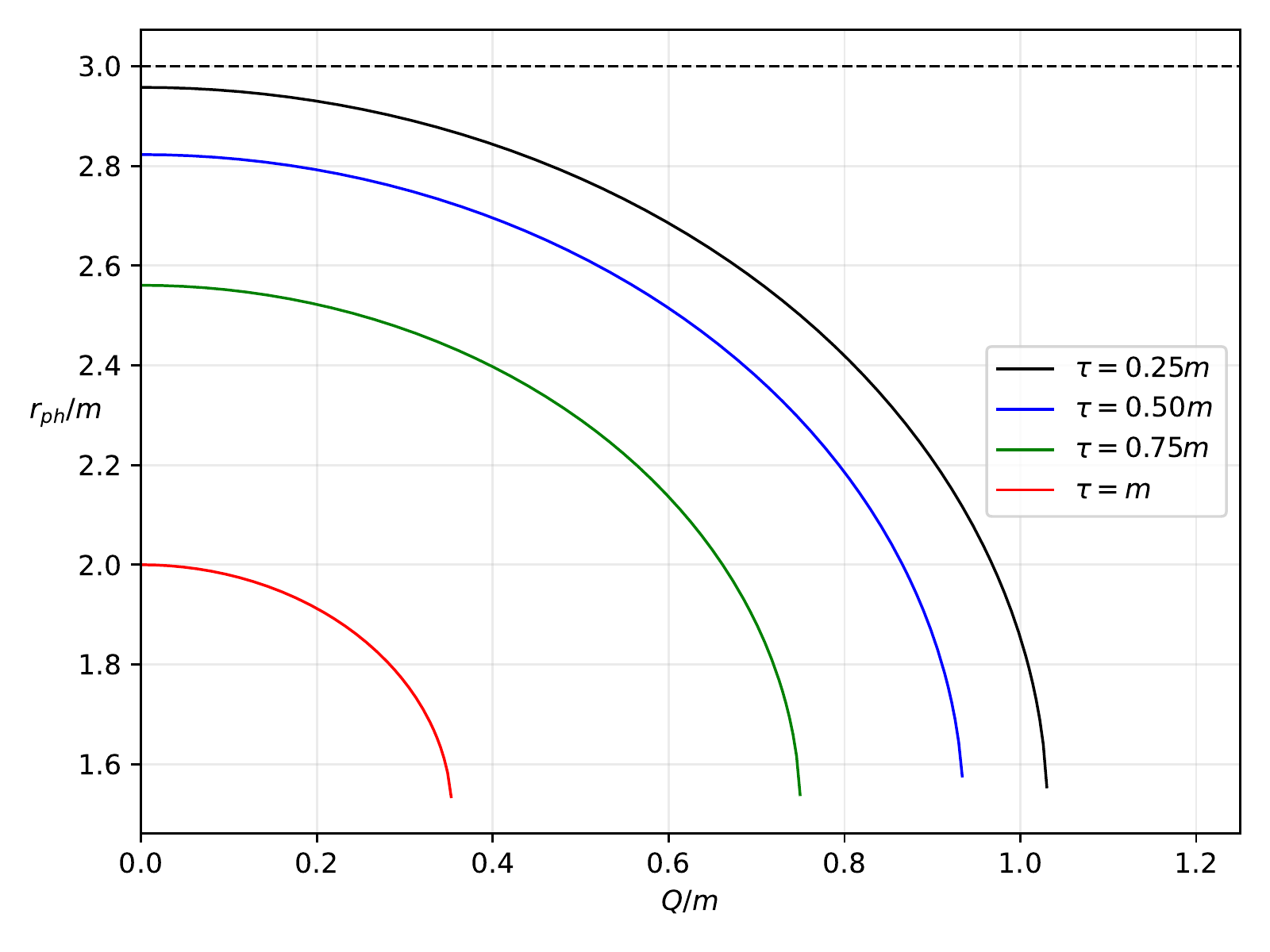}
    \caption{Location of the photonsphere as $Q$ varies. These deviations are independent of $\Lambda$.}
    \label{fig_rph}
\end{figure}

In this section, we study the shadow of the black hole using the method defined in Perlick et al.  \cite{Perlick_2015}. First, we calculate the null geodesic in the equatorial plane, by taking $D(r) = C(r)$ for spherically static and symmetric (SSS) spacetime, the Hamiltonian for light rays is in general given by
\begin{equation} \label{e28}
    H = \frac{1}{2} g^{ik} p_{i} p_{k} = \frac{1}{2} \left( -\frac{p_{t}^{2}}{A(r)} + \frac{p_{r}^{2}}{B(r)} + \frac{p_{\phi }^{2}}{C(r)} \right).
\end{equation}
Then we write the equations of motion (EOM) for null particles:
\begin{equation} \label{e29}
    \dot{x}^{i} = \frac{\partial H}{\partial p_{i}}, \quad \quad \dot{p}_{i} = -\frac{\partial H}{\partial x^{i}}.
\end{equation}
Here, $\dot{x}=dx/d\lambda$ and $\dot{p}$ stands for the conjugate momenta. Eq. \eqref{e29} gives
\begin{equation} \label{e30}
    \dot{t} = -\frac{p_{t}}{A(r)}, \quad \quad \dot{\phi } = \frac{p_{\phi }}{C(r)}, \quad \quad \dot{r} = -\frac{p_{r}}{B(r)},
\end{equation}
and
\begin{align} \label{e31}
    &\dot{p}_{t} = 0, \quad \quad \dot{p}_{\phi } = 0 \nonumber \\
    &\dot{p}_{r} = \frac{1}{2} \left( -\frac{p_{t}^{2} A'(r)}{A(r)^{2}} + \frac{p_{r}^{2} B'(r)}{B(r)^{2}} + \frac{p_{\phi }^{2}D'(r)}{C(r)} \right).
\end{align}
The first line in Eq. \eqref{e31} implies the existence of the two conserved quantities for null geodesic given as
\begin{equation}
    E = A(r)\frac{dt}{d\lambda}, \quad L = C(r)\frac{d\phi}{d\lambda},
\end{equation}
where $\lambda$ is the affine parameter along the light ray. The impact parameter, which is a constant of motion, is then defined as
\begin{equation} \label{eb}
    b \equiv \frac{L}{E} = \frac{C(r)}{A(r)}\frac{d\phi}{dt}.
\end{equation}
\begin{align} \label{eorb}
    \left(\frac{dr}{d\phi}\right)^2 =\frac{C(r)}{B(r)}\left(\frac{h(r)^2}{b^2}-1\right),
\end{align}
where $h(r)^2$ is defined as \cite{Perlick_2015}
\begin{equation}
    h(r)^2 = \frac{C(r)}{A(r)},
\end{equation}
which is useful because the radius of the photonsphere can be immediately calculated as
\begin{equation}
    \frac{d}{dr}\left(\frac{C(r)}{A(r)}\right) = 0,
\end{equation}
then in our case, we have
\begin{equation} \label{erph}
    r_\text{ph}=\frac{3m}{2}\pm\frac{1}{2}\sqrt{9m^{2}-8\left(\tau^2+Q^2\right)}.
\end{equation}
In Fig. \ref{fig_rph}, we plot the location of the photonsphere radius using the exact form in Eq. \eqref{erph}. We can see how the location of the photonsphere varies for different values of $Q$. In other words, these are only the allowed values for $r_\text{ph}$ for a given value of $Q$. We observe that the effect of the increasing torsion parameter value is to decrease the photonsphere radius.

For the black hole shadow, we consider first the static observer. The shadow depends on the initial direction of light rays that spiral toward the outermost photon sphere. We aim to calculate its radius. With the careful inspection of the line element, the definition \cite{Perlick:2018}
\begin{equation}
    \tan(\alpha_{\text{sh}}) = \lim_{\Delta x \to 0}\frac{\Delta y}{\Delta x} = \left(\frac{C(r)}{B(r)}\right)^{1/2} \frac{d\phi}{dr} \bigg|_{r=r_\text{o}},
\end{equation}
or
\begin{equation} \label{e44}
    \cot^2(\alpha_{\text{sh}}) = \left(\frac{B(r)}{C(r)}\right) \left. \left(\frac{dr}{d\phi}\right)^2 \right|_{r=r_\text{o}}.
\end{equation}
With the help of Eq. \eqref{eorb} and with simple trigonometry, we then find
\begin{equation} \label{eangrad}
    \sin^{2}(\alpha_\text{sh}) = \frac{b_\text{crit}^{2}}{h(r_\text{o})^{2}},
\end{equation}
where $ b_\text{crit} $ is associated with the photonsphere radius. It can be derived by satisfying the condition $dr/d\phi=0$ and imposing $r \to r_\text{ph}$:
\begin{align} \label{ebcrit}
    &b_\text{crit}^2 = \frac{h(r)}{\left[B'(r)C(r)-B(r)C'(r)\right]} \Bigg[h(r)B'(r)C(r) \nonumber \\
    &-h(r)B(r)C'(r) 
    -2h'(r)B(r)C(r) \Bigg],
\end{align}
where in our case yields
\begin{equation}
    b_\text{crit}^2 = \frac{6r_\text{ph}^3}{3r_\text{ph}-3m-2\Lambda r_\text{ph}^3}.
\end{equation}
Using the above equation and Eq. \eqref{eangrad}, we obtain the exact formula for the shadow radius:
\begin{equation} \label{eshadexact}
    \mathcal{R}_{\text{sh}} = r_\text{ph}\left[\frac{2\left(1-\frac{2m}{r_\text{o}}+\frac{q^{2}}{r_\text{o}^{2}}-\frac{\Lambda}{3}r_\text{o}^{2}\right)}{1-\frac{m}{r_\text{ph}}-\frac{2\Lambda}{3}r_\text{ph}^{2}}\right]^{1/2}.
\end{equation}
Although the photonsphere radius is independent of $\Lambda$'s influence, the shadow radius is not. The main reason for this is that the shadow radius is affected by the astrophysical environment induced by $\Lambda$ as the photons travel toward the observer. Afterward, we use the observational constraint of the torsion parameter using the obtained data from M87* \cite{EventHorizonTelescope:2019dse} and Sgr. A* \cite{EventHorizonTelescope:2022xnr}. For M87*, the shadow angular diameter is $\theta_\text{M87*} = 42 \pm 3 \:\mu$as, the distance of the M87* from the Earth is $D = 16.8$ Mpc, and the mass of the M87* is $m_\text{M87*} = 6.5 \pm 0.90$x$10^9 \: M_\odot$. For Sgr. A* the shadow angular diameter is $\theta_\text{Sgr. A*} = 48.7 \pm 7 \:\mu$as (EHT), the distance of the Sgr. A* from the Earth is $D = 8277\pm33$ pc and the mass of the black hole is $m_\text{Sgr. A*} = 4.3 \pm 0.013$x$10^6 \: M_\odot$ (VLTI). Indeed, knowing that these black holes are rotating, these empirical values for the shadow angular radius are a good estimate of the non-rotating case, since the spin parameter $a$ main feature is to show shadow deformations. In other words, the spin parameter gives the D-shaped contour of the shadow. Then we obtain the diameter of the shadow in units of the SMBH mass using
\begin{equation} \label{ed}
    d_\text{sh} = \frac{D \theta}{M}.
\end{equation}
Note that theoretically the diameter of the shadow can be calculated by $d_\text{sh}^\text{theo} = 2\mathcal{R}_\text{sh}$. Therefore, by using Eq. \eqref{ed}, we get the diameter of the shadow image of M87* and Sgr. A* as $d^\text{M87*}_\text{sh} = (11 \pm 1.5)m$, and $d^\text{Sgr. A*}_\text{sh} = (9.5 \pm 1.4)m$ respectively.
\begin{figure*}
    \centering
    \includegraphics[width=0.48\textwidth]{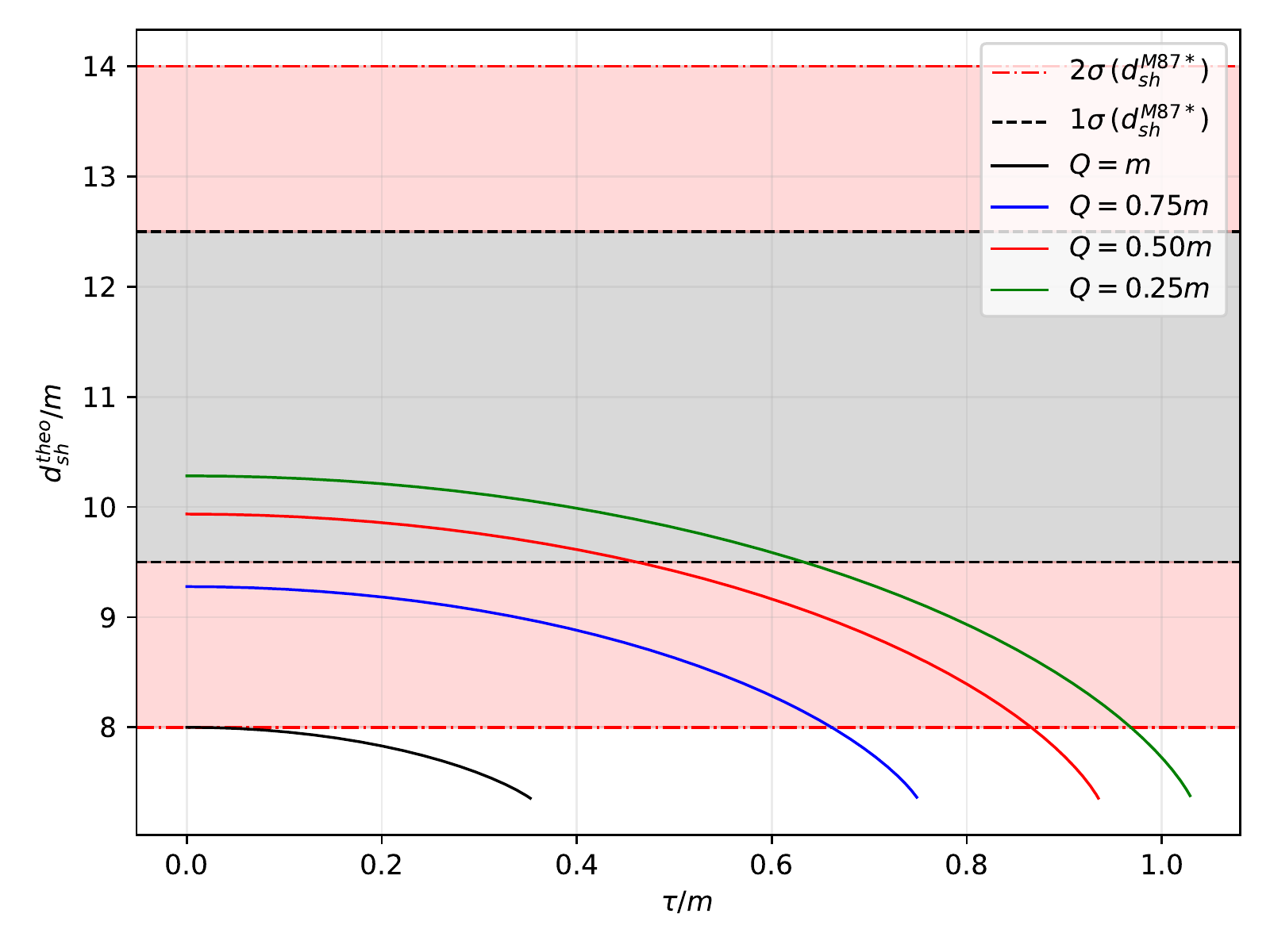}
    \includegraphics[width=0.48\textwidth]{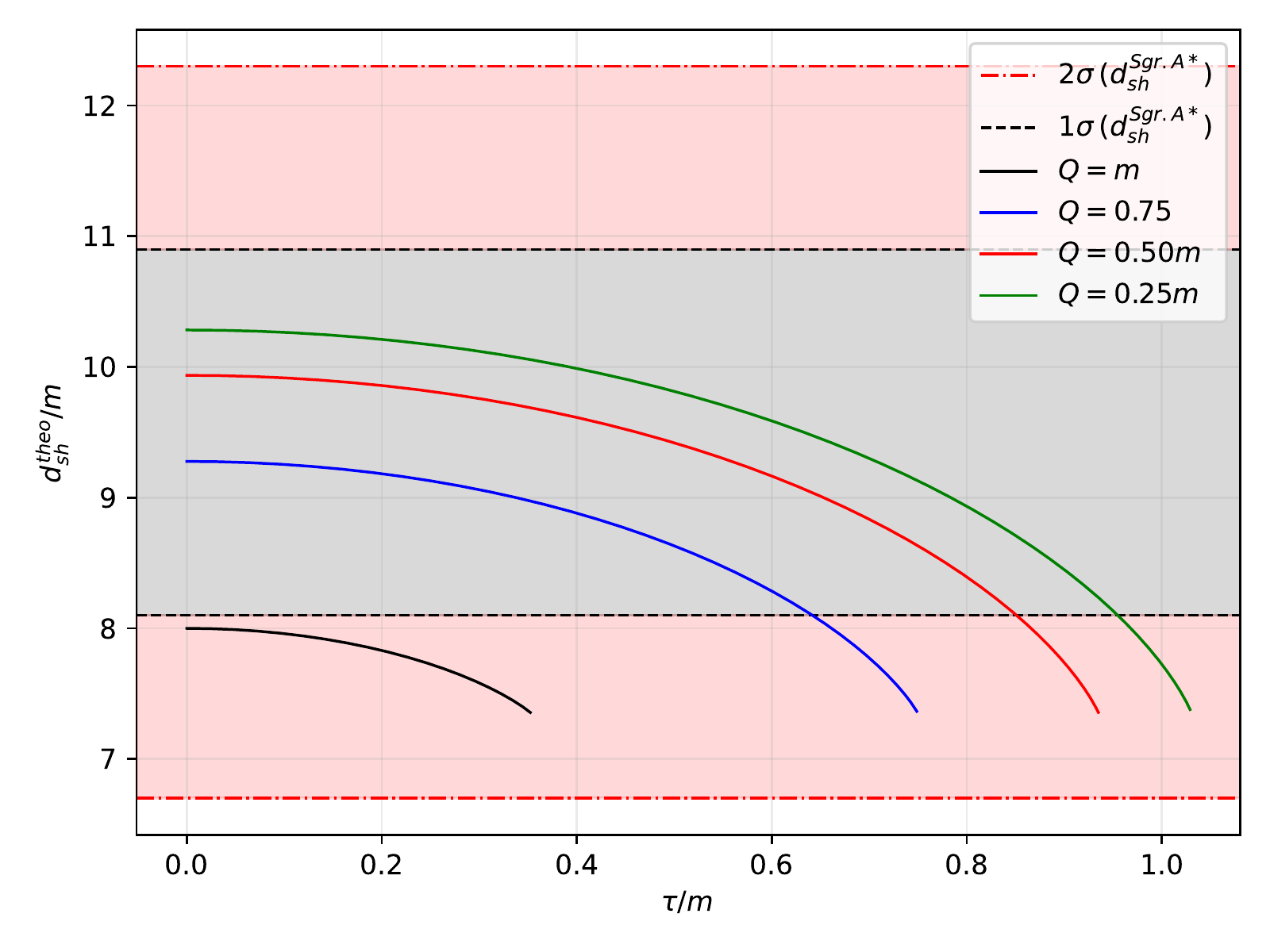}
    \caption{The variation of the shadow diameter with the torsion parameter $\tau$ for different values of charge $Q=m$(black), $Q=0.75m$(blue), $Q=0.50m$(red) and $Q=0.25m$(green) with $1\sigma$(dotted black) and $2\sigma$(dot-dashed red). Here, $m$ is the mass of the M87* or Sgr. A* black holes. Here, we have used the estimated value of the cosmological constant $\Lambda \sim -1.1^{-52}\text{ m}^{-2}$. For instance, $\Lambda m^2 = 9.217$x$10^{-27}$ for M87*, and $\Lambda m^2 = 4.034$x$10^{-33}$ for Sgr. A*.}
    \label{sha_cons}
\end{figure*}
In Fig. \ref{sha_cons}, we plotted the allowed values of the torsion parameter for a given value of $Q$ within $1\sigma$ and $2\sigma$ from the M87* and Sgr. A* data for the shadow diameter. For M87*, we see that the maximal charge $Q = m$ falls below the lower bound of $2\sigma$. At $2\sigma$, the lower bounds for $Q = 0.75m, 0.50m, 0.25m$ are $\tau = 0.66m, 0.87m, 0.97m$ respectively. At At $1\sigma$, the lower bounds for $\tau$ only occurs for $Q = 0.50m, 0.25m$, which are $\tau = 0.46m, 0.63m$ respectively. For Sgr. A*, we see that the case where $Q = m$ is within the region of $2\sigma$. All values of $Q$ have no lower bound in $2\sigma$. In $1\sigma$, the lower bounds for $Q = 0.75m, 0.50m, 0.25m$ are $\tau = 0.64m, 0.85m, 0.95m$ respectively. We note that even with torsion, the possibility of the SMBH at M87 galaxy to have an extremal charge is ruled out at 68\% confidence level, which is consistent with the result in Ref. \cite{Prashant2021}. Nevertheless, as we consider Sgr. A*, the possibility of having an extreme charge, in addition to torsion, is allowable within 68\% confidence level, which is one of the results in this study.

\begin{figure*}
    \centering
    \includegraphics[width=0.48\textwidth]{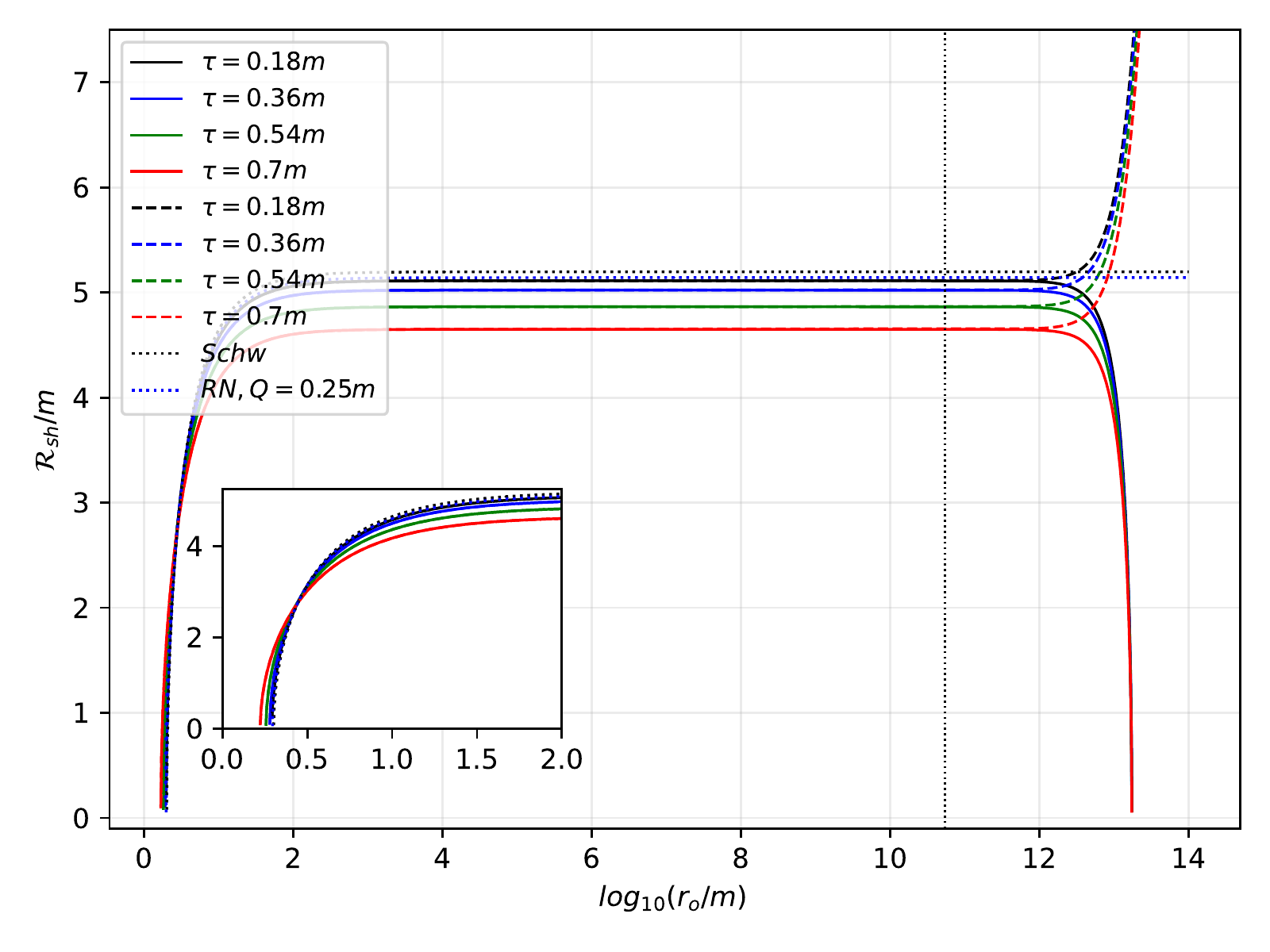}
    \includegraphics[width=0.48\textwidth]{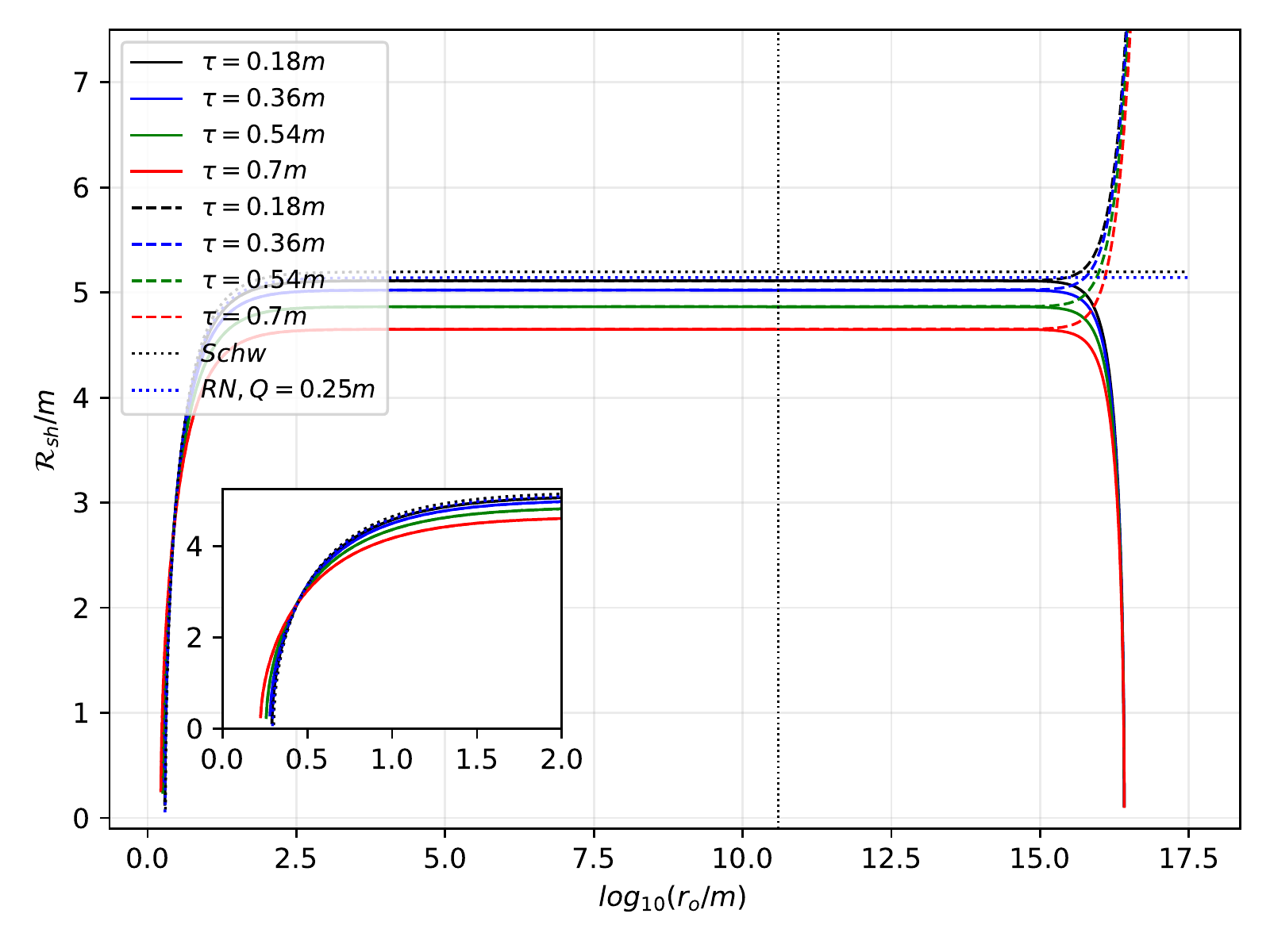}
    \caption{The behavior of the shadow radius under the effect of the torsion parameter and cosmological constant (M87* - left, Sgr. A* - right). The solid line is for the dS case ($\Lambda>0$), while the dashed line is for the AdS case ($\Lambda<0$). The dotted vertical line represents our location from the black hole. For M87* (left), $r_\text{o} = 5.40\text{x}10^{10}m$, while for Sgr. A* (right) $r_\text{o} = 4.02\text{x}10^{10}m$.}
    \label{sha_near}
\end{figure*}

Fig. \ref{sha_near} shows the plot of Eq. \eqref{eshadexact} for different values of $\tau$ following the bounds given in Fig. \ref{sha_cons}. While the charge $Q$ is always treated to be zero in the literature due to the neutralizing effect of ionized plasma, we will assume in this study that $Q \neq 0$. In this section, we assume a black hole charge of $Q = 0.25m$, although the charge of the SMBH in Ref. \cite{Zajacek:2018ycb} was found way lower than this. Nevertheless, the aim is to determine the deviation caused by the torsion parameter. Let us also take the observed value for the cosmological constant as $\Lambda \sim \pm1.1^{-52}\text{ m}^{-2}$ \cite{Perlick:2018}.
Overall, the effect of increasing the torsion parameter is to decrease the shadow size relative to the known values in Schwarzschild and RN cases. The vertical dotted line is Earth's location relative to the black holes (M87* and Sgr. A*) in units of its mass. The results in the plot suggest that such a vast distance is not enough to detect considerably the effects of the cosmological constant to the shadow radius, let alone to consider whether we are in a dS or AdS Universe. However, as can be gleaned, if the location of the observer is near the cosmological horizon, the deviations can be noticeable, especially the difference between the dS and AdS effects. We also remark that as we used the log-plot, the behavior of the shadow radius for observers near the black hole can also be gleaned (see the inset plot).

We can then see the overall shadow size behavior close and cosmologically far from the black hole. In the cosmological distance, the torsion effect in AdS does not seem to vanish and continues to rise relative to the cosmological horizon in the dS case. However, in the dS case, the shadow radius decreases, and the torsion effect is dominated by $\Lambda$. Near the black hole, the difference between dS and AdS type is not evident and can be difficult to distinguish, although the torsion parameter's effect can still be detected. The point where the curves intersect represents the observer's location where $\theta_\text{sh} = \pi/2$, where half of the sky is in total darkness.

Let us now approximate Eq. \eqref{eshadexact} due to observers located at the vertical dotted line in Fig. \ref{sha_near}. Indeed, at this location, $r_\text{o}>>m$ (same relation to charge and torsion parameters) and following the realistic value of $\Lambda$ implying $\Lambda <<0$, the approximation gives
\begin{equation}
    \mathcal{R}_\text{sh} = 3\sqrt{3} m -\frac{1}{8}\sqrt{3}\Lambda m\left[4k^{2}+51(\tau^{2}+Q^{2})\right],
\end{equation}
which reduces to the known result in Ref. \cite{Perlick:2018} when $\tau^{2}+Q^{2}=0$.

\section{Behavior of the shadow due to a co-moving observer} \label{sec5}
In the previous section, it is interesting how the shadow radii behaved due to a static observer at different locations. Here, since we are already incorporating the cosmological constant in this study, let us see how will the shadow radii behave relative to an observer that is co-moving with the cosmic expansion. We will follow the formalism presented in \cite{Tsupko2020}. It all begins with the McVittie metric, and with the inclusion of the dynamical torsion parameter $\tau$, we have \cite{Gao2004}:
\begin{align} \label{RNMV}
    &ds^2 = -\left[\frac{1-\frac{m^2}{4a(t)^2r^2}+\frac{\tau^2+Q^2}{4a(t)^2r^2}}{\left(1+\frac{m}{2a(t)r}\right)^2-\frac{\tau^2+Q^2}{4a(t)^2r^2}}\right]^2 dt^2 \nonumber\\
    &+ a(t)^2 \left[\left(1+\frac{m}{2a(t)r}\right)^2-\frac{\tau^2+Q^2}{4a(t)^2r^2}\right]^2 (dr^2 + r^2 d\Omega^2),
\end{align}
where $d\Omega^2 = \sin^2 \vartheta d\varphi^2 + d\vartheta^2$, $a(t) = e^{H_\text{0}t}$ as the scale factor, and $H_\text{0}$ is the present value of the Hubble constant. Note that Eq. \eqref{RNMV} is a time-dependent metric. However, if one is close to the black hole, this time dependence vanishes as $t_\text{0}-t<<H_\text{0}^{-1}$ since the expansion is negligible at such scale \cite{Tsupko2020}. Thus, as $a(t) \sim a(t_\text{0}) =$constant and using $x = a(t_\text{0}) r$, we can write Eq. \eqref{RNMV} as 
\begin{align} \label{RNiso}
    &ds^2 = -\left[\frac{1-\frac{m^2}{4x^2}+\frac{\tau^2+Q^2}{4x^2}}{\left(1+\frac{m}{2x}\right)^2-\frac{\tau^2+Q^2}{4x^2}}\right]^2 dt^2 \nonumber\\
    &+ \left[\left(1+\frac{m}{2x}\right)^2-\frac{\tau^2+Q^2}{4x^2}    \right]^2 (dx^2 + x^2 d\Omega^2),
\end{align}
which is the RN black hole with torsion in isotropic coordinates. Introducing
\begin{equation} \label{e60}
    R = x\left(1+ \frac{2m}{x} - \frac{\tau^2+Q^2}{x^2}\right)^2,
\end{equation}
where in the weak field limit ($m, \tau, Q \sim 0$), $R \sim x$, we recover a form similar to the RN metric:
\begin{align} \label{e61}
    ds^2&=-\left(1- \frac{2m}{R} + \frac{\tau^2+Q^2}{R^2}\right)dt^2 \nonumber \\
    &+ \frac{dR^2}{\left(1- \frac{2m}{R} + \frac{\tau^2+Q^2}{R^2}\right)} + R^2 d\Omega^2.
\end{align}

Let the present time be $t_\text{0}$ where the observer is at a radial position $r_\text{in}$, observing the shadow in a strong field limit. The subscript "in" denotes the inner region. Then $x_\text{in} = a(t_\text{0}) r_\text{in}$ and using these variables, we can rewrite  Eq. \eqref{e60} as
\begin{equation} \label{e62}
    R_\text{in} = x_\text{in}\left(1+ \frac{2m}{x_\text{in}} - \frac{\tau^2+Q^2}{x_\text{in}^2}\right)^2.
\end{equation}
Then using Eq. \eqref{e61}, an observer co-moving with the spacetime in Eq. \eqref{RNiso} will then observe the shadow radius $\mathcal{R}_\text{in}$ as
\begin{equation} \label{e63}
    \mathcal{R}_\text{in}=R_\text{in}\sin\alpha_\text{comov}=r_\text{ph}\left[\frac{2\left(1-\frac{2m}{R_\text{in}}+\frac{\tau^2+Q^2}{R_\text{in}^{2}}\right)}{1-\frac{m}{r_\text{ph}}}\right]^{1/2},
\end{equation}
where $r_\text{ph}$ is still given by Eq. \eqref{erph}. Eq. \eqref{e63} is the "inner" solution to the shadow radius where the expansion of the Universe is considered to be negligible.

Let us now consider the "outer" region, where the gravitational influence of the black hole is negligible, while the effect of the cosmological expansion is considerable (rapid expansion). In this case, then, the McVittie metric in Eq. \eqref{RNMV} will reduce to the FRW metric
\begin{equation} \label{FRW}
    ds^2 = -dt^2 + a(t)^2 (dr^2 + r^2 d\Omega^2).
\end{equation}
In this spacetime, effective linear shadow radius $L_\text{sh}$ is given in terms of the angular size of the black hole shadow $\Psi_\text{cosmo}$ as
\begin{equation} \label{e65}
    L_\text{sh} = \Psi_\text{cosmo} D_\text{A}(z),
\end{equation}
where
\begin{equation} \label{e66}
    D_\text{A}(z) = \frac{1}{1+z}\int_\text{0}^z \frac{dk}{H(k)},
\end{equation}
and
\begin{equation}
    H(k) = H_\text{0}[\Omega_\text{mat}(1+k)^3+\Omega_\text{rad}(1+k)^4+\Omega_\Lambda]^{1/2}.
\end{equation}
Here, $\Omega_\text{mat}, \Omega_\text{rad}, \Omega_\Lambda$ are present dimensionless density parameters for matter, radiation, and dark energy, respectively. Note here that $\Psi_\text{cosmo}$ is so small that the relation $\sin(\Psi_\text{cosmo}) \sim \Psi_\text{cosmo}$ applies. Eq. \eqref{e65} is indeed a function of $z$ and at such a large distance from the BH, $x_\text{out} \sim R_\text{out}$. Then one can find the connection between $z$ and $R_\text{out}$ defined by \cite{Tsupko2020}
\begin{equation} \label{e68}
    R_\text{out} = \int_\text{0}^z \frac{dk}{H(k)}.
\end{equation}

Now, there is a region between the inner and outer regions where the black hole begins to lose influence and at the same point, the cosmological expansion begins to gain influence. Let this "overlapping" region be in $R_\text{o}$ which, by definition, is still very far from the black hole. If the observer is located at this location at time $t_\text{0}$ where the scale factor is $a(t_\text{0})$, then it can be approximated that $z<<1$. Thus, in this case, $D_\text{A}(z) \sim R_\text{o}$ and we then have
\begin{equation} \label{e69}
    L_\text{o} = \Psi_\text{o} R_\text{o},
\end{equation}
which is still equal to the weak field approximation of Eq. \eqref{e63}:
\begin{equation} \label{e70}
    \mathcal{R}_\text{o} = L_\text{o}=3\sqrt{3}m+\frac{\sqrt{3}(\tau^2+Q^2)^{2}}{2}\left[\frac{1}{m}+\frac{1}{R_\text{o}}\right]+\mathcal{O}(R_\text{o}^{-2},R_\text{o}^{-3}).
\end{equation}
$L_\text{o}$ is equal to $L_\text{sh}$ by matching as pointed out in Ref. \cite{Tsupko2020}. Using $R_\text{out}$ instead of $R_\text{o}$, the effective shadow radius $\mathcal{R}_\text{cosmo}$ in the outer region of rapid expansion is then
\begin{align}
    &\mathcal{R}_\text{cosmo} = \Psi_\text{cosmo}R_\text{out} =\frac{R_\text{out}}{D_\text{A}(z)}\Bigg[3\sqrt{3}m \nonumber\\
    &+\frac{\sqrt{3}(\tau^2+Q^2)^{2}}{2}\left(\frac{1}{m}+\frac{1}{R_\text{out}}\right)+\mathcal{O}(R_\text{out}^{-2},R_\text{out}^{-3})\Bigg].
\end{align}
Finally, the perceived shadow radius $\mathcal{R}_\text{approx}$ by an observer co-moving with the cosmic expansion can be approximated through the composite solution \cite{Tsupko2020}:
\begin{equation} \label{e72}
    \mathcal{R}_\text{approx} = \mathcal{R}_\text{in} + \mathcal{R}_\text{cosmo} - \mathcal{R}_\text{o}.
\end{equation}

\begin{figure}
    \centering
    \includegraphics[width=0.48\textwidth]{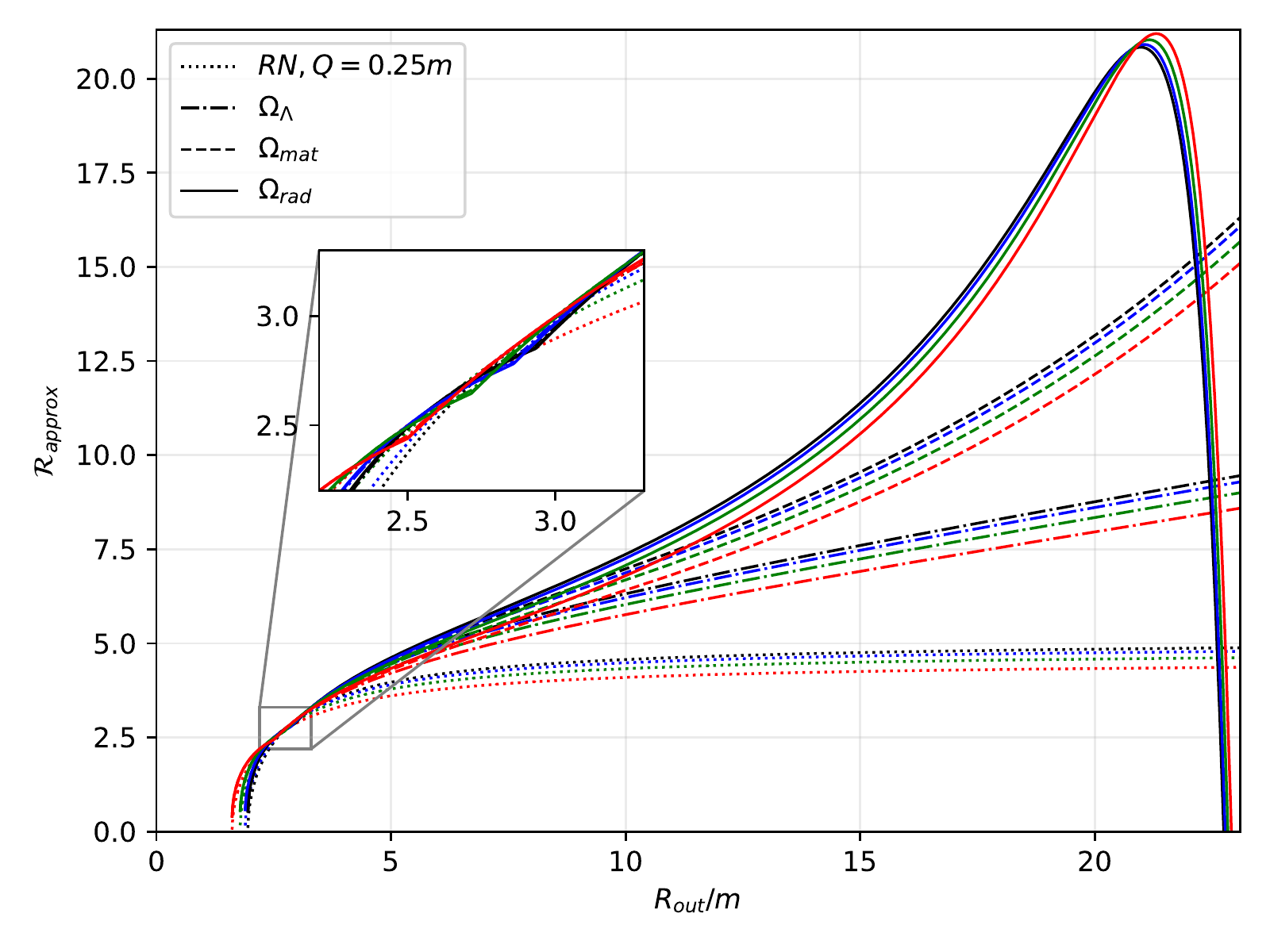}
    \caption{$\tau=0.19$ (black),$\tau=0.38$ (blue), $\tau=0.57$ (green), $\tau=0.75$ (red). Also, $Q = 0.75M$ and $H_\text{0} = 0.0408 \text{ m}^{-2}$.}
    \label{sha_com}
\end{figure}

For a Universe dominated by dark energy, we have the following after evaluating Eqs. \eqref{e66} and \eqref{e68}:
\begin{equation}
    D_\text{A}(z) = \frac{z}{(1+z)H_\text{0}}, \quad R_\text{out} = \frac{z}{H_\text{0}}.
\end{equation}
Let
\begin{align}
    &W = r_\text{ph}\left[\frac{2\left(1-\frac{2m}{R_\text{out}}+\frac{\tau^2+Q^2}{R_\text{out}^2}\right)}{1-\frac{m}{r_\text{ph}}} \right]^{1/2}, \nonumber \\
    &w = 3\sqrt{3}m+\frac{\sqrt{3}(\tau^2+Q^2)^{2}}{2}\left[\frac{1}{m}+\frac{1}{R_\text{out}}\right].
\end{align}
Thus, Eq. \eqref{e72} implies that the approximate shadow radius seen by an observer co-moving with the cosmic expansion is
\begin{equation} \label{e75}
    \mathcal{R}_\text{approx}^{\Lambda} = W + w H_\text{0} R_\text{out}.
\end{equation}
In a Universe dominated by matter, we have
\begin{equation}
    D_\text{A}(z) = \frac{2\left(\sqrt{z+1}-1\right)}{H_\text{0}(z+1)^{3/2}}, \quad R_\text{out} = \frac{2\left(\sqrt{z+1}-1\right)}{H_\text{0}\sqrt{z+1}},
\end{equation}
and
\begin{align} \label{e77}
    \mathcal{R}_\text{approx}^\text{mat} = W + w\left[\left(1-\frac{H_\text{0} R_\text{out}}{2}\right)^{-2}-1\right].
\end{align}
Finally, radiation dominated Universe gives
\begin{equation}
    D_\text{A}(z) = \frac{z}{(z+1)^2 H_\text{0}}, \quad R_\text{out} = \frac{z}{(z+1) H_\text{0}},
\end{equation}
and
\begin{align} \label{e79}
    \mathcal{R}_\text{approx}^\text{rad} = W - w \left[1 + (R_\text{out} H_\text{0} - 1)^{-1} \right].
\end{align}
We plotted Eqs. \eqref{e75}, \eqref{e77}, \eqref{e79} numerically for an immediate comparison shown in Fig. \ref{sha_com}. As can be gleaned from the plot, the overall effect of the torsion parameter even for the co-moving observer is to decrease the shadow radius. Next, the black hole shadow in the radiation-dominated Universe gives the largest deviation, especially near the cosmological horizon. Furthermore, a feature arises due to the existence of the peak value for the shadow radius. The value then drops significantly and intersects with the values for the matter and dark energy-dominated Universes. Thus, observers at this intersection cannot determine using deviations in the shadow radius, whether they are in the radiation-dominated or the matter-dominated (and dark-energy) Universes. Interestingly, for the region between $2.5m <R_\text{out} < 3.0m$ (see inset plot), the deviation caused by these types of Universes for the co-moving observer's perception of the shadow is nearly negligible since we can still see some tiny deviation to the pure RN type case. Thus, relating Fig. \ref{sha_com} to Figs. \ref{sha_near}, Earth's co-moving location is in the region where the effect of the cosmological constant is again small. Based on the inset plot, however, the tiny deviation caused by the effect of co-moving motion can still be possibly detected, and as to which type of Universe we are in is very hard to tell in the context of observing the black hole shadow. 

\section{Spherically infalling accretion} \label{sec6}
In this section, we study spherically free-falling accretion onto a black hole from infinity by applying a realistic model to visualize the shadow using the method defined in \cite{Jaroszynski:1997bw,Bambi:2012tg}.

First, we study the specific intensity of the BH observed at the photon frequency $\nu_\text{obs}$ by solving this integral along the light ray:
        \begin{equation}
            I(\nu_\text{obs},b_\gamma) = \int_\gamma g^3 j(\nu_e) dl_\text{prop}.
            \label{eq:bambiI}
        \end{equation}
It is noted that the emissivity per unit volume is given by $j(\nu_e)$ and the impact parameter by $b_{\gamma}$. Moreover, the infinitesimal proper length is shown by $dl_\text{prop}$ and the photon frequency of the emitter is $\nu_e$. For the infalling accretion we define the red-shift factor as follows:
        \begin{equation}
            g = \frac{k_\mu u^\mu_o}{k_\mu u^\mu_e},
        \end{equation}
where the 4-velocity of the photon is $k^\mu=\dot{x}_\mu$ and 4-velocity of the distant observer is $u^\mu_o=(1,0,0,0)$. Moreover, the $u^\mu_e$ stands for the 4-velocity of the infalling accretion
\begin{equation}
u_{\mathrm{e}}^{t}=\frac{1}{A(r)}, \quad u_{\mathrm{e}}^{r}=-\sqrt{\frac{1-A(r)}{A(r) B(r)}}, \quad u_{\mathrm{e}}^{\theta}=u_{\mathrm{e}}^{\phi}=0.
\end{equation}

Using the relation of  $k_{\alpha} k^{\alpha}=0$, one can derive $k_{r}$ and $k_{t}$ which is a constant of motion for the photons:
\begin{equation}
k_{r}=\pm k_{t} \sqrt{B(r)\left(\frac{1}{A(r)}-\frac{b^{2}}{r^{2}}\right)}.
\end{equation}
Here $\pm$ stands for the photon getting close to/away from the black hole. Then the red-shift factor $g$ and proper distance $dl_\gamma$ can be written as follows
   \begin{equation}
   g = \Big( u_e^t + \frac{k_r}{k_t}u_e^r \Big)^{-1},
  \end{equation}
  and
 \begin{equation}
  dl_\gamma = k_\mu u^\mu_e d\lambda = \frac{k^t}{g |k_r|}dr.
\end{equation}
For the specific emissivity, we consider only the monochromatic emission as follows:
        \begin{equation}
            j(\nu_e) \propto \frac{\delta(\nu_e - \nu_*)}{r^2},
        \end{equation}
with rest-frame frequency $\nu_*$.

Afterwards, the intensity equation given in \eqref{eq:bambiI} become
        
        \begin{equation}
            F(b_\gamma) \propto \int_\gamma \frac{g^3}{r^2} \frac{k_e^t}{k_e^r} dr.
        \end{equation}
        
\begin{figure}
    \centering
    \includegraphics[width=0.48\textwidth]{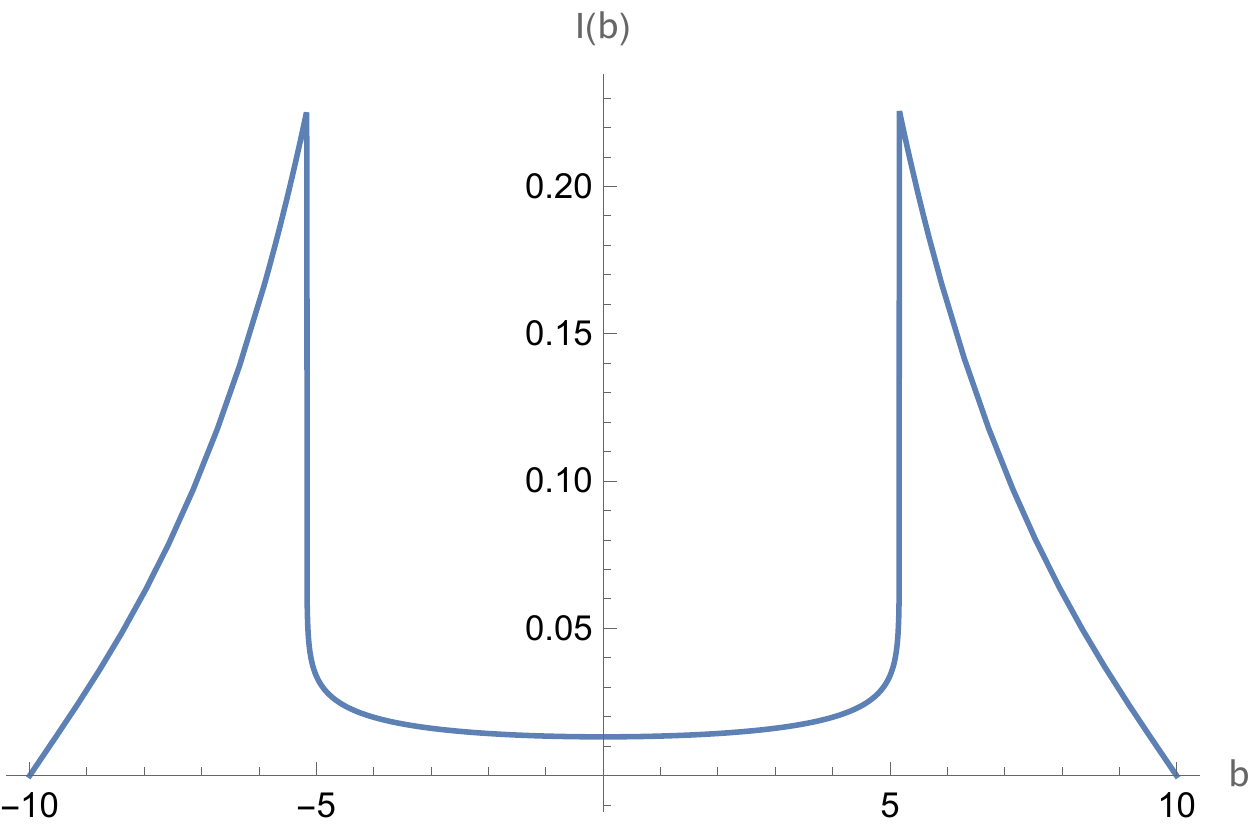}\hspace{1cm}
    \includegraphics[width=0.48\textwidth]{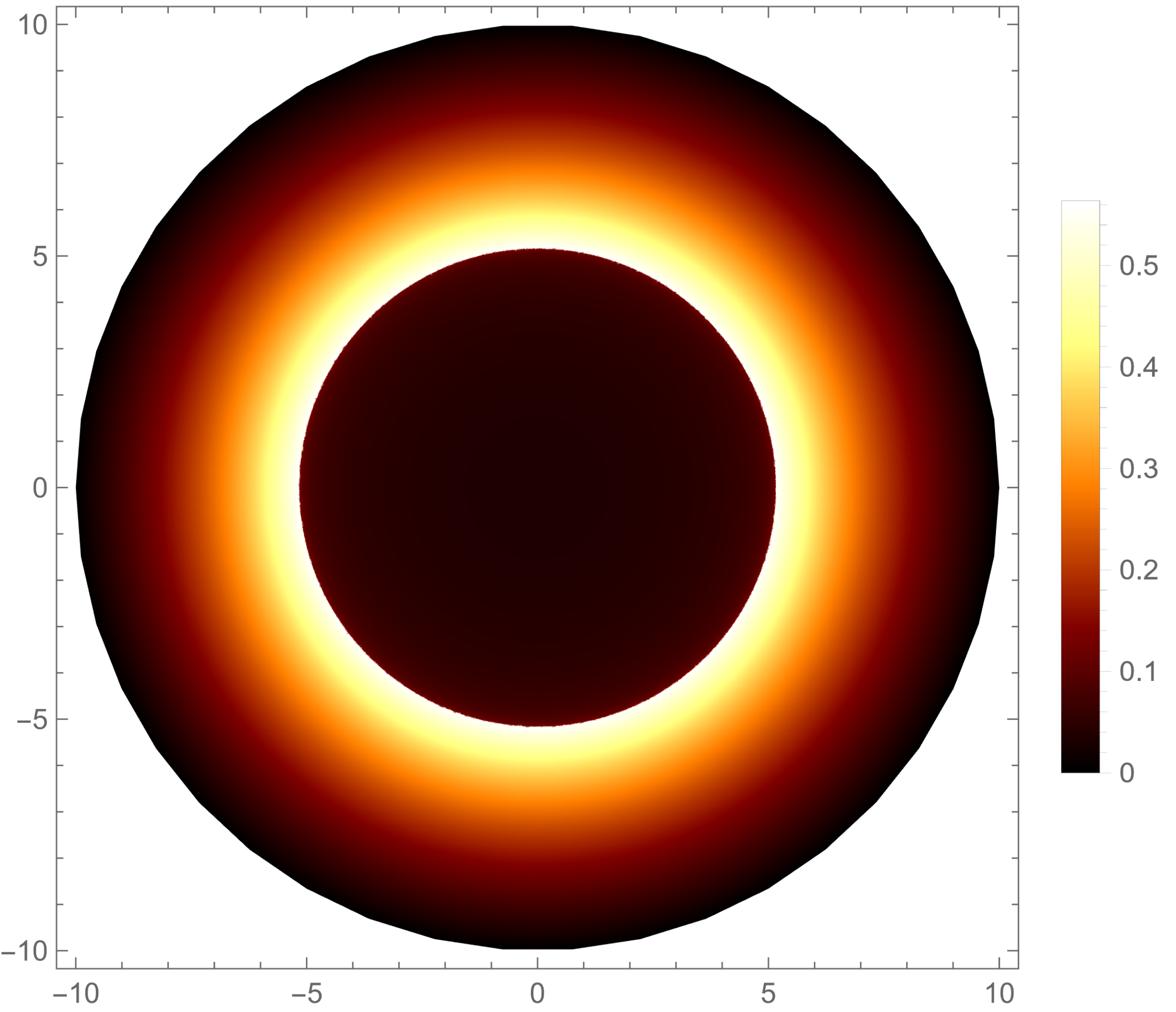}
    \caption{The specific intensity $I_\text{obs}$ seen by a distant observer for an infalling accretion at fixed $\Lambda=0$, and $\tau=0.19m$.}
    \label{fig:thinacc1}	
\end{figure}

\begin{figure}
    \centering
    \includegraphics[width=0.48\textwidth]{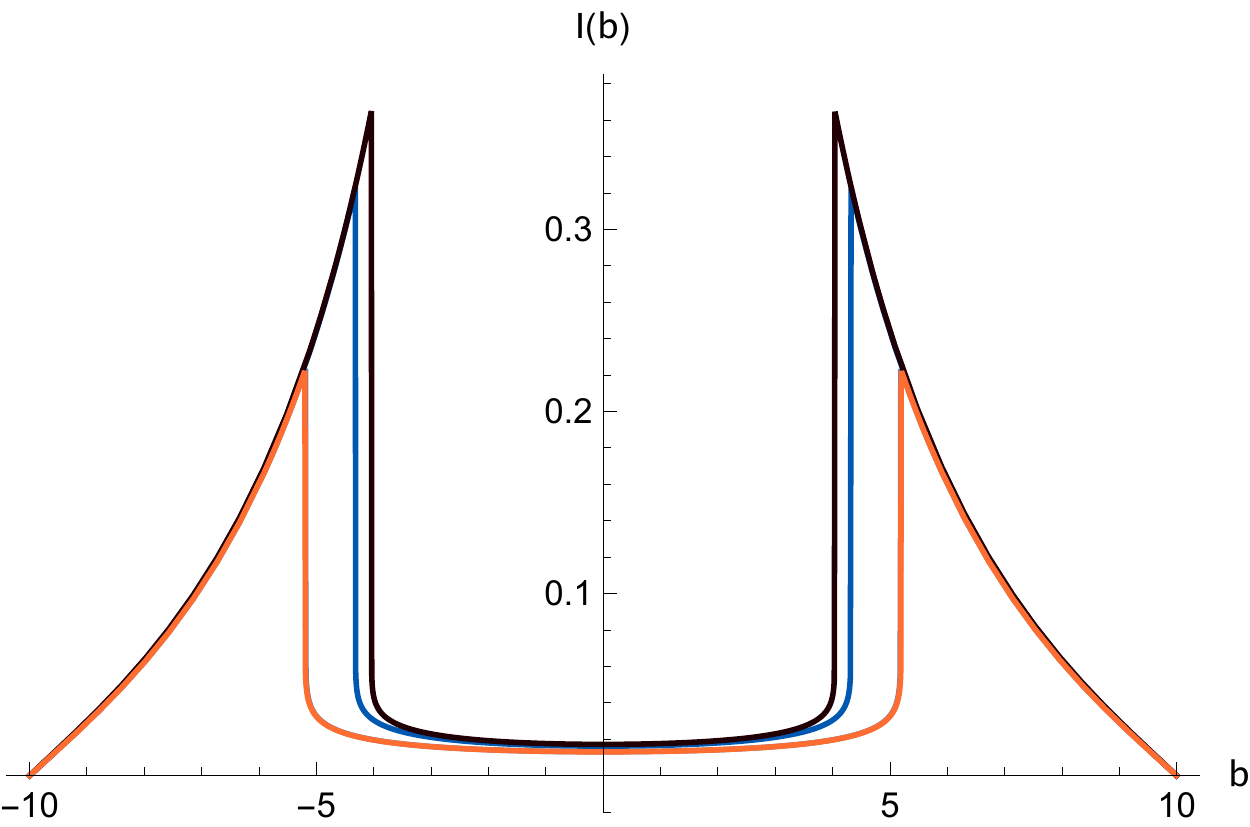}\hspace{1cm}
    \caption{The specific intensity $I_\text{obs}$ seen by a distant observer for an infalling accretion at fixed $\Lambda=0$, $\tau=0.99m $(black), $\tau=0.38m $(blue), and $\tau=0.10m $(orange).}
    \label{fig:thinacc2}	
\end{figure}

\begin{figure}
    \centering
    \includegraphics[width=0.48\textwidth]{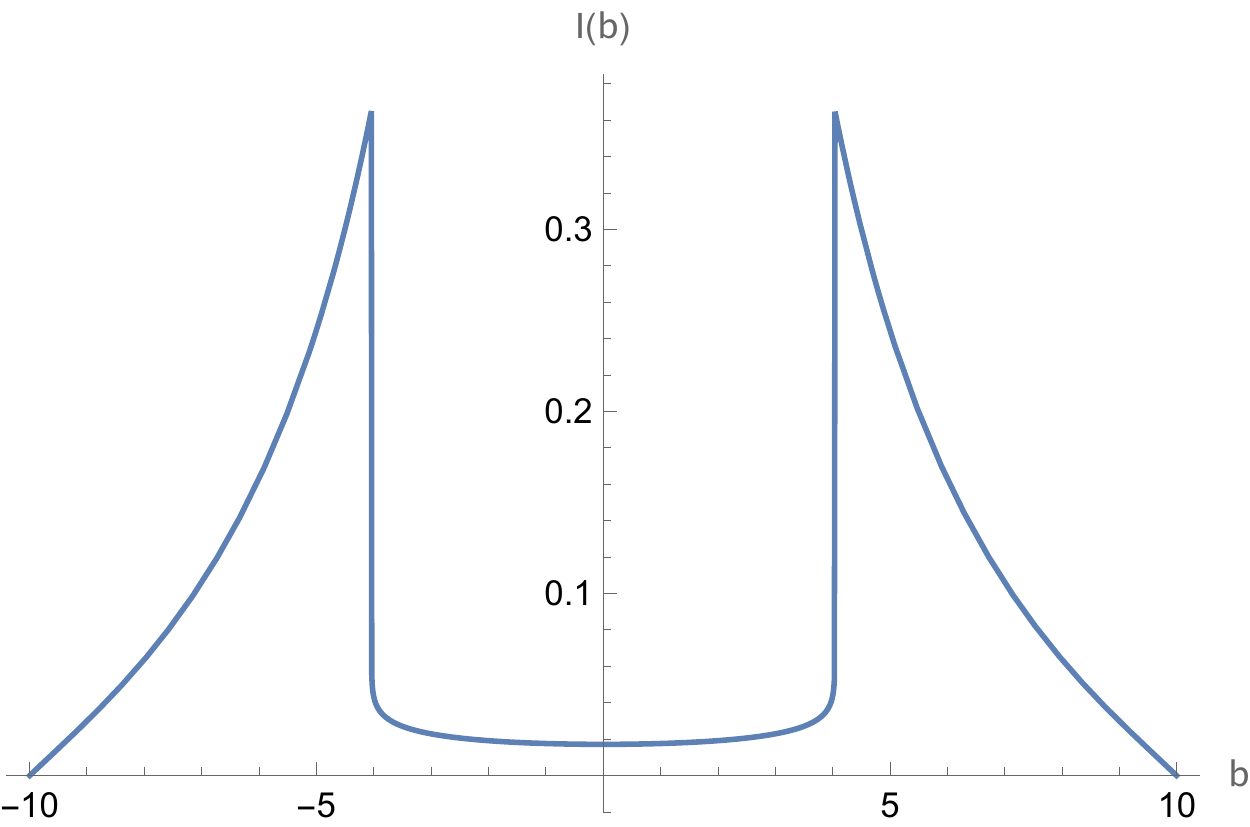}\hspace{1cm}
    \includegraphics[width=0.48\textwidth]{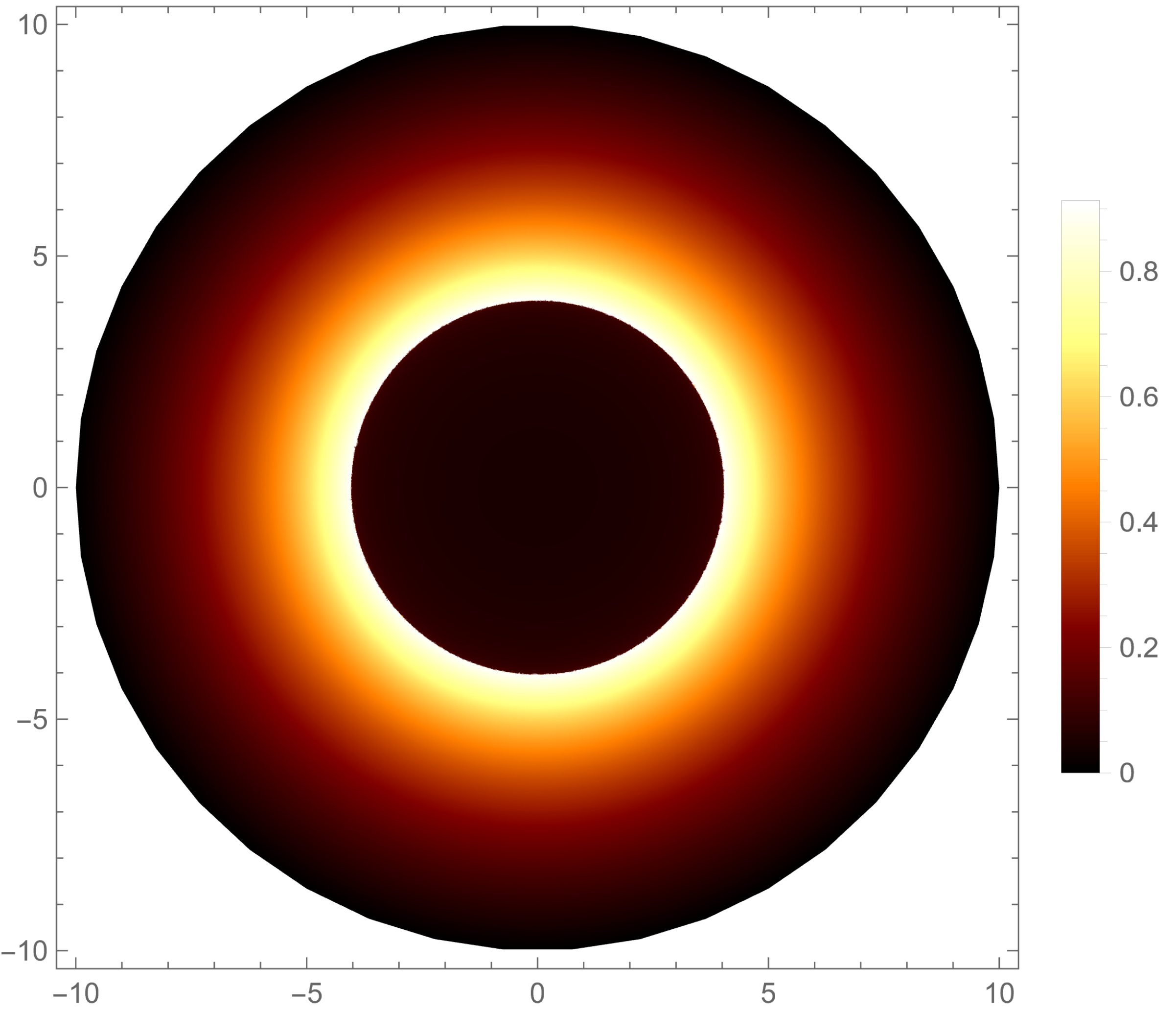}
    \caption{The specific intensity $I_\text{obs}$ seen by a distant observer for an infalling accretion at fixed $\Lambda=0$, and $\tau=0.99m$.}
    \label{fig:thinacc3}	
\end{figure}

We investigate the shadow cast with the thin accretion disk in BH. The above integral is solved numerically via \textit{Mathematica} package \cite{Okyay:2021nnh}, (also used in \cite{Chakhchi:2022fls,Kuang:2022xjp,Uniyal:2022vdu}). Calculations of the flux give us the hint about the impact of the dynamical torsion $\tau$, for this purpose, we plot the intensity for various values of the dynamical torsion parameter $\tau$ versus $b$ in Figs. (\ref{fig:thinacc1}, \ref{fig:thinacc2} and \ref{fig:thinacc3}). Here we observe that the luminosity of the accretion disk rises with increasing the value of the dynamical torsion parameter $\tau$ as seen in Fig. \ref{fig:thinacc3}.

\section{Quasinormal modes in eikonol limit} \label{sec7}
Since the first detection of gravitational waves (GWs) from the coalescence of two stellar-mass black holes in 2015 by the LIGO/VIRGO collaborations \cite{LIGOScientific:2016aoc}, gravitational wave physics has begun. Then wealth of data from gravitational waves is analyzed to test alternative theories of gravity and different models of compact objects. Perturbative analysis of black hole spacetimes dominated by `quasinormal ringing' is used to do this.  Quasi-normal modes (QNMs) are oscillations with complex frequencies with energy dissipation. The complex frequencies of QNMs have the characteristic properties of the BHs such as mass, charge, and angular momentum, independent of the initial perturbations. To do so, first, we use the following spherical symmetric spacetime:
\begin{eqnarray}
	\d s^2=-f(r) \d t^2 + \frac{\d r^2}{f(r)} + h^2(r)(\d 
	\theta^2+\sin^2\theta\d \varphi^2),\label{CBBmetric}
\end{eqnarray}
where $f(r)$ and $h^2(r)$. The general covariant equations for the scalar $\Phi$ and electromagnetic $A_{\mu}$ fields are written as follows:
\begin{eqnarray}
\label{scalar}
\frac{1}{\sqrt{-g}} \partial_{\mu}\left(\sqrt{-g} g^{\mu \nu} \partial_{\nu} \Phi\right)  &=0  \notag \\ 
\frac{1}{\sqrt{-g}} \partial_{\mu}\left(F_{\rho \sigma} g^{\rho \nu} g^{\sigma \mu} \sqrt{-g}\right) &=0,
\end{eqnarray}
where $F_{\mu \nu}=\partial_{\mu} A_{\nu}-\partial_{\nu} A_{\mu}$.

After solving the above equations on the background of black hole spacetime Eq. \eqref{CBBmetric}, and separating the variables, Eq. \eqref{scalar} reduce to the Schrodinger-like form with the tortoise coordinate $r_*$  defined by  $\d r_*/\d r=1/f(r)$. Note that $\Psi_s$ stands for the scalar or vector field oscillating and decaying at a complex frequency $\omega$, and $V_s$ is the spin-dependent Regge-Wheeler (RW) potential. First, let us  consider a massless scalar ($s=0$) perturbation field  with its wave equation:
\begin{eqnarray}
	\frac{1}{\sqrt{-g}}\partial_{\mu} \left(\sqrt{-g}g^{\mu \nu}\partial_{\nu} \phi \right)=0, \label{KGEq}
\end{eqnarray}
in which $g$ is the determinant. After we decompose $\phi (t,r,\theta,\varphi)$ into Fourier modes,
\begin{eqnarray}
	\phi(t,r,\theta,\varphi)=\sum_{\ell,m}e^{-iwt} \frac{\Psi_{s=0}(r)}{h(r)} Y_{\ell m}(\theta,\varphi),\label{decompose}
\end{eqnarray} 
and redefine $\Psi_{s=0}(r)$ as the perturbation field, where $Y_{\ell m}(\theta,\varphi)$ (for the spherical harmonics).  Note that near the event horizon, solutions must be purely ingoing, and at spatial infinity, solutions must be purely outgoing for asymptotically flat or de Sitter solutions, but different in AdS because AdS has a timelike boundary). Then we substitute the decomposition Eq. \eqref{decompose} into Eq. \eqref{KGEq}, and find master equation for $\Psi_{s=0}(r)$ with the RW potential,
\begin{eqnarray}
	V_{s=0}=f(r)\left\lbrace \frac{\ell(\ell+1)}{h^2(r)} +\frac{1}{h(r)}\frac{\d }{\d r}\left[ f(r)\frac{\d h(r)}{\d r}\right]  \right\rbrace.
\end{eqnarray}

For a linearized Maxwell ($s=1$) field perturbation, we can obtain the RW potential  in a  similar way to the scalar field, or one can derive the RW potential of the spin-1 field using the formalism developed in Ref. \cite{Boonserm:2013dua}:
\begin{equation}
	V_{s=1}=f(r)\left[ \frac{\ell(\ell+1)}{h^2(r)}  \right] .
\end{equation}
Hence, for the spacetime defined in (\ref{CBBmetric}), the master equation with the spin-dependent RW potential for the massless scalar ($s=0$) and the electromagnetic ($s=1$) field perturbations can be written in compact form as follows ($l=1,2,3$ is multipole number, and $\omega=\omega_{R}-i \omega_{I}$ is a complex quasinormal mode frequency) \cite{Berti:2009kk,Konoplya:2011qq,Kokkotas:1999bd}:
\begin{eqnarray}
	V_{s}=f(r)\left\lbrace \frac{\ell(\ell+1)}{h^2(r)} +\frac{(1-s)}{h(r)}\frac{\d }{\d r}\left[ f(r)\frac{\d h(r)}{\d r}\right]  \right\rbrace.\label{RWpotential}
\end{eqnarray}

Further in this section, we use the method of eikonal (geometric optics) limit (unstable circular null geodesic method) to derive the QNMs \cite{Cardoso:2008bp,Konoplya:2017wot}. The imaginary part of the quasinormal mode frequency (Im $\omega$=-$\omega_I$) which is responsible for the temporal, exponential decay can be calculated, in the large-$l$ limit ($l\to \infty$)
(only the $g_{tt}$ component is relevant) \cite{Glampedakis:2019dqh,Churilova:2019jqx} i.e. as the angular momentum number describing our mode solution becomes very large, as follows:
\begin{equation}
    \omega_{l \gg 1}=l \Omega_\text{ph}-i\left(n+\frac{1}{2}\right)\left|\lambda_{\mathrm{L}}\right|,
\end{equation}
with the angular velocity $\Omega_\text{ph}$:
\begin{equation}
    \Omega_\text{ph}= \frac{\sqrt{-g_{t t}\left(r_\text{ph}\right)}}{r_\text{ph}}=\frac{\sqrt{f\left(r_\text{ph}\right)}}{r_\text{ph}},
\end{equation} 
and Lyapunov exponent $\lambda_{\mathrm{L}}$:
\begin{equation}
    \lambda_{\mathrm{L}}=\sqrt{\frac{f\left(r_\text{ph}\right)\left[2 f\left(r_\text{ph}\right)-r_\text{ph}^{2} f^{\prime \prime}\left(r_\text{ph}\right)\right]}{2 r_\text{ph}^{2}}}, \label{Lyapunov}
\end{equation}
where $n$ is the overtone number and take values $n= 0, 1, 2,...$. Note that the eikonal limit is independent of the spin of the perturbation, so that scalar, electromagnetic, and gravitational perturbations of black holes give the same behavior in the eikonal limit \cite{Kodama:2003jz}. We show our results in Table \ref{tab:table2} that the real parts increase, but the imaginary part of the QNMs decrease with the increasing dynamical torsion parameter $\tau$. We can conclude that these modes are stable cause the imaginary parts of the QNMs frequencies (Im $\omega$=-$\omega_I$) are negative. We can say that when the dynamical torsion parameter $\tau$ increases, the scalar perturbations oscillate with lower frequency $\omega$ which means that oscillates decay slowly.
\begin{table}[ht!]
    \centering
    \begin{tabular}{ |p{1cm}||p{2cm}|p{2cm}||p{2cm}| }
    \hline
    $\tau/m$ &  $\omega_{R}$ & $\omega_{I}$ \\ [0.5ex] 
    \hline
    0.06 & 0.192572  &  0.865842  \\
    0.12 & 0.192922 & 0.865315 \\
    0.18 & 0.193511 & 0.864422  \\
    0.24 & 0.194299 &  0.863216  \\
    \hline
    \end{tabular}
    \caption{Effects of the torsion parameter $\tau$ on the quasinormal modes frequencies in eikonal limits for fixed $\Lambda=0$, $Q=0.02m$, $s=1$, $n=0$, and $l=1$.}
    \label{tab:table2}
\end{table}

\section{Bounds of Greybody factors and High-Energy Absorption cross-section via Sinc approximation} \label{sec8}
In this section, we calculate the bound for the greybody factor of the black hole with dynamical torsion by using the rigorous lower bound to probe the effect of $\tau$ on the bound. Authors in \cite{Visser:1998ke,Boonserm:2008zg} give the rigorous bound of greybody factor as follows \cite{Visser:1998ke,Boonserm:2008zg}: 
\begin{equation} \label{bound}
T_{b} \geq \operatorname{sech}^{2}\left(\frac{1}{2 \omega} \int_{-\infty}^{\infty}\left|V\right| \frac{d r}{f(r)} \right).
\end{equation}

\begin{figure}
    \centering
    \includegraphics[width=0.48\textwidth]{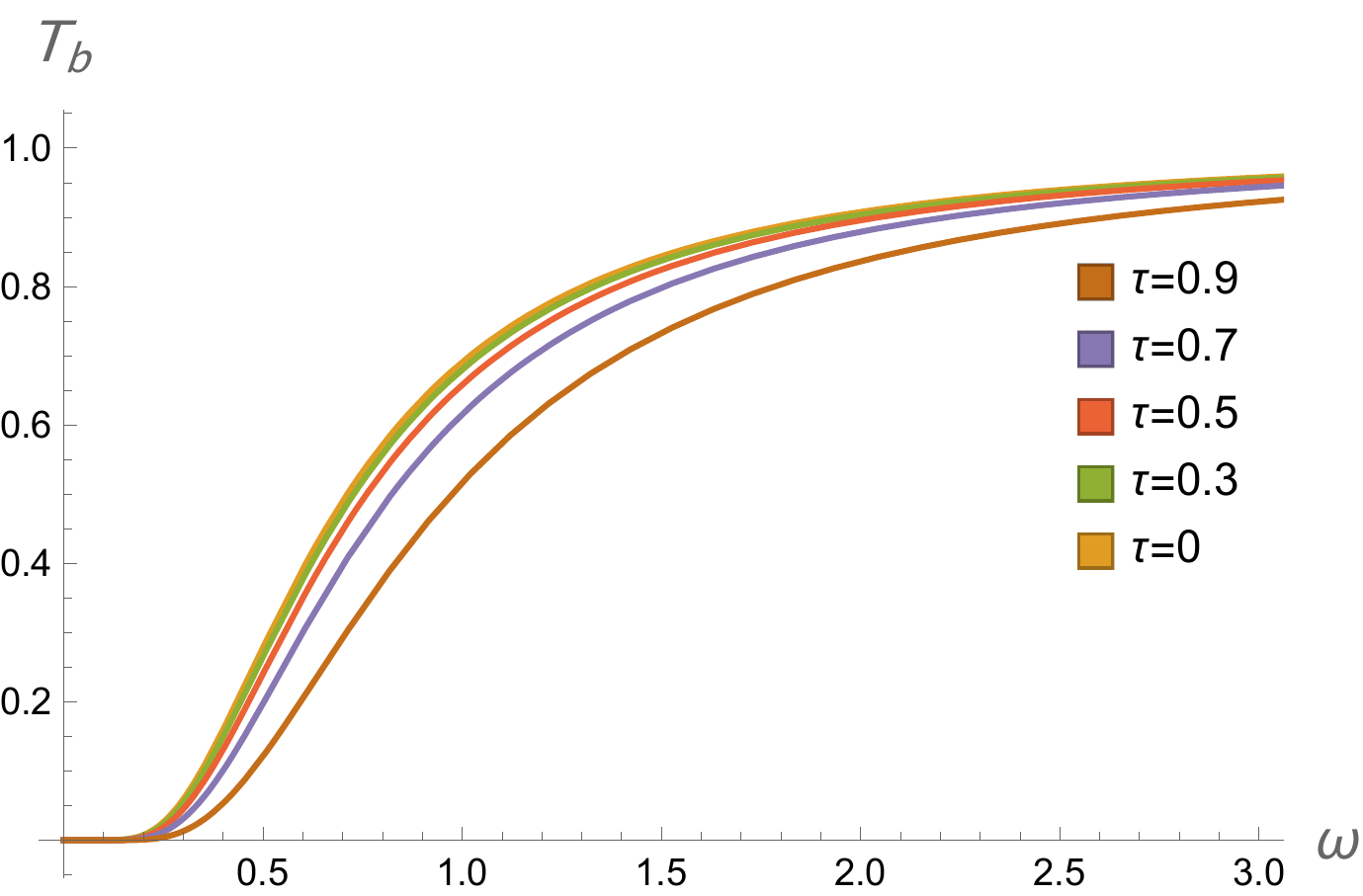}
    \includegraphics[width=0.48\textwidth]{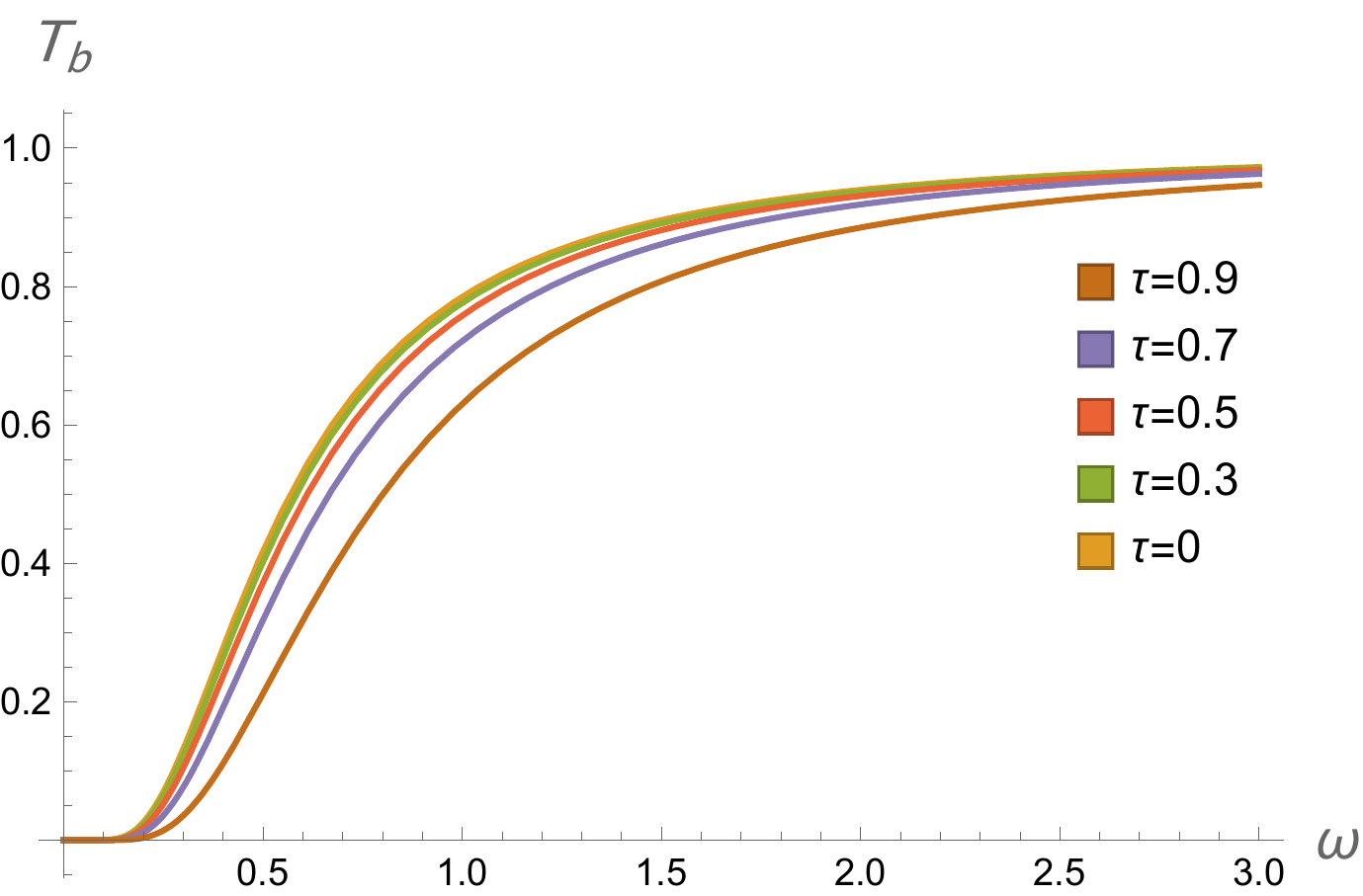}
    \caption{(Scalar Field) The Greybody Bound $T_b$ versus the $\omega$ for different values of torsion charge, with $m = 2$, $l=1$, $s=0$, and $Q = 0.18m$. (EM-field) The Greybody Bound $T_b$ versus the $\omega$ for different values of torsion charge$\tau$, with $m = 2$, $l=1$, $s=1$, and $Q = 0.18m$.}
    \label{fig:greybody}
\end{figure}

\begin{figure}
    \centering
	\includegraphics[width=0.48\textwidth]{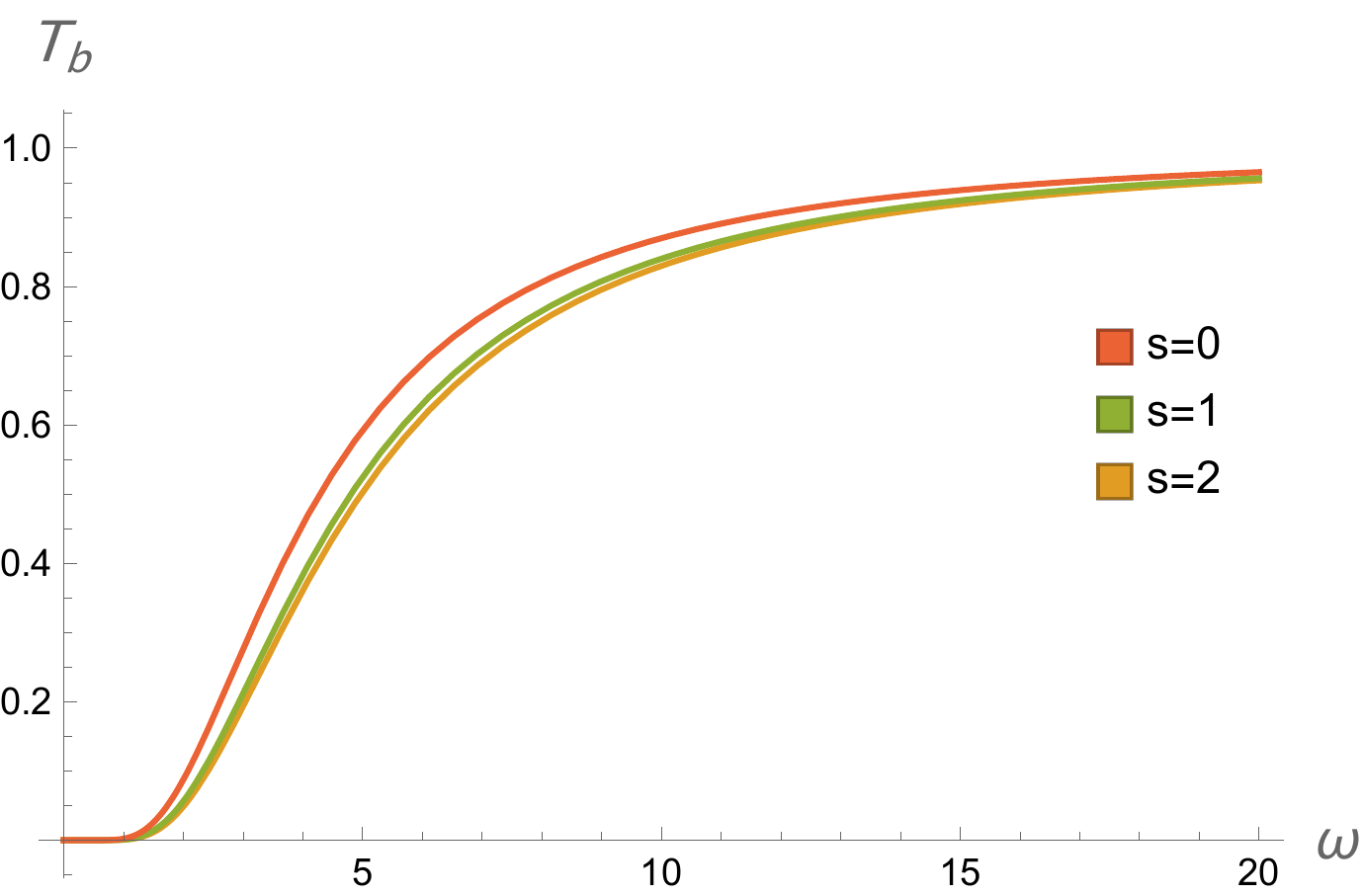}
    \caption{The Greybody Bound $T_b$ versus the $\omega$ for different values of the $s$, with $\tau=0.9m$ $m = 2$, $l=1$, $s=1$, and $Q = 0.18m$.}
    \label{fig:greybody3}
\end{figure}

In the previous section, we give the RW potential for the massless scalar field, and using it we can calculate the bound as  follows:
\begin{equation}
    T \geq T_b = \text{sech}^2\left[\frac{-\frac{2 \left(\tau ^2+Q^2\right)}{3 r_{\text{outer}}^3}+\frac{l (l+1)}{r_{\text{outer}}}+\frac{m}{r_{\text{outer}}^2}}{2 \omega }\right].
\end{equation}
Note that this bound reduces to the Schwarzschild case at $ (\tau,Q)\rightarrow 0$ correctly, as $T_\text{Sch} \geq\text{sech}^2\left(\frac{2 l (l+1)+1}{8 m \omega }\right)$. To show the effect of dynamical torsion on the greybody bound of the black hole, we plot it in Fig. \ref{fig:greybody} (Scalar Field). Similarly, using the RW potential for the EM-field from the last section, we calculate the bound as follows:
\begin{equation}
    T \geq T_b =   \text{sech}^2\left[\frac{l (l+1)}{2 \omega  \left(\sqrt{-\tau ^2-Q^2+m^2}+m\right)}\right],
\end{equation}
then we plot it to show the effect of the dynamical torsion parameter on the greybody factor in Fig. \ref{fig:greybody} (EM field). Then we plot the greybody bound versus the $\omega$ for different values of $s$ to show the effect of the spin where increase the bound when the spin number increases from zero to two in Fig. \ref{fig:greybody3}.

Lastly, we calculate the high-energy absorption cross-section via Sinc approximation. First, Sanchez studied the absorption cross-section for the Schwarzschild black hole in the high-frequency regime and it was shown that increasing the frequency for the ordinary material sphere, monotonically increase the absorption cross-section oscillated around the constant geometric-optics value for the black hole (related to the photon sphere) \cite{Sanchez:1977si}. Then the relation between the impact parameter and cross-section of the photon sphere is given at critical values and limiting the value of the absorption cross-section. One can conclude at low energy scales, it is the characteristic properties of BH and the cross-section equals to BH area \cite{Das:1996we}. On the other hand, at high energies using the geometrical cross-section of the photon sphere can be studied using the complex angular momentum technique \cite{Decanini:2011xi}. Decanini et al. use the Regge pole techniques to prove the oscillatory pattern of the high-energy absorption cross-section related to a Sinc(x) function including the photon sphere.

\begin{figure}
    \centering
	\includegraphics[width=0.48\textwidth]{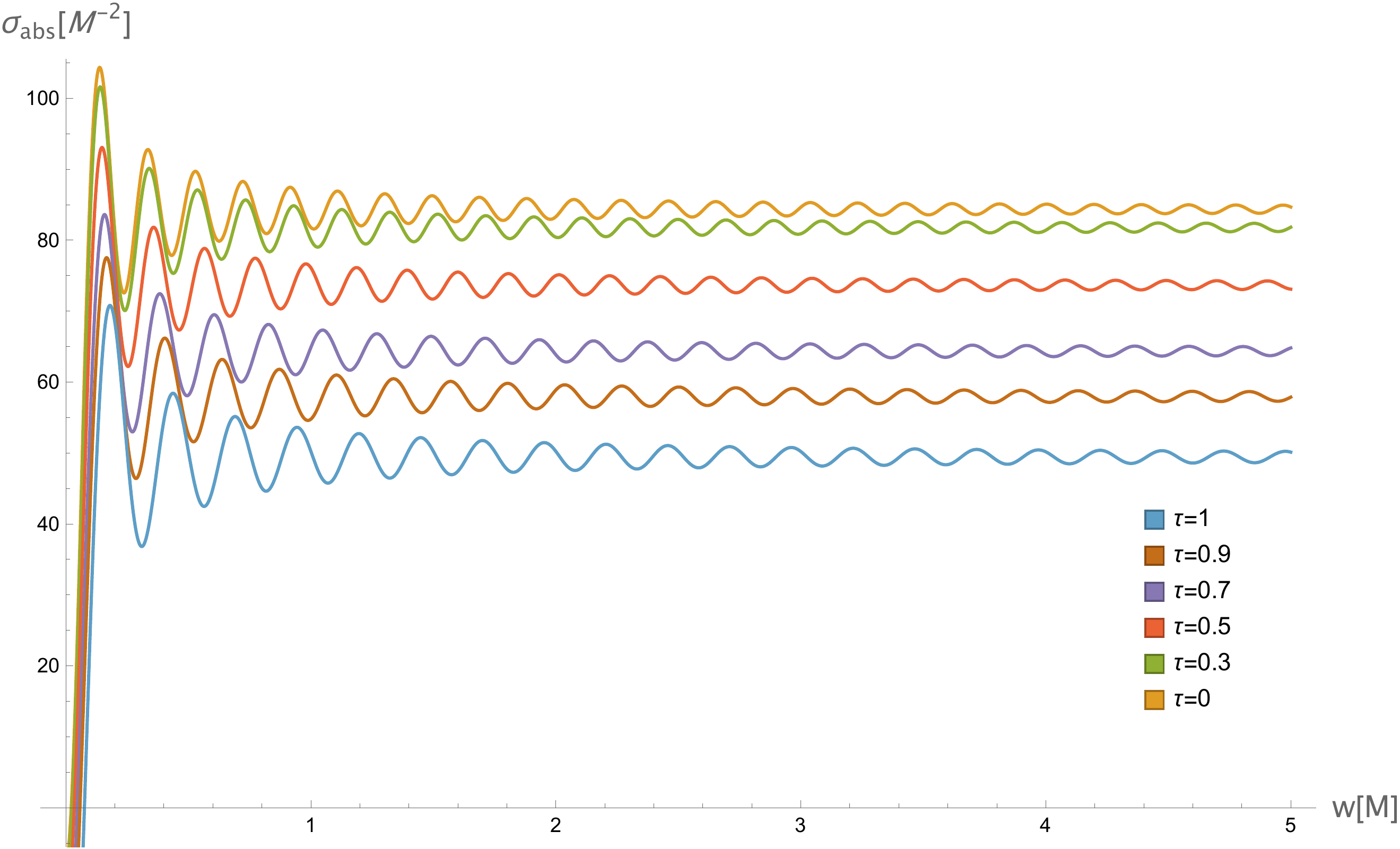}
    \caption{The total absorption cross section for various values of the torsion charge $\tau$, with $m=2$, and $Q = 0.18m$.}
    \label{fig:totabsorption}
\end{figure}

At high frequency, the absorption cross section is approximately equal to the classical capture cross-section of null geodesics: $\sigma_\text{geo} = \pi b_\text{crit}^2$ with the critical impact parameter $b_\text{crit}$. Then one can calculate the oscillatory part of the absorption cross section in the eikonal limit as follows \cite{Decanini:2011xi}:
\begin{equation}
    \sigma_\text{osc} = - 4 \pi \frac{\lambda b_\text{crit}^2}{w} e^{-\pi \lambda_L} \sin\frac{2 \pi w}{\Omega_\text{ph}},
\end{equation}
where $\lambda_L$ is known as the Lyapunov exponent given in Eq. \eqref{Lyapunov} and angular velocity is $\Omega_\text{ph}$ with the radius of the photon sphere $r_\text{ph}$. Hence, the Sinc approximation said that the total absorption cross section at the eikonal limit is $\sigma_\text{abs} \approx \sigma_\text{osc} + \sigma_\text{geo}$ \cite{Sanchez:1977si,Decanini:2011xi,Magalhaes:2020sea,Paula:2020yfr,Lima:2020seq,Boonserm:2019mon,Xavier:2021sje}.  In the Fig. \ref{fig:totabsorption} we plot the total absorption cross section for various values of $\tau$. It can be seen that as the values of dynamical torsion parameter increase, the total absorption cross section decreases. Moreover, there is a regular oscillatory behavior around the high-frequency limit.

\section{Conclusion} \label{sec9}
In the present work, we have analyzed the effect of the torsion in correspondence with the RN geometries with cosmological constant within the framework of PG theory with massive torsion, first, to the weak deflection angle of massive and null particles. Our findings indicate that, in general, increasing the value of the torsion parameter decreases the value of the weak deflection angle, and its effect diminishes at distances comparable to the cosmological horizon. In addition, both the dS and AdS cases gave interesting and unique differences in the behavior of the weak deflection angle. First, the deflection angle of massive particles can give more detectability due to a higher value of $\alpha$ as compared to the null particles, especially in the AdS case. For a receiver near the black hole (Fig. \ref{wda_new}), $\alpha$ is lesser compared to the receiver far from the black hole (Fig. \ref{wda_new2}). Such a higher value of $\alpha$ is compensated by the ability to detect positive deflection angles at larger distances. Intuitively, we should expect that the effect of the cosmological constant will manifest greatly at distances comparable to the cosmological horizon. However, in the case of the weak deflection angle, its effect already manifests greatly for receivers with a considerable distance from the BH.

Meanwhile, the behavior of the shadow radius was also examined. The main result of the study, for the torsion effect, is similar to the $\alpha$. More interestingly, considering the Earth's location from the known BHs in our galaxy and M87, we find that the black hole shadow does not react strongly to the cosmological background, unlike in $\hat{\alpha}$. Nonetheless, the effect is magnified near the cosmological horizon. We also see differing behavior when compared to the AdS case. We also considered the co-moving case, for dark energy, matter, and radiation-dominated Universes. We get similar observations at cosmological distances. Comparable to Earth's location, however, we can potentially detect a deviation in the shadow, but it will be difficult to differentiate between which type of Universe we live in.

Moreover, we have also analyzed the effects of the torsion parameter on the shadow with infalling spherical accretion. It is shown that the dynamical torsion parameter $\tau$ has some effects on the thin accretion disk. For increasing the dynamical torsion parameter $\tau$, the luminosity of the photon sphere rises gradually in Figs. (\ref{fig:thinacc1}, \ref{fig:thinacc2} and \ref{fig:thinacc3}). Moreover, we have calculated the QNMs in the eikonal limit and showed our results in Table \ref{tab:table2} that the real parts increase, but the imaginary part of the QNMs decrease with the increasing dynamical torsion parameter $\tau$. Last, we have obtained greybody bounds and the high-energy absorption cross-section via Sinc approximation in the high-frequency regime and have shown how the dynamical torsion parameter $\tau$ affects them in Figs. \ref{fig:greybody},\ref{fig:greybody3}, \ref{fig:totabsorption}. The difference in the dynamical torsion parameter $\tau$ is indeed visible.

Research prospects include extending the study to obtain the rotating solution of the metric in Eq. \eqref{metric}. It would also be interesting to find the exact form of the shadow radius expression for an observer that is co-moving with the cosmological expansion.

\acknowledgments
A. {\"O}. and R. P. would like to acknowledge networking support by the COST Action CA18108 - Quantum gravity phenomenology in the multi-messenger approach (QG-MM).

\textbf{Data Availability Statement:} This manuscript has no associated data or the data will not be deposited.

\bibliography{references}

\begin{thebibliography}{137}
\expandafter\ifx\csname natexlab\endcsname\relax\def\natexlab#1{#1}\fi
\expandafter\ifx\csname bibnamefont\endcsname\relax
  \def\bibnamefont#1{#1}\fi
\expandafter\ifx\csname bibfnamefont\endcsname\relax
  \def\bibfnamefont#1{#1}\fi
\expandafter\ifx\csname citenamefont\endcsname\relax
  \def\citenamefont#1{#1}\fi
\expandafter\ifx\csname url\endcsname\relax
  \def\url#1{\texttt{#1}}\fi
\expandafter\ifx\csname urlprefix\endcsname\relax\def\urlprefix{URL }\fi
\providecommand{\bibinfo}[2]{#2}
\providecommand{\eprint}[2][]{\url{#2}}

\bibitem[{\citenamefont{Einstein}(1916)}]{Einstein:1916vd}
\bibinfo{author}{\bibfnamefont{A.}~\bibnamefont{Einstein}},
  \bibinfo{journal}{Annalen Phys.} \textbf{\bibinfo{volume}{49}},
  \bibinfo{pages}{769} (\bibinfo{year}{1916}).

\bibitem[{\citenamefont{Dyson et~al.}(1920)\citenamefont{Dyson, Eddington, and
  Davidson}}]{Dyson:1920cwa}
\bibinfo{author}{\bibfnamefont{F.~W.} \bibnamefont{Dyson}},
  \bibinfo{author}{\bibfnamefont{A.~S.} \bibnamefont{Eddington}},
  \bibnamefont{and} \bibinfo{author}{\bibfnamefont{C.}~\bibnamefont{Davidson}},
  \bibinfo{journal}{Phil. Trans. Roy. Soc. Lond. A}
  \textbf{\bibinfo{volume}{220}}, \bibinfo{pages}{291} (\bibinfo{year}{1920}).

\bibitem[{\citenamefont{Abbott et~al.}(2016)}]{LIGOScientific:2016aoc}
\bibinfo{author}{\bibfnamefont{B.~P.} \bibnamefont{Abbott}}
  \bibnamefont{et~al.} (\bibinfo{collaboration}{LIGO Scientific, Virgo}),
  \bibinfo{journal}{Phys. Rev. Lett.} \textbf{\bibinfo{volume}{116}},
  \bibinfo{pages}{061102} (\bibinfo{year}{2016}), \eprint{1602.03837}.

\bibitem[{\citenamefont{Akiyama et~al.}(2019)}]{EventHorizonTelescope:2019dse}
\bibinfo{author}{\bibfnamefont{K.}~\bibnamefont{Akiyama}} \bibnamefont{et~al.}
  (\bibinfo{collaboration}{Event Horizon Telescope}),
  \bibinfo{journal}{Astrophys. J. Lett.} \textbf{\bibinfo{volume}{875}},
  \bibinfo{pages}{L1} (\bibinfo{year}{2019}), \eprint{1906.11238}.

\bibitem[{\citenamefont{Akiyama et~al.}(2022)}]{EventHorizonTelescope:2022xnr}
\bibinfo{author}{\bibfnamefont{K.}~\bibnamefont{Akiyama}} \bibnamefont{et~al.}
  (\bibinfo{collaboration}{Event Horizon Telescope}),
  \bibinfo{journal}{Astrophys. J. Lett.} \textbf{\bibinfo{volume}{930}},
  \bibinfo{pages}{L12} (\bibinfo{year}{2022}).

\bibitem[{\citenamefont{{Weyl}}(1929)}]{weyl}
\bibinfo{author}{\bibfnamefont{H.}~\bibnamefont{{Weyl}}},
  \bibinfo{journal}{Zeitschrift fur Physik} \textbf{\bibinfo{volume}{56}},
  \bibinfo{pages}{330} (\bibinfo{year}{1929}).

\bibitem[{\citenamefont{Cartan et~al.}(1986)\citenamefont{Cartan, Magnon,
  Ashtekar, and Trautman}}]{cartan1986manifolds}
\bibinfo{author}{\bibfnamefont{E.}~\bibnamefont{Cartan}},
  \bibinfo{author}{\bibfnamefont{A.}~\bibnamefont{Magnon}},
  \bibinfo{author}{\bibfnamefont{A.}~\bibnamefont{Ashtekar}}, \bibnamefont{and}
  \bibinfo{author}{\bibfnamefont{A.}~\bibnamefont{Trautman}},
  \emph{\bibinfo{title}{On Manifolds with an Affine Connection and the Theory
  of General Relativity}}, Lecture notes (\bibinfo{publisher}{Bibliopolis},
  \bibinfo{year}{1986}), ISBN \bibinfo{isbn}{9788870880861}.

\bibitem[{\citenamefont{{Fock}}(1929)}]{fock}
\bibinfo{author}{\bibfnamefont{V.}~\bibnamefont{{Fock}}},
  \bibinfo{journal}{Zeitschrift fur Physik} \textbf{\bibinfo{volume}{57}},
  \bibinfo{pages}{261} (\bibinfo{year}{1929}).

\bibitem[{\citenamefont{Utiyama}(1956)}]{Utiyama}
\bibinfo{author}{\bibfnamefont{R.}~\bibnamefont{Utiyama}},
  \bibinfo{journal}{Phys. Rev.} \textbf{\bibinfo{volume}{101}},
  \bibinfo{pages}{1597} (\bibinfo{year}{1956}).

\bibitem[{\citenamefont{Sciama}(1962)}]{sciama1962analogy}
\bibinfo{author}{\bibfnamefont{D.~W.} \bibnamefont{Sciama}},
  \bibinfo{journal}{Recent developments in general relativity} p.
  \bibinfo{pages}{415} (\bibinfo{year}{1962}).

\bibitem[{\citenamefont{Kibble}(1961)}]{Kibble:1961ba}
\bibinfo{author}{\bibfnamefont{T.~W.~B.} \bibnamefont{Kibble}},
  \bibinfo{journal}{J. Math. Phys.} \textbf{\bibinfo{volume}{2}},
  \bibinfo{pages}{212} (\bibinfo{year}{1961}).

\bibitem[{\citenamefont{Blagojevic and Hehl}(2012)}]{Blagojevic:2012bc}
\bibinfo{author}{\bibfnamefont{M.}~\bibnamefont{Blagojevic}} \bibnamefont{and}
  \bibinfo{author}{\bibfnamefont{F.~W.} \bibnamefont{Hehl}}
  (\bibinfo{year}{2012}), \eprint{1210.3775}.

\bibitem[{\citenamefont{Hehl et~al.}(1976)\citenamefont{Hehl, Von Der~Heyde,
  Kerlick, and Nester}}]{Hehltorsion}
\bibinfo{author}{\bibfnamefont{F.~W.} \bibnamefont{Hehl}},
  \bibinfo{author}{\bibfnamefont{P.}~\bibnamefont{Von Der~Heyde}},
  \bibinfo{author}{\bibfnamefont{G.~D.} \bibnamefont{Kerlick}},
  \bibnamefont{and} \bibinfo{author}{\bibfnamefont{J.~M.}
  \bibnamefont{Nester}}, \bibinfo{journal}{Rev. Mod. Phys.}
  \textbf{\bibinfo{volume}{48}}, \bibinfo{pages}{393} (\bibinfo{year}{1976}).

\bibitem[{\citenamefont{Cembranos and
  Gigante~Valcarcel}(2018)}]{Cembranos:2017pcs}
\bibinfo{author}{\bibfnamefont{J.~A.~R.} \bibnamefont{Cembranos}}
  \bibnamefont{and}
  \bibinfo{author}{\bibfnamefont{J.}~\bibnamefont{Gigante~Valcarcel}},
  \bibinfo{journal}{Phys. Lett. B} \textbf{\bibinfo{volume}{779}},
  \bibinfo{pages}{143} (\bibinfo{year}{2018}), \eprint{1708.00374}.

\bibitem[{\citenamefont{Blagojevi\'c and
  Cvetkovi\'c}(2022)}]{Blagojevic:2021pqp}
\bibinfo{author}{\bibfnamefont{M.}~\bibnamefont{Blagojevi\'c}}
  \bibnamefont{and}
  \bibinfo{author}{\bibfnamefont{B.}~\bibnamefont{Cvetkovi\'c}},
  \bibinfo{journal}{Phys. Lett. B} \textbf{\bibinfo{volume}{824}},
  \bibinfo{pages}{136815} (\bibinfo{year}{2022}), \eprint{2112.02099}.

\bibitem[{\citenamefont{Obukhov}(2020)}]{Obukhov:2020hlp}
\bibinfo{author}{\bibfnamefont{Y.~N.} \bibnamefont{Obukhov}},
  \bibinfo{journal}{Phys. Rev. D} \textbf{\bibinfo{volume}{102}},
  \bibinfo{pages}{104059} (\bibinfo{year}{2020}), \eprint{2009.00284}.

\bibitem[{\citenamefont{Bahamonde and Valcarcel}(2020)}]{Bahamonde:2020fnq}
\bibinfo{author}{\bibfnamefont{S.}~\bibnamefont{Bahamonde}} \bibnamefont{and}
  \bibinfo{author}{\bibfnamefont{J.~G.} \bibnamefont{Valcarcel}},
  \bibinfo{journal}{JCAP} \textbf{\bibinfo{volume}{09}}, \bibinfo{pages}{057}
  (\bibinfo{year}{2020}), \eprint{2006.06749}.

\bibitem[{\citenamefont{He et~al.}(2022{\natexlab{a}})\citenamefont{He, Tan,
  and Li}}]{he2022}
\bibinfo{author}{\bibfnamefont{K.-J.} \bibnamefont{He}},
  \bibinfo{author}{\bibfnamefont{S.-C.} \bibnamefont{Tan}}, \bibnamefont{and}
  \bibinfo{author}{\bibfnamefont{G.-P.} \bibnamefont{Li}},
  \bibinfo{journal}{Eur. Phys. J. C} \textbf{\bibinfo{volume}{82}},
  \bibinfo{pages}{81} (\bibinfo{year}{2022}{\natexlab{a}}).

\bibitem[{\citenamefont{Gkigkitzis et~al.}(2019)\citenamefont{Gkigkitzis,
  Haranas, and Cavan}}]{gkigkitzis2019}
\bibinfo{author}{\bibfnamefont{I.}~\bibnamefont{Gkigkitzis}},
  \bibinfo{author}{\bibfnamefont{I.}~\bibnamefont{Haranas}}, \bibnamefont{and}
  \bibinfo{author}{\bibfnamefont{E.}~\bibnamefont{Cavan}},
  \bibinfo{journal}{Ukr. J. Phys.} \textbf{\bibinfo{volume}{64}},
  \bibinfo{pages}{683} (\bibinfo{year}{2019}).

\bibitem[{\citenamefont{Nikiforova}(2020)}]{nikiforova2020}
\bibinfo{author}{\bibfnamefont{V.}~\bibnamefont{Nikiforova}},
  \bibinfo{journal}{Phys. Rev. D} \textbf{\bibinfo{volume}{102}},
  \bibinfo{pages}{124007} (\bibinfo{year}{2020}), \eprint{2010.05910}.

\bibitem[{\citenamefont{Kottler}(1918)}]{kottler1918}
\bibinfo{author}{\bibfnamefont{F.}~\bibnamefont{Kottler}},
  \bibinfo{journal}{Annalen der Physik} \textbf{\bibinfo{volume}{361}},
  \bibinfo{pages}{401} (\bibinfo{year}{1918}).

\bibitem[{\citenamefont{Perlick
  et~al.}(2018{\natexlab{a}})\citenamefont{Perlick, Tsupko, and
  Bisnovatyi-Kogan}}]{Perlick:2018}
\bibinfo{author}{\bibfnamefont{V.}~\bibnamefont{Perlick}},
  \bibinfo{author}{\bibfnamefont{O.~Y.} \bibnamefont{Tsupko}},
  \bibnamefont{and} \bibinfo{author}{\bibfnamefont{G.~S.}
  \bibnamefont{Bisnovatyi-Kogan}}, \bibinfo{journal}{Phys. Rev. D}
  \textbf{\bibinfo{volume}{97}}, \bibinfo{pages}{104062}
  (\bibinfo{year}{2018}{\natexlab{a}}).

\bibitem[{\citenamefont{Bisnovatyi-Kogan and Tsupko}(2018)}]{bisnovatyi2018}
\bibinfo{author}{\bibfnamefont{G.~S.} \bibnamefont{Bisnovatyi-Kogan}}
  \bibnamefont{and} \bibinfo{author}{\bibfnamefont{O.~Y.}
  \bibnamefont{Tsupko}}, \bibinfo{journal}{Phys. Rev. D}
  \textbf{\bibinfo{volume}{98}}, \bibinfo{pages}{084020}
  (\bibinfo{year}{2018}), \eprint{1805.03311}.

\bibitem[{\citenamefont{Firouzjaee and Allahyari}(2019)}]{firouzjaee2019}
\bibinfo{author}{\bibfnamefont{J.~T.} \bibnamefont{Firouzjaee}}
  \bibnamefont{and}
  \bibinfo{author}{\bibfnamefont{A.}~\bibnamefont{Allahyari}},
  \bibinfo{journal}{Eur. Phys. J. C} \textbf{\bibinfo{volume}{79}},
  \bibinfo{pages}{930} (\bibinfo{year}{2019}), \eprint{1905.07378}.

\bibitem[{\citenamefont{Tsupko and Bisnovatyi-Kogan}(2020)}]{Tsupko2020}
\bibinfo{author}{\bibfnamefont{O.~Y.} \bibnamefont{Tsupko}} \bibnamefont{and}
  \bibinfo{author}{\bibfnamefont{G.~S.} \bibnamefont{Bisnovatyi-Kogan}},
  \bibinfo{journal}{International Journal of Modern Physics D}
  \textbf{\bibinfo{volume}{29}}, \bibinfo{pages}{2050062}
  (\bibinfo{year}{2020}).

\bibitem[{\citenamefont{Maluf and Neves}(2021)}]{maluf2021}
\bibinfo{author}{\bibfnamefont{R.~V.} \bibnamefont{Maluf}} \bibnamefont{and}
  \bibinfo{author}{\bibfnamefont{J.~C.~S.} \bibnamefont{Neves}},
  \bibinfo{journal}{Phys. Rev. D} \textbf{\bibinfo{volume}{103}},
  \bibinfo{pages}{044002} (\bibinfo{year}{2021}), \eprint{2011.12841}.

\bibitem[{\citenamefont{Synge}(1966)}]{Synge1966}
\bibinfo{author}{\bibfnamefont{J.~L.} \bibnamefont{Synge}},
  \bibinfo{journal}{Mon. Not. R. Astron. Soc.} \textbf{\bibinfo{volume}{131}},
  \bibinfo{pages}{463} (\bibinfo{year}{1966}), ISSN \bibinfo{issn}{0035-8711}.

\bibitem[{\citenamefont{Luminet}(1979)}]{Luminet1979}
\bibinfo{author}{\bibfnamefont{J.~P.} \bibnamefont{Luminet}},
  \bibinfo{journal}{Astron. Astrophys.} \textbf{\bibinfo{volume}{75}},
  \bibinfo{pages}{228} (\bibinfo{year}{1979}).

\bibitem[{\citenamefont{Herdeiro et~al.}(2021)\citenamefont{Herdeiro, Pombo,
  Radu, Cunha, and Sanchis-Gual}}]{Herdeiro:2021lwl}
\bibinfo{author}{\bibfnamefont{C.~A.~R.} \bibnamefont{Herdeiro}},
  \bibinfo{author}{\bibfnamefont{A.~M.} \bibnamefont{Pombo}},
  \bibinfo{author}{\bibfnamefont{E.}~\bibnamefont{Radu}},
  \bibinfo{author}{\bibfnamefont{P.~V.~P.} \bibnamefont{Cunha}},
  \bibnamefont{and}
  \bibinfo{author}{\bibfnamefont{N.}~\bibnamefont{Sanchis-Gual}},
  \bibinfo{journal}{JCAP} \textbf{\bibinfo{volume}{04}}, \bibinfo{pages}{051}
  (\bibinfo{year}{2021}), \eprint{2102.01703}.

\bibitem[{\citenamefont{Cunha et~al.}(2020)\citenamefont{Cunha, Eir\'o,
  Herdeiro, and Lemos}}]{Cunha:2019hzj}
\bibinfo{author}{\bibfnamefont{P.~V.~P.} \bibnamefont{Cunha}},
  \bibinfo{author}{\bibfnamefont{N.~A.} \bibnamefont{Eir\'o}},
  \bibinfo{author}{\bibfnamefont{C.~A.~R.} \bibnamefont{Herdeiro}},
  \bibnamefont{and} \bibinfo{author}{\bibfnamefont{J.~P.~S.}
  \bibnamefont{Lemos}}, \bibinfo{journal}{JCAP} \textbf{\bibinfo{volume}{03}},
  \bibinfo{pages}{035} (\bibinfo{year}{2020}), \eprint{1912.08833}.

\bibitem[{\citenamefont{Cunha et~al.}(2019)\citenamefont{Cunha, Herdeiro, and
  Radu}}]{Cunha:2019ikd}
\bibinfo{author}{\bibfnamefont{P.~V.~P.} \bibnamefont{Cunha}},
  \bibinfo{author}{\bibfnamefont{C.~A.~R.} \bibnamefont{Herdeiro}},
  \bibnamefont{and} \bibinfo{author}{\bibfnamefont{E.}~\bibnamefont{Radu}},
  \bibinfo{journal}{Universe} \textbf{\bibinfo{volume}{5}},
  \bibinfo{pages}{220} (\bibinfo{year}{2019}), \eprint{1909.08039}.

\bibitem[{\citenamefont{Cunha and Herdeiro}(2018)}]{Cunha:2018acu}
\bibinfo{author}{\bibfnamefont{P.~V.~P.} \bibnamefont{Cunha}} \bibnamefont{and}
  \bibinfo{author}{\bibfnamefont{C.~A.~R.} \bibnamefont{Herdeiro}},
  \bibinfo{journal}{Gen. Rel. Grav.} \textbf{\bibinfo{volume}{50}},
  \bibinfo{pages}{42} (\bibinfo{year}{2018}), \eprint{1801.00860}.

\bibitem[{\citenamefont{Cunha et~al.}(2017)\citenamefont{Cunha, Herdeiro,
  Kleihaus, Kunz, and Radu}}]{Cunha:2016wzk}
\bibinfo{author}{\bibfnamefont{P.~V.~P.} \bibnamefont{Cunha}},
  \bibinfo{author}{\bibfnamefont{C.~A.~R.} \bibnamefont{Herdeiro}},
  \bibinfo{author}{\bibfnamefont{B.}~\bibnamefont{Kleihaus}},
  \bibinfo{author}{\bibfnamefont{J.}~\bibnamefont{Kunz}}, \bibnamefont{and}
  \bibinfo{author}{\bibfnamefont{E.}~\bibnamefont{Radu}},
  \bibinfo{journal}{Phys. Lett. B} \textbf{\bibinfo{volume}{768}},
  \bibinfo{pages}{373} (\bibinfo{year}{2017}), \eprint{1701.00079}.

\bibitem[{\citenamefont{Vincent et~al.}(2016)\citenamefont{Vincent,
  Gourgoulhon, Herdeiro, and Radu}}]{Vincent:2016sjq}
\bibinfo{author}{\bibfnamefont{F.~H.} \bibnamefont{Vincent}},
  \bibinfo{author}{\bibfnamefont{E.}~\bibnamefont{Gourgoulhon}},
  \bibinfo{author}{\bibfnamefont{C.}~\bibnamefont{Herdeiro}}, \bibnamefont{and}
  \bibinfo{author}{\bibfnamefont{E.}~\bibnamefont{Radu}},
  \bibinfo{journal}{Phys. Rev. D} \textbf{\bibinfo{volume}{94}},
  \bibinfo{pages}{084045} (\bibinfo{year}{2016}), \eprint{1606.04246}.

\bibitem[{\citenamefont{Afrin et~al.}(2021)\citenamefont{Afrin, Kumar, and
  Ghosh}}]{Afrin:2021imp}
\bibinfo{author}{\bibfnamefont{M.}~\bibnamefont{Afrin}},
  \bibinfo{author}{\bibfnamefont{R.}~\bibnamefont{Kumar}}, \bibnamefont{and}
  \bibinfo{author}{\bibfnamefont{S.~G.} \bibnamefont{Ghosh}},
  \bibinfo{journal}{Mon. Not. Roy. Astron. Soc.}
  \textbf{\bibinfo{volume}{504}}, \bibinfo{pages}{5927} (\bibinfo{year}{2021}),
  \eprint{2103.11417}.

\bibitem[{\citenamefont{Jha and Rahaman}(2021)}]{Jha:2021bue}
\bibinfo{author}{\bibfnamefont{S.~K.} \bibnamefont{Jha}} \bibnamefont{and}
  \bibinfo{author}{\bibfnamefont{A.}~\bibnamefont{Rahaman}}
  (\bibinfo{year}{2021}), \eprint{2111.02817}.

\bibitem[{\citenamefont{Khodadi et~al.}(2021)\citenamefont{Khodadi, Lambiase,
  and Mota}}]{Khodadi:2021gbc}
\bibinfo{author}{\bibfnamefont{M.}~\bibnamefont{Khodadi}},
  \bibinfo{author}{\bibfnamefont{G.}~\bibnamefont{Lambiase}}, \bibnamefont{and}
  \bibinfo{author}{\bibfnamefont{D.~F.} \bibnamefont{Mota}},
  \bibinfo{journal}{JCAP} \textbf{\bibinfo{volume}{09}}, \bibinfo{pages}{028}
  (\bibinfo{year}{2021}), \eprint{2107.00834}.

\bibitem[{\citenamefont{Khodadi et~al.}(2020)\citenamefont{Khodadi, Allahyari,
  Vagnozzi, and Mota}}]{Khodadi:2020jij}
\bibinfo{author}{\bibfnamefont{M.}~\bibnamefont{Khodadi}},
  \bibinfo{author}{\bibfnamefont{A.}~\bibnamefont{Allahyari}},
  \bibinfo{author}{\bibfnamefont{S.}~\bibnamefont{Vagnozzi}}, \bibnamefont{and}
  \bibinfo{author}{\bibfnamefont{D.~F.} \bibnamefont{Mota}},
  \bibinfo{journal}{JCAP} \textbf{\bibinfo{volume}{09}}, \bibinfo{pages}{026}
  (\bibinfo{year}{2020}), \eprint{2005.05992}.

\bibitem[{\citenamefont{Kumar and Ghosh}(2020{\natexlab{a}})}]{Kumar:2018ple}
\bibinfo{author}{\bibfnamefont{R.}~\bibnamefont{Kumar}} \bibnamefont{and}
  \bibinfo{author}{\bibfnamefont{S.~G.} \bibnamefont{Ghosh}},
  \bibinfo{journal}{Astrophys. J.} \textbf{\bibinfo{volume}{892}},
  \bibinfo{pages}{78} (\bibinfo{year}{2020}{\natexlab{a}}),
  \eprint{1811.01260}.

\bibitem[{\citenamefont{Kumar and Ghosh}(2020{\natexlab{b}})}]{Kumar:2020owy}
\bibinfo{author}{\bibfnamefont{R.}~\bibnamefont{Kumar}} \bibnamefont{and}
  \bibinfo{author}{\bibfnamefont{S.~G.} \bibnamefont{Ghosh}},
  \bibinfo{journal}{JCAP} \textbf{\bibinfo{volume}{07}}, \bibinfo{pages}{053}
  (\bibinfo{year}{2020}{\natexlab{b}}), \eprint{2003.08927}.

\bibitem[{\citenamefont{Zeng et~al.}(2020)\citenamefont{Zeng, Zhang, and
  Zhang}}]{Zeng:2020dco}
\bibinfo{author}{\bibfnamefont{X.-X.} \bibnamefont{Zeng}},
  \bibinfo{author}{\bibfnamefont{H.-Q.} \bibnamefont{Zhang}}, \bibnamefont{and}
  \bibinfo{author}{\bibfnamefont{H.}~\bibnamefont{Zhang}},
  \bibinfo{journal}{Eur. Phys. J. C} \textbf{\bibinfo{volume}{80}},
  \bibinfo{pages}{872} (\bibinfo{year}{2020}), \eprint{2004.12074}.

\bibitem[{\citenamefont{He et~al.}(2022{\natexlab{b}})\citenamefont{He, Tan,
  and Li}}]{He:2022yse}
\bibinfo{author}{\bibfnamefont{K.-J.} \bibnamefont{He}},
  \bibinfo{author}{\bibfnamefont{S.-C.} \bibnamefont{Tan}}, \bibnamefont{and}
  \bibinfo{author}{\bibfnamefont{G.-P.} \bibnamefont{Li}},
  \bibinfo{journal}{Eur. Phys. J. C} \textbf{\bibinfo{volume}{82}},
  \bibinfo{pages}{81} (\bibinfo{year}{2022}{\natexlab{b}}).

\bibitem[{\citenamefont{Dokuchaev and Nazarova}(2020)}]{Dokuchaev:2020wqk}
\bibinfo{author}{\bibfnamefont{V.~I.} \bibnamefont{Dokuchaev}}
  \bibnamefont{and} \bibinfo{author}{\bibfnamefont{N.~O.}
  \bibnamefont{Nazarova}}, \bibinfo{journal}{Universe}
  \textbf{\bibinfo{volume}{6}}, \bibinfo{pages}{154} (\bibinfo{year}{2020}),
  \eprint{2007.14121}.

\bibitem[{\citenamefont{Vagnozzi et~al.}(2022)}]{Vagnozzi:2022moj}
\bibinfo{author}{\bibfnamefont{S.}~\bibnamefont{Vagnozzi}} \bibnamefont{et~al.}
  (\bibinfo{year}{2022}), \eprint{2205.07787}.

\bibitem[{\citenamefont{Roy et~al.}(2022)\citenamefont{Roy, Vagnozzi, and
  Visinelli}}]{Roy:2021uye}
\bibinfo{author}{\bibfnamefont{R.}~\bibnamefont{Roy}},
  \bibinfo{author}{\bibfnamefont{S.}~\bibnamefont{Vagnozzi}}, \bibnamefont{and}
  \bibinfo{author}{\bibfnamefont{L.}~\bibnamefont{Visinelli}},
  \bibinfo{journal}{Phys. Rev. D} \textbf{\bibinfo{volume}{105}},
  \bibinfo{pages}{083002} (\bibinfo{year}{2022}), \eprint{2112.06932}.

\bibitem[{\citenamefont{Vagnozzi and Visinelli}(2019)}]{Vagnozzi:2019apd}
\bibinfo{author}{\bibfnamefont{S.}~\bibnamefont{Vagnozzi}} \bibnamefont{and}
  \bibinfo{author}{\bibfnamefont{L.}~\bibnamefont{Visinelli}},
  \bibinfo{journal}{Phys. Rev. D} \textbf{\bibinfo{volume}{100}},
  \bibinfo{pages}{024020} (\bibinfo{year}{2019}), \eprint{1905.12421}.

\bibitem[{\citenamefont{Allahyari et~al.}(2020)\citenamefont{Allahyari,
  Khodadi, Vagnozzi, and Mota}}]{Allahyari:2019jqz}
\bibinfo{author}{\bibfnamefont{A.}~\bibnamefont{Allahyari}},
  \bibinfo{author}{\bibfnamefont{M.}~\bibnamefont{Khodadi}},
  \bibinfo{author}{\bibfnamefont{S.}~\bibnamefont{Vagnozzi}}, \bibnamefont{and}
  \bibinfo{author}{\bibfnamefont{D.~F.} \bibnamefont{Mota}},
  \bibinfo{journal}{JCAP} \textbf{\bibinfo{volume}{02}}, \bibinfo{pages}{003}
  (\bibinfo{year}{2020}), \eprint{1912.08231}.

\bibitem[{\citenamefont{Bambi et~al.}(2019)\citenamefont{Bambi, Freese,
  Vagnozzi, and Visinelli}}]{Bambi:2019tjh}
\bibinfo{author}{\bibfnamefont{C.}~\bibnamefont{Bambi}},
  \bibinfo{author}{\bibfnamefont{K.}~\bibnamefont{Freese}},
  \bibinfo{author}{\bibfnamefont{S.}~\bibnamefont{Vagnozzi}}, \bibnamefont{and}
  \bibinfo{author}{\bibfnamefont{L.}~\bibnamefont{Visinelli}},
  \bibinfo{journal}{Phys. Rev. D} \textbf{\bibinfo{volume}{100}},
  \bibinfo{pages}{044057} (\bibinfo{year}{2019}), \eprint{1904.12983}.

\bibitem[{\citenamefont{Meng et~al.}(2022)\citenamefont{Meng, Kuang, and
  Tang}}]{Meng:2022kjs}
\bibinfo{author}{\bibfnamefont{Y.}~\bibnamefont{Meng}},
  \bibinfo{author}{\bibfnamefont{X.-M.} \bibnamefont{Kuang}}, \bibnamefont{and}
  \bibinfo{author}{\bibfnamefont{Z.-Y.} \bibnamefont{Tang}}
  (\bibinfo{year}{2022}), \eprint{2204.00897}.

\bibitem[{\citenamefont{Chen}(2022)}]{Chen:2022lct}
\bibinfo{author}{\bibfnamefont{C.-Y.} \bibnamefont{Chen}}
  (\bibinfo{year}{2022}), \eprint{2205.06962}.

\bibitem[{\citenamefont{Chen et~al.}(2022)\citenamefont{Chen, Roy, Vagnozzi,
  and Visinelli}}]{Chen:2022nbb}
\bibinfo{author}{\bibfnamefont{Y.}~\bibnamefont{Chen}},
  \bibinfo{author}{\bibfnamefont{R.}~\bibnamefont{Roy}},
  \bibinfo{author}{\bibfnamefont{S.}~\bibnamefont{Vagnozzi}}, \bibnamefont{and}
  \bibinfo{author}{\bibfnamefont{L.}~\bibnamefont{Visinelli}}
  (\bibinfo{year}{2022}), \eprint{2205.06238}.

\bibitem[{\citenamefont{Wang et~al.}(2022)\citenamefont{Wang, Chen, and
  Jing}}]{Wang:2022kvg}
\bibinfo{author}{\bibfnamefont{M.}~\bibnamefont{Wang}},
  \bibinfo{author}{\bibfnamefont{S.}~\bibnamefont{Chen}}, \bibnamefont{and}
  \bibinfo{author}{\bibfnamefont{J.}~\bibnamefont{Jing}}
  (\bibinfo{year}{2022}), \eprint{2205.05855}.

\bibitem[{\citenamefont{Bronzwaer and Falcke}(2021)}]{Bronzwaer:2021lzo}
\bibinfo{author}{\bibfnamefont{T.}~\bibnamefont{Bronzwaer}} \bibnamefont{and}
  \bibinfo{author}{\bibfnamefont{H.}~\bibnamefont{Falcke}},
  \bibinfo{journal}{Astrophys. J.} \textbf{\bibinfo{volume}{920}},
  \bibinfo{pages}{155} (\bibinfo{year}{2021}), \eprint{2108.03966}.

\bibitem[{\citenamefont{Falcke et~al.}(2000)\citenamefont{Falcke, Melia, and
  Agol}}]{Falcke:1999pj}
\bibinfo{author}{\bibfnamefont{H.}~\bibnamefont{Falcke}},
  \bibinfo{author}{\bibfnamefont{F.}~\bibnamefont{Melia}}, \bibnamefont{and}
  \bibinfo{author}{\bibfnamefont{E.}~\bibnamefont{Agol}},
  \bibinfo{journal}{Astrophys. J. Lett.} \textbf{\bibinfo{volume}{528}},
  \bibinfo{pages}{L13} (\bibinfo{year}{2000}), \eprint{astro-ph/9912263}.

\bibitem[{\citenamefont{Atamurotov et~al.}(2013)\citenamefont{Atamurotov,
  Abdujabbarov, and Ahmedov}}]{Atamurotov:2013sca}
\bibinfo{author}{\bibfnamefont{F.}~\bibnamefont{Atamurotov}},
  \bibinfo{author}{\bibfnamefont{A.}~\bibnamefont{Abdujabbarov}},
  \bibnamefont{and} \bibinfo{author}{\bibfnamefont{B.}~\bibnamefont{Ahmedov}},
  \bibinfo{journal}{Phys. Rev. D} \textbf{\bibinfo{volume}{88}},
  \bibinfo{pages}{064004} (\bibinfo{year}{2013}).

\bibitem[{\citenamefont{Abdujabbarov et~al.}(2015)\citenamefont{Abdujabbarov,
  Rezzolla, and Ahmedov}}]{Abdujabbarov:2015xqa}
\bibinfo{author}{\bibfnamefont{A.~A.} \bibnamefont{Abdujabbarov}},
  \bibinfo{author}{\bibfnamefont{L.}~\bibnamefont{Rezzolla}}, \bibnamefont{and}
  \bibinfo{author}{\bibfnamefont{B.~J.} \bibnamefont{Ahmedov}},
  \bibinfo{journal}{Mon. Not. Roy. Astron. Soc.}
  \textbf{\bibinfo{volume}{454}}, \bibinfo{pages}{2423} (\bibinfo{year}{2015}),
  \eprint{1503.09054}.

\bibitem[{\citenamefont{Wei et~al.}(2019{\natexlab{a}})\citenamefont{Wei, Zou,
  Liu, and Mann}}]{Wei:2019pjf}
\bibinfo{author}{\bibfnamefont{S.-W.} \bibnamefont{Wei}},
  \bibinfo{author}{\bibfnamefont{Y.-C.} \bibnamefont{Zou}},
  \bibinfo{author}{\bibfnamefont{Y.-X.} \bibnamefont{Liu}}, \bibnamefont{and}
  \bibinfo{author}{\bibfnamefont{R.~B.} \bibnamefont{Mann}},
  \bibinfo{journal}{JCAP} \textbf{\bibinfo{volume}{08}}, \bibinfo{pages}{030}
  (\bibinfo{year}{2019}{\natexlab{a}}), \eprint{1904.07710}.

\bibitem[{\citenamefont{Wei et~al.}(2019{\natexlab{b}})\citenamefont{Wei, Liu,
  and Mann}}]{Wei:2018xks}
\bibinfo{author}{\bibfnamefont{S.-W.} \bibnamefont{Wei}},
  \bibinfo{author}{\bibfnamefont{Y.-X.} \bibnamefont{Liu}}, \bibnamefont{and}
  \bibinfo{author}{\bibfnamefont{R.~B.} \bibnamefont{Mann}},
  \bibinfo{journal}{Phys. Rev. D} \textbf{\bibinfo{volume}{99}},
  \bibinfo{pages}{041303} (\bibinfo{year}{2019}{\natexlab{b}}),
  \eprint{1811.00047}.

\bibitem[{\citenamefont{Abdolrahimi
  et~al.}(2015{\natexlab{a}})\citenamefont{Abdolrahimi, Mann, and
  Tzounis}}]{Abdolrahimi:2015rua}
\bibinfo{author}{\bibfnamefont{S.}~\bibnamefont{Abdolrahimi}},
  \bibinfo{author}{\bibfnamefont{R.~B.} \bibnamefont{Mann}}, \bibnamefont{and}
  \bibinfo{author}{\bibfnamefont{C.}~\bibnamefont{Tzounis}},
  \bibinfo{journal}{Phys. Rev. D} \textbf{\bibinfo{volume}{91}},
  \bibinfo{pages}{084052} (\bibinfo{year}{2015}{\natexlab{a}}),
  \eprint{1502.00073}.

\bibitem[{\citenamefont{Adair et~al.}(2020)\citenamefont{Adair, Bueno, Cano,
  Hennigar, and Mann}}]{Adair:2020vso}
\bibinfo{author}{\bibfnamefont{C.}~\bibnamefont{Adair}},
  \bibinfo{author}{\bibfnamefont{P.}~\bibnamefont{Bueno}},
  \bibinfo{author}{\bibfnamefont{P.~A.} \bibnamefont{Cano}},
  \bibinfo{author}{\bibfnamefont{R.~A.} \bibnamefont{Hennigar}},
  \bibnamefont{and} \bibinfo{author}{\bibfnamefont{R.~B.} \bibnamefont{Mann}},
  \bibinfo{journal}{Phys. Rev. D} \textbf{\bibinfo{volume}{102}},
  \bibinfo{pages}{084001} (\bibinfo{year}{2020}), \eprint{2004.09598}.

\bibitem[{\citenamefont{Abdolrahimi
  et~al.}(2015{\natexlab{b}})\citenamefont{Abdolrahimi, Mann, and
  Tzounis}}]{Abdolrahimi:2015kma}
\bibinfo{author}{\bibfnamefont{S.}~\bibnamefont{Abdolrahimi}},
  \bibinfo{author}{\bibfnamefont{R.~B.} \bibnamefont{Mann}}, \bibnamefont{and}
  \bibinfo{author}{\bibfnamefont{C.}~\bibnamefont{Tzounis}},
  \bibinfo{journal}{Phys. Rev. D} \textbf{\bibinfo{volume}{92}},
  \bibinfo{pages}{124011} (\bibinfo{year}{2015}{\natexlab{b}}),
  \eprint{1510.03530}.

\bibitem[{\citenamefont{Konoplya and Zinhailo}(2020)}]{Konoplya:2020bxa}
\bibinfo{author}{\bibfnamefont{R.~A.} \bibnamefont{Konoplya}} \bibnamefont{and}
  \bibinfo{author}{\bibfnamefont{A.~F.} \bibnamefont{Zinhailo}},
  \bibinfo{journal}{Eur. Phys. J. C} \textbf{\bibinfo{volume}{80}},
  \bibinfo{pages}{1049} (\bibinfo{year}{2020}), \eprint{2003.01188}.

\bibitem[{\citenamefont{Konoplya}(2019)}]{Konoplya:2019sns}
\bibinfo{author}{\bibfnamefont{R.~A.} \bibnamefont{Konoplya}},
  \bibinfo{journal}{Phys. Lett. B} \textbf{\bibinfo{volume}{795}},
  \bibinfo{pages}{1} (\bibinfo{year}{2019}), \eprint{1905.00064}.

\bibitem[{\citenamefont{Konoplya}(2020)}]{Konoplya:2019xmn}
\bibinfo{author}{\bibfnamefont{R.~A.} \bibnamefont{Konoplya}},
  \bibinfo{journal}{Phys. Lett. B} \textbf{\bibinfo{volume}{804}},
  \bibinfo{pages}{135363} (\bibinfo{year}{2020}), \eprint{1912.10582}.

\bibitem[{\citenamefont{Shaikh et~al.}(2019)\citenamefont{Shaikh, Kocherlakota,
  Narayan, and Joshi}}]{Shaikh:2018lcc}
\bibinfo{author}{\bibfnamefont{R.}~\bibnamefont{Shaikh}},
  \bibinfo{author}{\bibfnamefont{P.}~\bibnamefont{Kocherlakota}},
  \bibinfo{author}{\bibfnamefont{R.}~\bibnamefont{Narayan}}, \bibnamefont{and}
  \bibinfo{author}{\bibfnamefont{P.~S.} \bibnamefont{Joshi}},
  \bibinfo{journal}{Mon. Not. Roy. Astron. Soc.}
  \textbf{\bibinfo{volume}{482}}, \bibinfo{pages}{52} (\bibinfo{year}{2019}),
  \eprint{1802.08060}.

\bibitem[{\citenamefont{Shaikh}(2019)}]{Shaikh:2019fpu}
\bibinfo{author}{\bibfnamefont{R.}~\bibnamefont{Shaikh}},
  \bibinfo{journal}{Phys. Rev. D} \textbf{\bibinfo{volume}{100}},
  \bibinfo{pages}{024028} (\bibinfo{year}{2019}), \eprint{1904.08322}.

\bibitem[{\citenamefont{Rahaman et~al.}(2021)\citenamefont{Rahaman, Singh,
  Shaikh, Manna, and Aktar}}]{Rahaman:2021web}
\bibinfo{author}{\bibfnamefont{F.}~\bibnamefont{Rahaman}},
  \bibinfo{author}{\bibfnamefont{K.~N.} \bibnamefont{Singh}},
  \bibinfo{author}{\bibfnamefont{R.}~\bibnamefont{Shaikh}},
  \bibinfo{author}{\bibfnamefont{T.}~\bibnamefont{Manna}}, \bibnamefont{and}
  \bibinfo{author}{\bibfnamefont{S.}~\bibnamefont{Aktar}},
  \bibinfo{journal}{Class. Quant. Grav.} \textbf{\bibinfo{volume}{38}},
  \bibinfo{pages}{215007} (\bibinfo{year}{2021}), \eprint{2108.09930}.

\bibitem[{\citenamefont{Belhaj et~al.}(2020)\citenamefont{Belhaj, Benali,
  El~Balali, El~Moumni, and Ennadifi}}]{Belhaj:2020rdb}
\bibinfo{author}{\bibfnamefont{A.}~\bibnamefont{Belhaj}},
  \bibinfo{author}{\bibfnamefont{M.}~\bibnamefont{Benali}},
  \bibinfo{author}{\bibfnamefont{A.}~\bibnamefont{El~Balali}},
  \bibinfo{author}{\bibfnamefont{H.}~\bibnamefont{El~Moumni}},
  \bibnamefont{and} \bibinfo{author}{\bibfnamefont{S.~E.}
  \bibnamefont{Ennadifi}}, \bibinfo{journal}{Class. Quant. Grav.}
  \textbf{\bibinfo{volume}{37}}, \bibinfo{pages}{215004}
  (\bibinfo{year}{2020}), \eprint{2006.01078}.

\bibitem[{\citenamefont{Belhaj et~al.}(2021{\natexlab{a}})\citenamefont{Belhaj,
  Belmahi, Benali, El~Hadri, El~Moumni, and Torrente-Lujan}}]{Belhaj:2020okh}
\bibinfo{author}{\bibfnamefont{A.}~\bibnamefont{Belhaj}},
  \bibinfo{author}{\bibfnamefont{H.}~\bibnamefont{Belmahi}},
  \bibinfo{author}{\bibfnamefont{M.}~\bibnamefont{Benali}},
  \bibinfo{author}{\bibfnamefont{W.}~\bibnamefont{El~Hadri}},
  \bibinfo{author}{\bibfnamefont{H.}~\bibnamefont{El~Moumni}},
  \bibnamefont{and}
  \bibinfo{author}{\bibfnamefont{E.}~\bibnamefont{Torrente-Lujan}},
  \bibinfo{journal}{Phys. Lett. B} \textbf{\bibinfo{volume}{812}},
  \bibinfo{pages}{136025} (\bibinfo{year}{2021}{\natexlab{a}}),
  \eprint{2008.13478}.

\bibitem[{\citenamefont{Belhaj et~al.}(2021{\natexlab{b}})\citenamefont{Belhaj,
  Belmahi, and Benali}}]{Belhaj:2021rae}
\bibinfo{author}{\bibfnamefont{A.}~\bibnamefont{Belhaj}},
  \bibinfo{author}{\bibfnamefont{H.}~\bibnamefont{Belmahi}}, \bibnamefont{and}
  \bibinfo{author}{\bibfnamefont{M.}~\bibnamefont{Benali}},
  \bibinfo{journal}{Phys. Lett. B} \textbf{\bibinfo{volume}{821}},
  \bibinfo{pages}{136619} (\bibinfo{year}{2021}{\natexlab{b}}),
  \eprint{2110.06771}.

\bibitem[{\citenamefont{Chakhchi et~al.}(2022)\citenamefont{Chakhchi,
  El~Moumni, and Masmar}}]{Chakhchi:2022fls}
\bibinfo{author}{\bibfnamefont{L.}~\bibnamefont{Chakhchi}},
  \bibinfo{author}{\bibfnamefont{H.}~\bibnamefont{El~Moumni}},
  \bibnamefont{and} \bibinfo{author}{\bibfnamefont{K.}~\bibnamefont{Masmar}},
  \bibinfo{journal}{Phys. Rev. D} \textbf{\bibinfo{volume}{105}},
  \bibinfo{pages}{064031} (\bibinfo{year}{2022}).

\bibitem[{\citenamefont{Perlick
  et~al.}(2018{\natexlab{b}})\citenamefont{Perlick, Tsupko, and
  Bisnovatyi-Kogan}}]{Perlick:2018iye}
\bibinfo{author}{\bibfnamefont{V.}~\bibnamefont{Perlick}},
  \bibinfo{author}{\bibfnamefont{O.~Y.} \bibnamefont{Tsupko}},
  \bibnamefont{and} \bibinfo{author}{\bibfnamefont{G.~S.}
  \bibnamefont{Bisnovatyi-Kogan}}, \bibinfo{journal}{Phys. Rev. D}
  \textbf{\bibinfo{volume}{97}}, \bibinfo{pages}{104062}
  (\bibinfo{year}{2018}{\natexlab{b}}), \eprint{1804.04898}.

\bibitem[{\citenamefont{Perlick and Tsupko}(2022)}]{Perlick:2021aok}
\bibinfo{author}{\bibfnamefont{V.}~\bibnamefont{Perlick}} \bibnamefont{and}
  \bibinfo{author}{\bibfnamefont{O.~Y.} \bibnamefont{Tsupko}},
  \bibinfo{journal}{Phys. Rept.} \textbf{\bibinfo{volume}{947}},
  \bibinfo{pages}{1} (\bibinfo{year}{2022}), \eprint{2105.07101}.

\bibitem[{\citenamefont{Pantig and Rodulfo}(2020{\natexlab{a}})}]{Pantig2020b}
\bibinfo{author}{\bibfnamefont{R.~C.} \bibnamefont{Pantig}} \bibnamefont{and}
  \bibinfo{author}{\bibfnamefont{E.~T.} \bibnamefont{Rodulfo}},
  \bibinfo{journal}{Chinese J. Phys.} \textbf{\bibinfo{volume}{68}},
  \bibinfo{pages}{236} (\bibinfo{year}{2020}{\natexlab{a}}).

\bibitem[{\citenamefont{Sokoliuk et~al.}(2022)\citenamefont{Sokoliuk, Praharaj,
  Baransky, and Sahoo}}]{Sokoliuk:2022owk}
\bibinfo{author}{\bibfnamefont{O.}~\bibnamefont{Sokoliuk}},
  \bibinfo{author}{\bibfnamefont{S.}~\bibnamefont{Praharaj}},
  \bibinfo{author}{\bibfnamefont{A.}~\bibnamefont{Baransky}}, \bibnamefont{and}
  \bibinfo{author}{\bibfnamefont{P.~K.} \bibnamefont{Sahoo}},
  \bibinfo{journal}{Astron. Astrophys.} \textbf{\bibinfo{volume}{665}},
  \bibinfo{pages}{A139} (\bibinfo{year}{2022}), \eprint{2207.07193}.

\bibitem[{\citenamefont{Pantig Reggie~C. and
  {\"O}vg{\"u}n}(2022)}]{PANTIG2022168722}
\bibinfo{author}{\bibfnamefont{R.~E.~T.} \bibnamefont{Pantig Reggie~C.},
  \bibfnamefont{Yu~Paul~K.}} \bibnamefont{and}
  \bibinfo{author}{\bibfnamefont{A.}~\bibnamefont{{\"O}vg{\"u}n}},
  \bibinfo{journal}{Annals of Physics} \textbf{\bibinfo{volume}{436}},
  \bibinfo{pages}{168722} (\bibinfo{year}{2022}).

\bibitem[{\citenamefont{Pantig and
  \"Ovg\"un}(2022{\natexlab{a}})}]{Pantig:2022toh}
\bibinfo{author}{\bibfnamefont{R.~C.} \bibnamefont{Pantig}} \bibnamefont{and}
  \bibinfo{author}{\bibfnamefont{A.}~\bibnamefont{\"Ovg\"un}},
  \bibinfo{journal}{Eur. Phys. J. C} \textbf{\bibinfo{volume}{82}},
  \bibinfo{pages}{391} (\bibinfo{year}{2022}{\natexlab{a}}),
  \eprint{2201.03365}.

\bibitem[{\citenamefont{\"Ovg\"un}(2021)}]{Ovgun:2021ttv}
\bibinfo{author}{\bibfnamefont{A.}~\bibnamefont{\"Ovg\"un}},
  \bibinfo{journal}{Phys. Lett. B} \textbf{\bibinfo{volume}{820}},
  \bibinfo{pages}{136517} (\bibinfo{year}{2021}), \eprint{2105.05035}.

\bibitem[{\citenamefont{Okyay and \"Ovg\"un}(2022)}]{Okyay:2021nnh}
\bibinfo{author}{\bibfnamefont{M.}~\bibnamefont{Okyay}} \bibnamefont{and}
  \bibinfo{author}{\bibfnamefont{A.}~\bibnamefont{\"Ovg\"un}},
  \bibinfo{journal}{JCAP} \textbf{\bibinfo{volume}{01}}, \bibinfo{pages}{009}
  (\bibinfo{year}{2022}), \eprint{2108.07766}.

\bibitem[{\citenamefont{\c{C}imdiker et~al.}(2021)\citenamefont{\c{C}imdiker,
  Demir, and \"Ovg\"un}}]{Cimdiker:2021cpz}
\bibinfo{author}{\bibfnamefont{I.}~\bibnamefont{\c{C}imdiker}},
  \bibinfo{author}{\bibfnamefont{D.}~\bibnamefont{Demir}}, \bibnamefont{and}
  \bibinfo{author}{\bibfnamefont{A.}~\bibnamefont{\"Ovg\"un}},
  \bibinfo{journal}{Phys. Dark Univ.} \textbf{\bibinfo{volume}{34}},
  \bibinfo{pages}{100900} (\bibinfo{year}{2021}), \eprint{2110.11904}.

\bibitem[{\citenamefont{Pantig and
  \"Ovg\"un}(2022{\natexlab{b}})}]{Pantig:2022whj}
\bibinfo{author}{\bibfnamefont{R.~C.} \bibnamefont{Pantig}} \bibnamefont{and}
  \bibinfo{author}{\bibfnamefont{A.}~\bibnamefont{\"Ovg\"un}},
  \bibinfo{journal}{JCAP} \textbf{\bibinfo{volume}{08}}, \bibinfo{pages}{056}
  (\bibinfo{year}{2022}{\natexlab{b}}), \eprint{2202.07404}.

\bibitem[{\citenamefont{Kuang and \"Ovg\"un}(2022)}]{Kuang:2022xjp}
\bibinfo{author}{\bibfnamefont{X.-M.} \bibnamefont{Kuang}} \bibnamefont{and}
  \bibinfo{author}{\bibfnamefont{A.}~\bibnamefont{\"Ovg\"un}},
  \bibinfo{journal}{Annals Phys.} \textbf{\bibinfo{volume}{447}},
  \bibinfo{pages}{169147} (\bibinfo{year}{2022}).

\bibitem[{\citenamefont{Uniyal et~al.}(2022)\citenamefont{Uniyal, Pantig, and
  \"Ovg\"un}}]{Uniyal:2022vdu}
\bibinfo{author}{\bibfnamefont{A.}~\bibnamefont{Uniyal}},
  \bibinfo{author}{\bibfnamefont{R.~C.} \bibnamefont{Pantig}},
  \bibnamefont{and} \bibinfo{author}{\bibfnamefont{A.}~\bibnamefont{\"Ovg\"un}}
  (\bibinfo{year}{2022}), \eprint{2205.11072}.

\bibitem[{\citenamefont{\"Ovg\"un et~al.}(2018)\citenamefont{\"Ovg\"un,
  Sakall\i{}, and Saavedra}}]{Ovgun:2018tua}
\bibinfo{author}{\bibfnamefont{A.}~\bibnamefont{\"Ovg\"un}},
  \bibinfo{author}{\bibfnamefont{I.}~\bibnamefont{Sakall\i{}}},
  \bibnamefont{and} \bibinfo{author}{\bibfnamefont{J.}~\bibnamefont{Saavedra}},
  \bibinfo{journal}{JCAP} \textbf{\bibinfo{volume}{10}}, \bibinfo{pages}{041}
  (\bibinfo{year}{2018}), \eprint{1807.00388}.

\bibitem[{\citenamefont{\"Ovg\"un and Sakall\i{}}(2020)}]{Ovgun:2020gjz}
\bibinfo{author}{\bibfnamefont{A.}~\bibnamefont{\"Ovg\"un}} \bibnamefont{and}
  \bibinfo{author}{\bibfnamefont{I.}~\bibnamefont{Sakall\i{}}},
  \bibinfo{journal}{Class. Quant. Grav.} \textbf{\bibinfo{volume}{37}},
  \bibinfo{pages}{225003} (\bibinfo{year}{2020}), \eprint{2005.00982}.

\bibitem[{\citenamefont{Virbhadra and Ellis}(2000)}]{Virbhadra:1999nm}
\bibinfo{author}{\bibfnamefont{K.~S.} \bibnamefont{Virbhadra}}
  \bibnamefont{and} \bibinfo{author}{\bibfnamefont{G.~F.~R.}
  \bibnamefont{Ellis}}, \bibinfo{journal}{Phys. Rev. D}
  \textbf{\bibinfo{volume}{62}}, \bibinfo{pages}{084003}
  (\bibinfo{year}{2000}), \eprint{astro-ph/9904193}.

\bibitem[{\citenamefont{Bozza}(2002)}]{Bozza:2002zj}
\bibinfo{author}{\bibfnamefont{V.}~\bibnamefont{Bozza}},
  \bibinfo{journal}{Phys. Rev. D} \textbf{\bibinfo{volume}{66}},
  \bibinfo{pages}{103001} (\bibinfo{year}{2002}), \eprint{gr-qc/0208075}.

\bibitem[{\citenamefont{Hasse and Perlick}(2002)}]{Hasse:2001by}
\bibinfo{author}{\bibfnamefont{W.}~\bibnamefont{Hasse}} \bibnamefont{and}
  \bibinfo{author}{\bibfnamefont{V.}~\bibnamefont{Perlick}},
  \bibinfo{journal}{Gen. Rel. Grav.} \textbf{\bibinfo{volume}{34}},
  \bibinfo{pages}{415} (\bibinfo{year}{2002}), \eprint{gr-qc/0108002}.

\bibitem[{\citenamefont{Perlick}(2004)}]{Perlick:2003vg}
\bibinfo{author}{\bibfnamefont{V.}~\bibnamefont{Perlick}},
  \bibinfo{journal}{Phys. Rev. D} \textbf{\bibinfo{volume}{69}},
  \bibinfo{pages}{064017} (\bibinfo{year}{2004}), \eprint{gr-qc/0307072}.

\bibitem[{\citenamefont{He et~al.}(2020)\citenamefont{He, Zhou, Feng, Mu, Wang,
  Li, Pan, and Lin}}]{He:2020eah}
\bibinfo{author}{\bibfnamefont{G.}~\bibnamefont{He}},
  \bibinfo{author}{\bibfnamefont{X.}~\bibnamefont{Zhou}},
  \bibinfo{author}{\bibfnamefont{Z.}~\bibnamefont{Feng}},
  \bibinfo{author}{\bibfnamefont{X.}~\bibnamefont{Mu}},
  \bibinfo{author}{\bibfnamefont{H.}~\bibnamefont{Wang}},
  \bibinfo{author}{\bibfnamefont{W.}~\bibnamefont{Li}},
  \bibinfo{author}{\bibfnamefont{C.}~\bibnamefont{Pan}}, \bibnamefont{and}
  \bibinfo{author}{\bibfnamefont{W.}~\bibnamefont{Lin}}, \bibinfo{journal}{Eur.
  Phys. J. C} \textbf{\bibinfo{volume}{80}}, \bibinfo{pages}{835}
  (\bibinfo{year}{2020}).

\bibitem[{\citenamefont{Gibbons and Werner}(2008)}]{Gibbons:2008rj}
\bibinfo{author}{\bibfnamefont{G.~W.} \bibnamefont{Gibbons}} \bibnamefont{and}
  \bibinfo{author}{\bibfnamefont{M.~C.} \bibnamefont{Werner}},
  \bibinfo{journal}{Class. Quant. Grav.} \textbf{\bibinfo{volume}{25}},
  \bibinfo{pages}{235009} (\bibinfo{year}{2008}), \eprint{0807.0854}.

\bibitem[{\citenamefont{Werner}(2012)}]{Werner_2012}
\bibinfo{author}{\bibfnamefont{M.~C.} \bibnamefont{Werner}},
  \bibinfo{journal}{Gen. Relativ. Gravit.} \textbf{\bibinfo{volume}{44}},
  \bibinfo{pages}{3047} (\bibinfo{year}{2012}).

\bibitem[{\citenamefont{\"Ovg\"un}(2018)}]{Ovgun:2018fnk}
\bibinfo{author}{\bibfnamefont{A.}~\bibnamefont{\"Ovg\"un}},
  \bibinfo{journal}{Phys. Rev. D} \textbf{\bibinfo{volume}{98}},
  \bibinfo{pages}{044033} (\bibinfo{year}{2018}), \eprint{1805.06296}.

\bibitem[{\citenamefont{\"Ovg\"un}(2019{\natexlab{a}})}]{Ovgun:2019wej}
\bibinfo{author}{\bibfnamefont{A.}~\bibnamefont{\"Ovg\"un}},
  \bibinfo{journal}{Phys. Rev. D} \textbf{\bibinfo{volume}{99}},
  \bibinfo{pages}{104075} (\bibinfo{year}{2019}{\natexlab{a}}),
  \eprint{1902.04411}.

\bibitem[{\citenamefont{\"Ovg\"un}(2019{\natexlab{b}})}]{Ovgun:2018oxk}
\bibinfo{author}{\bibfnamefont{A.}~\bibnamefont{\"Ovg\"un}},
  \bibinfo{journal}{Universe} \textbf{\bibinfo{volume}{5}},
  \bibinfo{pages}{115} (\bibinfo{year}{2019}{\natexlab{b}}),
  \eprint{1806.05549}.

\bibitem[{\citenamefont{Javed et~al.}(2019{\natexlab{a}})\citenamefont{Javed,
  Abbas, and \"Ovg\"un}}]{Javed:2019kon}
\bibinfo{author}{\bibfnamefont{W.}~\bibnamefont{Javed}},
  \bibinfo{author}{\bibfnamefont{J.}~\bibnamefont{Abbas}}, \bibnamefont{and}
  \bibinfo{author}{\bibfnamefont{A.}~\bibnamefont{\"Ovg\"un}},
  \bibinfo{journal}{Eur. Phys. J. C} \textbf{\bibinfo{volume}{79}},
  \bibinfo{pages}{694} (\bibinfo{year}{2019}{\natexlab{a}}),
  \eprint{1908.09632}.

\bibitem[{\citenamefont{Javed et~al.}(2019{\natexlab{b}})\citenamefont{Javed,
  Abbas, and \"Ovg\"un}}]{Javed:2019rrg}
\bibinfo{author}{\bibfnamefont{W.}~\bibnamefont{Javed}},
  \bibinfo{author}{\bibfnamefont{j.}~\bibnamefont{Abbas}}, \bibnamefont{and}
  \bibinfo{author}{\bibfnamefont{A.}~\bibnamefont{\"Ovg\"un}},
  \bibinfo{journal}{Phys. Rev. D} \textbf{\bibinfo{volume}{100}},
  \bibinfo{pages}{044052} (\bibinfo{year}{2019}{\natexlab{b}}),
  \eprint{1908.05241}.

\bibitem[{\citenamefont{Javed et~al.}(2019{\natexlab{c}})\citenamefont{Javed,
  Babar, and \"Ovg\"un}}]{Javed:2019ynm}
\bibinfo{author}{\bibfnamefont{W.}~\bibnamefont{Javed}},
  \bibinfo{author}{\bibfnamefont{R.}~\bibnamefont{Babar}}, \bibnamefont{and}
  \bibinfo{author}{\bibfnamefont{A.}~\bibnamefont{\"Ovg\"un}},
  \bibinfo{journal}{Phys. Rev. D} \textbf{\bibinfo{volume}{100}},
  \bibinfo{pages}{104032} (\bibinfo{year}{2019}{\natexlab{c}}),
  \eprint{1910.11697}.

\bibitem[{\citenamefont{Javed et~al.}(2020{\natexlab{a}})\citenamefont{Javed,
  Hamza, and \"Ovg\"un}}]{Javed:2020lsg}
\bibinfo{author}{\bibfnamefont{W.}~\bibnamefont{Javed}},
  \bibinfo{author}{\bibfnamefont{A.}~\bibnamefont{Hamza}}, \bibnamefont{and}
  \bibinfo{author}{\bibfnamefont{A.}~\bibnamefont{\"Ovg\"un}},
  \bibinfo{journal}{Phys. Rev. D} \textbf{\bibinfo{volume}{101}},
  \bibinfo{pages}{103521} (\bibinfo{year}{2020}{\natexlab{a}}),
  \eprint{2005.09464}.

\bibitem[{\citenamefont{Javed et~al.}(2019{\natexlab{d}})\citenamefont{Javed,
  Babar, and \"Ovg\"un}}]{Javed:2019qyg}
\bibinfo{author}{\bibfnamefont{W.}~\bibnamefont{Javed}},
  \bibinfo{author}{\bibfnamefont{R.}~\bibnamefont{Babar}}, \bibnamefont{and}
  \bibinfo{author}{\bibfnamefont{A.}~\bibnamefont{\"Ovg\"un}},
  \bibinfo{journal}{Phys. Rev. D} \textbf{\bibinfo{volume}{99}},
  \bibinfo{pages}{084012} (\bibinfo{year}{2019}{\natexlab{d}}),
  \eprint{1903.11657}.

\bibitem[{\citenamefont{\"Ovg\"un et~al.}(2019)\citenamefont{\"Ovg\"un,
  Sakall\i{}, and Saavedra}}]{Ovgun:2018fte}
\bibinfo{author}{\bibfnamefont{A.}~\bibnamefont{\"Ovg\"un}},
  \bibinfo{author}{\bibfnamefont{I.}~\bibnamefont{Sakall\i{}}},
  \bibnamefont{and} \bibinfo{author}{\bibfnamefont{J.}~\bibnamefont{Saavedra}},
  \bibinfo{journal}{Annals Phys.} \textbf{\bibinfo{volume}{411}},
  \bibinfo{pages}{167978} (\bibinfo{year}{2019}), \eprint{1806.06453}.

\bibitem[{\citenamefont{Javed et~al.}(2020{\natexlab{b}})\citenamefont{Javed,
  Abbas, and \"Ovg\"un}}]{Javed:2019jag}
\bibinfo{author}{\bibfnamefont{W.}~\bibnamefont{Javed}},
  \bibinfo{author}{\bibfnamefont{J.}~\bibnamefont{Abbas}}, \bibnamefont{and}
  \bibinfo{author}{\bibfnamefont{A.}~\bibnamefont{\"Ovg\"un}},
  \bibinfo{journal}{Annals Phys.} \textbf{\bibinfo{volume}{418}},
  \bibinfo{pages}{168183} (\bibinfo{year}{2020}{\natexlab{b}}),
  \eprint{2007.16027}.

\bibitem[{\citenamefont{Ishihara
  et~al.}(2016{\natexlab{a}})\citenamefont{Ishihara, Suzuki, Ono, Kitamura, and
  Asada}}]{Ishihara:2016vdc}
\bibinfo{author}{\bibfnamefont{A.}~\bibnamefont{Ishihara}},
  \bibinfo{author}{\bibfnamefont{Y.}~\bibnamefont{Suzuki}},
  \bibinfo{author}{\bibfnamefont{T.}~\bibnamefont{Ono}},
  \bibinfo{author}{\bibfnamefont{T.}~\bibnamefont{Kitamura}}, \bibnamefont{and}
  \bibinfo{author}{\bibfnamefont{H.}~\bibnamefont{Asada}},
  \bibinfo{journal}{Phys. Rev. D} \textbf{\bibinfo{volume}{94}},
  \bibinfo{pages}{084015} (\bibinfo{year}{2016}{\natexlab{a}}),
  \eprint{1604.08308}.

\bibitem[{\citenamefont{Takizawa et~al.}(2020)\citenamefont{Takizawa, Ono, and
  Asada}}]{Takizawa:2020egm}
\bibinfo{author}{\bibfnamefont{K.}~\bibnamefont{Takizawa}},
  \bibinfo{author}{\bibfnamefont{T.}~\bibnamefont{Ono}}, \bibnamefont{and}
  \bibinfo{author}{\bibfnamefont{H.}~\bibnamefont{Asada}},
  \bibinfo{journal}{Phys. Rev. D} \textbf{\bibinfo{volume}{101}},
  \bibinfo{pages}{104032} (\bibinfo{year}{2020}), \eprint{2001.03290}.

\bibitem[{\citenamefont{Ono and Asada}(2019)}]{Ono:2019hkw}
\bibinfo{author}{\bibfnamefont{T.}~\bibnamefont{Ono}} \bibnamefont{and}
  \bibinfo{author}{\bibfnamefont{H.}~\bibnamefont{Asada}},
  \bibinfo{journal}{Universe} \textbf{\bibinfo{volume}{5}},
  \bibinfo{pages}{218} (\bibinfo{year}{2019}), \eprint{1906.02414}.

\bibitem[{\citenamefont{Ishihara et~al.}(2017)\citenamefont{Ishihara, Suzuki,
  Ono, and Asada}}]{Ishihara:2016sfv}
\bibinfo{author}{\bibfnamefont{A.}~\bibnamefont{Ishihara}},
  \bibinfo{author}{\bibfnamefont{Y.}~\bibnamefont{Suzuki}},
  \bibinfo{author}{\bibfnamefont{T.}~\bibnamefont{Ono}}, \bibnamefont{and}
  \bibinfo{author}{\bibfnamefont{H.}~\bibnamefont{Asada}},
  \bibinfo{journal}{Phys. Rev. D} \textbf{\bibinfo{volume}{95}},
  \bibinfo{pages}{044017} (\bibinfo{year}{2017}).

\bibitem[{\citenamefont{Ono et~al.}(2017)\citenamefont{Ono, Ishihara, and
  Asada}}]{Ono:2017pie}
\bibinfo{author}{\bibfnamefont{T.}~\bibnamefont{Ono}},
  \bibinfo{author}{\bibfnamefont{A.}~\bibnamefont{Ishihara}}, \bibnamefont{and}
  \bibinfo{author}{\bibfnamefont{H.}~\bibnamefont{Asada}},
  \bibinfo{journal}{Phys. Rev. D} \textbf{\bibinfo{volume}{96}},
  \bibinfo{pages}{104037} (\bibinfo{year}{2017}).

\bibitem[{\citenamefont{Li and \"Ovg\"un}(2020)}]{Li:2020dln}
\bibinfo{author}{\bibfnamefont{Z.}~\bibnamefont{Li}} \bibnamefont{and}
  \bibinfo{author}{\bibfnamefont{A.}~\bibnamefont{\"Ovg\"un}},
  \bibinfo{journal}{Phys. Rev. D} \textbf{\bibinfo{volume}{101}},
  \bibinfo{pages}{024040} (\bibinfo{year}{2020}).

\bibitem[{\citenamefont{Li et~al.}(2020)\citenamefont{Li, Zhang, and
  \"Ovg\"un}}]{Li:2020wvn}
\bibinfo{author}{\bibfnamefont{Z.}~\bibnamefont{Li}},
  \bibinfo{author}{\bibfnamefont{G.}~\bibnamefont{Zhang}}, \bibnamefont{and}
  \bibinfo{author}{\bibfnamefont{A.}~\bibnamefont{\"Ovg\"un}},
  \bibinfo{journal}{Phys. Rev. D} \textbf{\bibinfo{volume}{101}},
  \bibinfo{pages}{124058} (\bibinfo{year}{2020}).

\bibitem[{\citenamefont{Pantig and
  Rodulfo}(2020{\natexlab{b}})}]{Pantig:2020odu}
\bibinfo{author}{\bibfnamefont{R.~C.} \bibnamefont{Pantig}} \bibnamefont{and}
  \bibinfo{author}{\bibfnamefont{E.~T.} \bibnamefont{Rodulfo}},
  \bibinfo{journal}{Chin. J. Phys.} \textbf{\bibinfo{volume}{66}},
  \bibinfo{pages}{691} (\bibinfo{year}{2020}{\natexlab{b}}).

\bibitem[{\citenamefont{Cembranos and Valcarcel}(2017)}]{Cembranos:2016gdt}
\bibinfo{author}{\bibfnamefont{J.~A.~R.} \bibnamefont{Cembranos}}
  \bibnamefont{and} \bibinfo{author}{\bibfnamefont{J.~G.}
  \bibnamefont{Valcarcel}}, \bibinfo{journal}{JCAP}
  \textbf{\bibinfo{volume}{01}}, \bibinfo{pages}{014} (\bibinfo{year}{2017}),
  \eprint{1608.00062}.

\bibitem[{\citenamefont{Ishihara
  et~al.}(2016{\natexlab{b}})\citenamefont{Ishihara, Suzuki, Ono
  et~al.}}]{Ishihara2016}
\bibinfo{author}{\bibfnamefont{A.}~\bibnamefont{Ishihara}},
  \bibinfo{author}{\bibfnamefont{Y.}~\bibnamefont{Suzuki}},
  \bibinfo{author}{\bibfnamefont{T.}~\bibnamefont{Ono}}, \bibnamefont{et~al.},
  \bibinfo{journal}{Phys. Rev. D} \textbf{\bibinfo{volume}{94}}
  (\bibinfo{year}{2016}{\natexlab{b}}).

\bibitem[{\citenamefont{Perlick et~al.}(2015)\citenamefont{Perlick, Tsupko, and
  Bisnovatyi-Kogan}}]{Perlick_2015}
\bibinfo{author}{\bibfnamefont{V.}~\bibnamefont{Perlick}},
  \bibinfo{author}{\bibfnamefont{O.~Y.} \bibnamefont{Tsupko}},
  \bibnamefont{and} \bibinfo{author}{\bibfnamefont{G.~S.}
  \bibnamefont{Bisnovatyi-Kogan}}, \bibinfo{journal}{Phys. Rev. D}
  \textbf{\bibinfo{volume}{92}}, \bibinfo{pages}{104031}
  (\bibinfo{year}{2015}).

\bibitem[{\citenamefont{Kocherlakota et~al.}(2021)\citenamefont{Kocherlakota,
  Rezzolla, Falcke et~al.}}]{Prashant2021}
\bibinfo{author}{\bibfnamefont{P.}~\bibnamefont{Kocherlakota}},
  \bibinfo{author}{\bibfnamefont{L.}~\bibnamefont{Rezzolla}},
  \bibinfo{author}{\bibfnamefont{H.}~\bibnamefont{Falcke}},
  \bibnamefont{et~al.} (\bibinfo{collaboration}{EHT Collaboration}),
  \bibinfo{journal}{Phys. Rev. D} \textbf{\bibinfo{volume}{103}},
  \bibinfo{pages}{104047} (\bibinfo{year}{2021}).

\bibitem[{\citenamefont{Zaja\v{c}ek et~al.}(2018)\citenamefont{Zaja\v{c}ek,
  Tursunov, Eckart, and Britzen}}]{Zajacek:2018ycb}
\bibinfo{author}{\bibfnamefont{M.}~\bibnamefont{Zaja\v{c}ek}},
  \bibinfo{author}{\bibfnamefont{A.}~\bibnamefont{Tursunov}},
  \bibinfo{author}{\bibfnamefont{A.}~\bibnamefont{Eckart}}, \bibnamefont{and}
  \bibinfo{author}{\bibfnamefont{S.}~\bibnamefont{Britzen}},
  \bibinfo{journal}{Mon. Not. Roy. Astron. Soc.}
  \textbf{\bibinfo{volume}{480}}, \bibinfo{pages}{4408} (\bibinfo{year}{2018}),
  \eprint{1808.07327}.

\bibitem[{\citenamefont{Gao and Zhang}(2004)}]{Gao2004}
\bibinfo{author}{\bibfnamefont{C.~J.} \bibnamefont{Gao}} \bibnamefont{and}
  \bibinfo{author}{\bibfnamefont{S.~N.} \bibnamefont{Zhang}},
  \bibinfo{journal}{Phys. Lett. B} \textbf{\bibinfo{volume}{595}},
  \bibinfo{pages}{28} (\bibinfo{year}{2004}), \eprint{gr-qc/0407045}.

\bibitem[{\citenamefont{Jaroszynski and Kurpiewski}(1997)}]{Jaroszynski:1997bw}
\bibinfo{author}{\bibfnamefont{M.}~\bibnamefont{Jaroszynski}} \bibnamefont{and}
  \bibinfo{author}{\bibfnamefont{A.}~\bibnamefont{Kurpiewski}},
  \bibinfo{journal}{Astron. Astrophys.} \textbf{\bibinfo{volume}{326}},
  \bibinfo{pages}{419} (\bibinfo{year}{1997}), \eprint{astro-ph/9705044}.

\bibitem[{\citenamefont{Bambi}(2012)}]{Bambi:2012tg}
\bibinfo{author}{\bibfnamefont{C.}~\bibnamefont{Bambi}},
  \bibinfo{journal}{Astrophys. J.} \textbf{\bibinfo{volume}{761}},
  \bibinfo{pages}{174} (\bibinfo{year}{2012}), \eprint{1210.5679}.

\bibitem[{\citenamefont{Boonserm et~al.}(2013)\citenamefont{Boonserm,
  Ngampitipan, and Visser}}]{Boonserm:2013dua}
\bibinfo{author}{\bibfnamefont{P.}~\bibnamefont{Boonserm}},
  \bibinfo{author}{\bibfnamefont{T.}~\bibnamefont{Ngampitipan}},
  \bibnamefont{and} \bibinfo{author}{\bibfnamefont{M.}~\bibnamefont{Visser}},
  \bibinfo{journal}{Phys. Rev. D} \textbf{\bibinfo{volume}{88}},
  \bibinfo{pages}{041502} (\bibinfo{year}{2013}), \eprint{1305.1416}.

\bibitem[{\citenamefont{Berti et~al.}(2009)\citenamefont{Berti, Cardoso, and
  Starinets}}]{Berti:2009kk}
\bibinfo{author}{\bibfnamefont{E.}~\bibnamefont{Berti}},
  \bibinfo{author}{\bibfnamefont{V.}~\bibnamefont{Cardoso}}, \bibnamefont{and}
  \bibinfo{author}{\bibfnamefont{A.~O.} \bibnamefont{Starinets}},
  \bibinfo{journal}{Class. Quant. Grav.} \textbf{\bibinfo{volume}{26}},
  \bibinfo{pages}{163001} (\bibinfo{year}{2009}), \eprint{0905.2975}.

\bibitem[{\citenamefont{Konoplya and Zhidenko}(2011)}]{Konoplya:2011qq}
\bibinfo{author}{\bibfnamefont{R.~A.} \bibnamefont{Konoplya}} \bibnamefont{and}
  \bibinfo{author}{\bibfnamefont{A.}~\bibnamefont{Zhidenko}},
  \bibinfo{journal}{Rev. Mod. Phys.} \textbf{\bibinfo{volume}{83}},
  \bibinfo{pages}{793} (\bibinfo{year}{2011}), \eprint{1102.4014}.

\bibitem[{\citenamefont{Kokkotas and Schmidt}(1999)}]{Kokkotas:1999bd}
\bibinfo{author}{\bibfnamefont{K.~D.} \bibnamefont{Kokkotas}} \bibnamefont{and}
  \bibinfo{author}{\bibfnamefont{B.~G.} \bibnamefont{Schmidt}},
  \bibinfo{journal}{Living Rev. Rel.} \textbf{\bibinfo{volume}{2}},
  \bibinfo{pages}{2} (\bibinfo{year}{1999}), \eprint{gr-qc/9909058}.

\bibitem[{\citenamefont{Cardoso et~al.}(2009)\citenamefont{Cardoso, Miranda,
  Berti, Witek, and Zanchin}}]{Cardoso:2008bp}
\bibinfo{author}{\bibfnamefont{V.}~\bibnamefont{Cardoso}},
  \bibinfo{author}{\bibfnamefont{A.~S.} \bibnamefont{Miranda}},
  \bibinfo{author}{\bibfnamefont{E.}~\bibnamefont{Berti}},
  \bibinfo{author}{\bibfnamefont{H.}~\bibnamefont{Witek}}, \bibnamefont{and}
  \bibinfo{author}{\bibfnamefont{V.~T.} \bibnamefont{Zanchin}},
  \bibinfo{journal}{Phys. Rev. D} \textbf{\bibinfo{volume}{79}},
  \bibinfo{pages}{064016} (\bibinfo{year}{2009}), \eprint{0812.1806}.

\bibitem[{\citenamefont{Konoplya and Stuchl\'\i{}k}(2017)}]{Konoplya:2017wot}
\bibinfo{author}{\bibfnamefont{R.~A.} \bibnamefont{Konoplya}} \bibnamefont{and}
  \bibinfo{author}{\bibfnamefont{Z.}~\bibnamefont{Stuchl\'\i{}k}},
  \bibinfo{journal}{Phys. Lett. B} \textbf{\bibinfo{volume}{771}},
  \bibinfo{pages}{597} (\bibinfo{year}{2017}), \eprint{1705.05928}.

\bibitem[{\citenamefont{Glampedakis and Silva}(2019)}]{Glampedakis:2019dqh}
\bibinfo{author}{\bibfnamefont{K.}~\bibnamefont{Glampedakis}} \bibnamefont{and}
  \bibinfo{author}{\bibfnamefont{H.~O.} \bibnamefont{Silva}},
  \bibinfo{journal}{Phys. Rev. D} \textbf{\bibinfo{volume}{100}},
  \bibinfo{pages}{044040} (\bibinfo{year}{2019}), \eprint{1906.05455}.

\bibitem[{\citenamefont{Churilova}(2019)}]{Churilova:2019jqx}
\bibinfo{author}{\bibfnamefont{M.~S.} \bibnamefont{Churilova}},
  \bibinfo{journal}{Eur. Phys. J. C} \textbf{\bibinfo{volume}{79}},
  \bibinfo{pages}{629} (\bibinfo{year}{2019}), \eprint{1905.04536}.

\bibitem[{\citenamefont{Kodama and Ishibashi}(2003)}]{Kodama:2003jz}
\bibinfo{author}{\bibfnamefont{H.}~\bibnamefont{Kodama}} \bibnamefont{and}
  \bibinfo{author}{\bibfnamefont{A.}~\bibnamefont{Ishibashi}},
  \bibinfo{journal}{Prog. Theor. Phys.} \textbf{\bibinfo{volume}{110}},
  \bibinfo{pages}{701} (\bibinfo{year}{2003}), \eprint{hep-th/0305147}.

\bibitem[{\citenamefont{Visser}(1999)}]{Visser:1998ke}
\bibinfo{author}{\bibfnamefont{M.}~\bibnamefont{Visser}},
  \bibinfo{journal}{Phys. Rev. A} \textbf{\bibinfo{volume}{59}},
  \bibinfo{pages}{427} (\bibinfo{year}{1999}), \eprint{quant-ph/9901030}.

\bibitem[{\citenamefont{Boonserm and Visser}(2008)}]{Boonserm:2008zg}
\bibinfo{author}{\bibfnamefont{P.}~\bibnamefont{Boonserm}} \bibnamefont{and}
  \bibinfo{author}{\bibfnamefont{M.}~\bibnamefont{Visser}},
  \bibinfo{journal}{Phys. Rev. D} \textbf{\bibinfo{volume}{78}},
  \bibinfo{pages}{101502} (\bibinfo{year}{2008}), \eprint{0806.2209}.

\bibitem[{\citenamefont{Sanchez}(1978)}]{Sanchez:1977si}
\bibinfo{author}{\bibfnamefont{N.~G.} \bibnamefont{Sanchez}},
  \bibinfo{journal}{Phys. Rev. D} \textbf{\bibinfo{volume}{18}},
  \bibinfo{pages}{1030} (\bibinfo{year}{1978}).

\bibitem[{\citenamefont{Das et~al.}(1997)\citenamefont{Das, Gibbons, and
  Mathur}}]{Das:1996we}
\bibinfo{author}{\bibfnamefont{S.~R.} \bibnamefont{Das}},
  \bibinfo{author}{\bibfnamefont{G.~W.} \bibnamefont{Gibbons}},
  \bibnamefont{and} \bibinfo{author}{\bibfnamefont{S.~D.}
  \bibnamefont{Mathur}}, \bibinfo{journal}{Phys. Rev. Lett.}
  \textbf{\bibinfo{volume}{78}}, \bibinfo{pages}{417} (\bibinfo{year}{1997}),
  \eprint{hep-th/9609052}.

\bibitem[{\citenamefont{Decanini et~al.}(2011)\citenamefont{Decanini,
  Esposito-Farese, and Folacci}}]{Decanini:2011xi}
\bibinfo{author}{\bibfnamefont{Y.}~\bibnamefont{Decanini}},
  \bibinfo{author}{\bibfnamefont{G.}~\bibnamefont{Esposito-Farese}},
  \bibnamefont{and} \bibinfo{author}{\bibfnamefont{A.}~\bibnamefont{Folacci}},
  \bibinfo{journal}{Phys. Rev. D} \textbf{\bibinfo{volume}{83}},
  \bibinfo{pages}{044032} (\bibinfo{year}{2011}), \eprint{1101.0781}.

\bibitem[{\citenamefont{Magalh\~aes et~al.}(2020)\citenamefont{Magalh\~aes,
  Leite, and Crispino}}]{Magalhaes:2020sea}
\bibinfo{author}{\bibfnamefont{R.~B.} \bibnamefont{Magalh\~aes}},
  \bibinfo{author}{\bibfnamefont{L.~C.~S.} \bibnamefont{Leite}},
  \bibnamefont{and} \bibinfo{author}{\bibfnamefont{L.~C.~B.}
  \bibnamefont{Crispino}}, \bibinfo{journal}{Eur. Phys. J. C}
  \textbf{\bibinfo{volume}{80}}, \bibinfo{pages}{386} (\bibinfo{year}{2020}),
  \eprint{2005.04515}.

\bibitem[{\citenamefont{Paula et~al.}(2020)\citenamefont{Paula, Leite, and
  Crispino}}]{Paula:2020yfr}
\bibinfo{author}{\bibfnamefont{M.~A.~A.} \bibnamefont{Paula}},
  \bibinfo{author}{\bibfnamefont{L.~C.~S.} \bibnamefont{Leite}},
  \bibnamefont{and} \bibinfo{author}{\bibfnamefont{L.~C.~B.}
  \bibnamefont{Crispino}}, \bibinfo{journal}{Phys. Rev. D}
  \textbf{\bibinfo{volume}{102}}, \bibinfo{pages}{104033}
  (\bibinfo{year}{2020}), \eprint{2011.08633}.

\bibitem[{\citenamefont{Lima et~al.}(2020)\citenamefont{Lima, Benone, and
  Crispino}}]{Lima:2020seq}
\bibinfo{author}{\bibfnamefont{H.~C.~D.} \bibnamefont{Lima}},
  \bibinfo{author}{\bibfnamefont{C.~L.} \bibnamefont{Benone}},
  \bibnamefont{and} \bibinfo{author}{\bibfnamefont{L.~C.~B.}
  \bibnamefont{Crispino}}, \bibinfo{journal}{Phys. Lett. B}
  \textbf{\bibinfo{volume}{811}}, \bibinfo{pages}{135921}
  (\bibinfo{year}{2020}), \eprint{2011.13446}.

\bibitem[{\citenamefont{Boonserm et~al.}(2019)\citenamefont{Boonserm,
  Ngampitipan, and Wongjun}}]{Boonserm:2019mon}
\bibinfo{author}{\bibfnamefont{P.}~\bibnamefont{Boonserm}},
  \bibinfo{author}{\bibfnamefont{T.}~\bibnamefont{Ngampitipan}},
  \bibnamefont{and} \bibinfo{author}{\bibfnamefont{P.}~\bibnamefont{Wongjun}},
  \bibinfo{journal}{Eur. Phys. J. C} \textbf{\bibinfo{volume}{79}},
  \bibinfo{pages}{330} (\bibinfo{year}{2019}), \eprint{1902.05215}.

\bibitem[{\citenamefont{Xavier et~al.}(2021)\citenamefont{Xavier, Benone, and
  Crispino}}]{Xavier:2021sje}
\bibinfo{author}{\bibfnamefont{S.~V. M. C.~B.} \bibnamefont{Xavier}},
  \bibinfo{author}{\bibfnamefont{C.~L.} \bibnamefont{Benone}},
  \bibnamefont{and} \bibinfo{author}{\bibfnamefont{L.~C.~B.}
  \bibnamefont{Crispino}}, \bibinfo{journal}{Eur. Phys. J. C}
  \textbf{\bibinfo{volume}{81}}, \bibinfo{pages}{1127} (\bibinfo{year}{2021}),
  \eprint{2112.12865}.

\end{thebibliography}
\bibliographystyle{apsrev}

\end{document}